\def\al{ {\it et al}.}
\def\eqn#1{\eqno(#1)}
\def\xrf#1{Fig.~\ref{#1}}

\documentclass[3p]{elsarticle}
\usepackage{epsfig}
\usepackage{amssymb}
\begin{document}
\title{Dependence of magnetic field generation by thermal convection\\
on the rotation rate: a case study}

\author[up,mitpan]{R.~Chertovskih\corref{cor}}
\ead{roman@fc.up.pt, tel.: +351 220 100 858, fax.: +351 220 402 209}
\cortext[cor]{Corresponding author}

\author[up]{S.M.A.~Gama}
\ead{smgama@fc.up.pt}

\author[mitpan,oca]{O.~Podvigina}
\ead{olgap@mitp.ru}

\author[mitpan,oca]{V.~Zheligovsky}
\ead{vlad@mitp.ru}

\address[up]{University of Porto, Faculty of Sciences,
Department of Applied Mathematics,\\Rua Campo Alegre 687, 4169-007 Porto, Portugal}

\address[mitpan]{International Institute of Earthquake Prediction Theory and
Mathematical Geophysics,\\84/32 Profsoyuznaya St, 117997 Moscow, Russian Federation}

\address[oca]{Observatoire de la C\^ote d'Azur,
BP~4229, 06304 Nice Cedex 4, France}

\begin{abstract}

Dependence of magnetic field generation on the rotation rate is explored
by direct numerical simulation of magnetohydrodynamic convective attractors
in a plane layer of conducting fluid with square periodicity cells for
the Taylor number varied from zero to 2000, for which the convective fluid
motion halts (other parameters of the system are fixed). We observe 5 types of
hydrodynamic (amagnetic) attractors: two families of two-dimensional (i.e.
depending on two spatial variables) rolls parallel to sides of periodicity
boxes of different widths and parallel to the diagonal, travelling waves and
three-dimensional ``wavy'' rolls. All types of attractors, except for
one family of rolls, are capable of kinematic magnetic field generation.
We have found 21 distinct nonlinear convective MHD attractors (13 steady
states and 8 periodic regimes) and identified bifurcations in which they
emerge. In addition, we have observed a family of periodic, two-frequency
quasiperiodic and chaotic regimes, as well as an incomplete Feigenbaum period
doubling sequence of bifurcations of a torus followed by a chaotic regime and
subsequently by a torus with 1/3 of the cascade frequency. The system
is highly symmetric. We have found two novel global bifurcations reminiscent of
the SNIC bifurcation, which are only possible in the presence of symmetries.
The universally accepted paradigm, whereby an increase of the rotation rate
below a certain level is beneficial for magnetic field generation, while
a further increase inhibits it (and halts the motion of fluid on continuing
the increase) remains unaltered, but we demonstrate that this ``large-scale''
picture lacks many significant details.

\end{abstract}


\maketitle

\section{Introduction}

Magnetic field of stars, planets and other astrophysical objects is often
attributed to the motion of electrically conducting melted substance in their
interior (Parker, 1979; Priest, 1984; Soward\al, 2005; Hughes\al, 2007;
Dormy and Soward 2007), which is usually sustained by compositional and thermal
convection. Convective flows in a plane layer, ranging from very simple
(Matthews, 1999) to turbulent (Meneguzzi and Pouquet, 1989; Cattaneo\al\ 2003)
ones, are capable of magnetic field generation. In simulations of the dynamo
in the Earth's liquid core (Glatzmaier and Roberts, 1995; Roberts and
Glatzmaier, 2000), a magnetic field of the approximately correct strength was
produced, which had the dipole structure and exhibited reversals similar
to the natural ones. Dynamos in spherical shells were also simulated by Grote
and Busse (2001), Ishinara and Kida (2002), Takahashi and Matsushima (2005) and
other authors.

Magnetohydrodynamic thermal convection is characterised in
the dimensionless form by the Rayleigh number, $R$ (indicating
the magnitude of thermal buoyancy forces); the Prandtl number, $P$ (the ratio
of kinematic viscosity to thermal diffusivity), the magnetic Prandtl number,
$P_m$ (the ratio of kinematic viscosity to magnetic diffusivity), and
the Taylor number, $Ta$ (measuring the speed of rotation). Usually no-slip or
stress-free boundary conditions for the flow velocity and perfectly conducting
or insulating boundaries for magnetic field are considered.

How magnetic field generation by convection depends on parameter values, is not
yet explored in full. Large Rayleigh numbers are beneficial for generation
(the critical $P_m$ decreases monotonically with the increasing $R$) in
spherical shells (Busse, 2000). Podvigina (2006) found a similar dependence
in plane layer dynamos on increasing $R$ over the critical value, but for $R$
over a certain threshold the behaviour of the critical $P_m$ ceased to be
monotonic; Podvigina (2008) also examined the influence of the Prandtl number.

Rotation, which is a common feature of the majority of astrophysical bodies,
can also assist magnetic field generation. Rotation of the Earth is relatively
rapid; it is believed, that geodynamo operates in the outer core in
the magnetostrophic regime, in which the strength of the Coriolis force is
comparable to that of other primary forces - the Lorentz force, pressure and
buoyancy. Rotation is also an important factor in physics of the Solar tachocline
(Christensen-Dalsgaard and Thompson, 2007), where apparently the solar dynamo
is located. Differential rotation gives rise to $\omega-$effect dynamos
(Moffatt, 1978). Although they are slow, as opposed to fast ones, which are
believed to operate in stars, this mechanism is regarded as a key element of
the dynamo of the Sun (Tobias and Weiss, 2007). Boundary layers and shear flows
developing in rapidly rotating fluids are the structures controlling the dynamics
of fluid and planetary dynamo processes (Busse\al, 2007). Therefore, how
the processes of generation are affected by the rate of rotation
is an interesting question for astrophysical applications.

Dependence of magnetic field generation on the rate of rotation was explored
in a number of papers. Meneguzzi and Pouquet (1989) observed that rotation
could be beneficial for nonlinear dynamos -- the critical magnetic Reynolds
number decreased and the ratio of magnetic to kinetic energies significantly
increased, when rotation was on. By contrast, Cattaneo and Hughes (2006) found
that rotation was not a significant factor: they reported ``similar growth rates
and similar saturation levels'' in rotating and non-rotating systems. Only
several runs were presented in each of the two papers, and hence the observations
were inconclusive. Whilst turbulent convective nonlinear dynamos do not require
rotation, near the onset of convection in a plane layer rotation is essential
for both kinematic (Matthews, 1999) and nonlinear dynamo action (Demircan and
Seehafer, 2002) -- in both regimes considered by these authors dynamos failed
in the absence of rotation.

Our paper is devoted to investigation of the dependence of magnetic
field generation on $Ta$. A fluid heated from below in a plane horizontal layer
rotating about the vertical axis is considered in the Boussinesq approximation,
whereby the buoyancy depends linearly on temperature, density variation is
neglected in the mass conservation equation and the flow is
incompressible. Perfectly electrically conducting stress-free horizontal
boundaries of the layer are held at constant temperatures; periodicity
in horizontal directions with the same period $L$ (measured in the units
of the layer depth) is assumed.

We fix all parameter values except for the Taylor number and investigate
numerically attractors of the system for $Ta$ increasing from zero, and
bifurcations delimiting branches of the attractors. Along the branches we trace
average magnetic, $E_m$, and kinetic, $E_k$, energies, as well as their ratio,
$E_m/E_k$, in saturated regimes; the latter quantity is a measure of dynamo
efficiency of prime concern in astrophysics (although we are clearly not in
the astrophysical range of parameter values). The system is also of interest
from the point of view of equivariant bifurcation theory, because it has a
large symmetry group.

Which parameter values are optimal for such an investigation? Naturally,
we need to employ the values, for which the structure of attractors of
the dynamical system and the geometry of their branches in the parameter space
are relatively simple. Behaviour of the dynamical system, which one encounters
in the geo- and astrophysical environments or in experimental dynamos, results
from a large number of bifurcations of the trivial steady state, and it
is prohibitively complex to serve as a starting point for such a study.
These difficult cases can be approached later by continuation
in parameters starting at the regimes determined in the initial study.
Also, in the initial investigation we cannot employ relatively
small values of $P_m$, because for them only vigorous turbulent flows would
yield magnetic dynamos: Many hydrodynamic bifurcations would be required
to bring the convective MHD system into the turbulent state and, consequently,
for such parameter values the system will also be prohibitively complex.
By contrast, for relatively large $P_m$ already laminar convective flows
of simple structure can act as dynamos; moreover, if $P_m$ is not too large,
magnetic field will not be too strong to enable the Lorentz force
to destabilise the convective flows. This suits ideally our purposes.

Following these considerations, we employed the same values as in Podvigina
(2006): $P=1$, $P_m=8$, $R=2300$ and $L=2\sqrt{2}$. The Rayleigh number is not
far from the critical value for the onset of convection; hence, convective
attractors have a simple roll structure, which simplifies a detailed
investigation of the attractors and bifurcations. The $P_m$ value is not far
from its critical values for the kinematic dynamo problem for convective
attractors, hence flows in convective MHD attractors (when dynamos operate)
are qualitatively similar to flows in non-magnetic ones. (Note, that in steady
MHD states flows can be reconstructed from the structure of the generated
magnetic field, see Zheligovsky, 2009a.) The assumed aspect ratio, $L$, is
equal to the spatial period (measured in the units of the layer width) of
the hydrodynamic mode becoming unstable the first when the onset of convective
motion occurs in a non-rotating layer on increasing the Rayleigh number.

Although these values are chosen on the basis of mathematical convenience,
they are not unphysical: $P_m$ are of the order of unity in accretion disks
(von Rekowski\al, 2001); magnetic Prandtl numbers $P_m$ in excess of unity are
typical for intergalactic and interstellar gases (Shukurov and Sokoloff, 2008;
see also Table~1 in Brandenburg and Subramanian, 2005). A plasma experiment
for the study of astrophysical dynamos, where $3\cdot10^{-4}\le P_m\le5$, was
proposed by Spence\al (2009). We also note that typically galaxies are thin
disks (see Ruzmaikin\al, 1988), and hence some features of the problem at hand
(an infinite plane layer with stress-free boundaries) are appropriate to model
galactic dynamos. We stress nevertheless, that we solve the problem in
an abstract setting and do not pretend to simulate any specific physical system.

\section{Statement of the problem}

The system is governed by the Navier-Stokes equation
$${\partial{\bf v}\over\partial t}={\bf v}\times(\nabla\times{\bf v})
+P\Delta{\bf v}+PR\theta{\bf e}_z+P\tau{\bf v}\times{\bf e}_z-\nabla p
-{\bf b}\times(\nabla\times{\bf b}),\eqn{1.a}$$
the magnetic induction equation
$${\partial{\bf b}\over\partial t}=\nabla\times({\bf v}\times{\bf b})
+PP_m^{-1}\Delta{\bf b}\eqn{1.b}$$
for solenoidal fields
$$\nabla\cdot{\bf v}=0,\ \nabla\cdot{\bf b}=0,\eqn{1.c}$$
and the heat transfer equation
$${\partial\theta\over\partial t}=-({\bf v}\cdot\nabla)\theta+v_z
+\Delta\theta.\eqn{1.d}$$
Here $\bf v$ denotes the flow velocity, $\bf b$ the magnetic field,
$\theta$ the difference between the temperature of fluid and the linear
temperature profile, and $\tau=\sqrt{Ta}$ is twice the angular speed
of rotation of the fluid.

The following boundary conditions on the horizontal boundaries:
$${\partial v_x\over\partial z}={\partial v_y\over\partial z}=v_z=0,
\quad\theta=0\quad\hbox{ at }z=0,1\eqn{2.a}$$
$${\partial b_x\over\partial z}={\partial b_y\over\partial z}=b_z=0
\quad\hbox{ at }z=0,1\eqn{2.b}$$
and periodicity in horizontal directions with the same period
$${\bf v}(x,y,z)={\bf v}(x+mL,y+nL,z),\quad\theta(x,y,z)=\theta(x+mL,y+nL,z),\eqn{3.a}$$
$${\bf b}(x,y,z)={\bf b}(x+mL,y+nL,z)\eqn{3.b}$$
$$\forall m,n\in\bf Z$$
are assumed.

\pagebreak
The equations are solved numerically by the standard pseudospectral methods
(Boyd, 2001; Peyret, 2002). Fields are represented as Fourier series
satisfying the boundary conditions (2):
$${\bf v}=\sum_{\bf n}\left(
\begin{tabular}{c}
$\hat{v}_{\bf n}^x\cos(\pi n_3z)$\\
$\hat{v}_{\bf n}^y\cos(\pi n_3z)$\\
$\hat{v}_{\bf n}^z\sin(\pi n_3z)$
\end{tabular}
\right){\rm e}^{{2\pi{\rm i}\over L}(n_1x+n_2y)},$$
$${\bf b}=\sum_{\bf n}\left(
\begin{tabular}{c}
$\hat{b}_{\bf n}^x\cos(\pi n_3z)$\\
$\hat{b}_{\bf n}^y\cos(\pi n_3z)$\\
$\hat{b}_{\bf n}^z\sin(\pi n_3z)$
\end{tabular}
\right){\rm e}^{{2\pi{\rm i}\over L}(n_1x+n_2y)},$$
$$\theta=\sum_{\bf n}\hat{\theta}_{\bf n}\sin(\pi n_3z)
{\rm e}^{{2\pi{\rm i}\over L}(n_1x+n_2y)}.$$
The resolution of $31\times31\times17$ Fourier harmonics has been employed
for computation of hydrodynamic convective attractors, and
\hbox{$63\times63\times33$} in simulations of kinematic dynamos (without
dealiasing in both cases). These simulations have been checked against runs
with the double resolution without dealiasing, and computations of kinematic
magnetic modes also against runs with the resolution of \hbox{$41\times41\times21$}
harmonics with dealiasing. We have been employing the standard method of
dealiasing in computation of all products in (1) (requiring to evaluate
fields on a uniform \hbox{$64\times64\times33$} mesh in the physical space in order
to compute \hbox{$41\times41\times21$} Fourier harmonics of the products).

Simulation of nonlinear hydromagnetic convective regimes has been performed
with the resolution of \hbox{$41\times41\times21$} Fourier harmonics with
dealiasing. We note that computation with a coarser resolution of
\hbox{$31\times31\times16$} Fourier harmonics without dealiasing
in earlier MHD convection simulations by Gertsenshtein\al\ (2007, 2008)
yielded a wrong classification of some regimes (a typical error consisted
of obtaining a periodic regime instead of a convective MHD steady state).
Although from the conservative point of view it thus may be also unsafe
to employ the resolution of \hbox{$41\times41\times21$} harmonics,
the huge amount of runs has forced us to use it in the main bulk
of computations. For this resolution magnetic energy spectrum
decreases\footnote{The ratio of energy in the spherical shell of width one
in the Fourier space with the largest energy content to the energy in the
last considered spherical shell is reported.} by at least four orders of
magnitude and the flow and temperature energy spectra decrease by about
nine orders of magnitude. Some computations of nonlinear hydromagnetic regimes
(including at least one attractor on each MHD branch) have been checked against
runs with the resolution of \hbox{$127\times127\times65$} Fourier harmonics
without dealiasing, the results remaining visibly unaffected (the energies
being reproduced with the accuracy better than 0.001\%).

By $E_k$ and $E_m$ we denote kinetic and magnetic energies, respectively,
averaged over the fluid layer:
$$E_k={1\over{L^2}}\int_0^1\int_0^L\int_0^L{|{\bf v}|^2\over2}\,dx\,dy\,dz,\quad
E_m={1\over{L^2}}\int_0^1\int_0^L\int_0^L{|{\bf b}|^2\over2}\,dx\,dy\,dz.$$
For non-steady convective MHD regimes, time averaging is also performed.

Branches of attractors are traced by continuation in parameter: computations
are done for initial conditions, which are a point (in the phase space)
on the attractor for a ``nearby'' $Ta$. Typically the distances between such
``nearby'' $Ta$ vary between 0.1 and 100. Also, for some $Ta$ runs have been
performed for ``random'' initial conditions, comprised of fields with
pseudorandomly generated Fourier coefficients and an exponentially decaying
spectrum, with either small ($\sim10^{-6}$), or large ($\sim10^2-10^4$) initial
kinetic, magnetic and thermal energies, in order to check whether multiple
attractors coexist for the value of the Taylor number under consideration.
Bifurcations of steady states are located by solving the eigenvalue problem
near the endpoints of the branches and extrapolating the eigenvalues or their
real parts to zero.

\section{Symmetries}

The symmetries of the rotating hydromagnetic convective system under
consideration constitute a subgroup of the group of symmetries of the system
in the absence of rotation. We list them
here for reader's convenience, using the notation of Podvigina (2006).

The symmetry group of the convective system (1) with the boundary conditions
(2), (3) is ${\bf Z}_4\ltimes{\bf T}^2\times{\bf Z}_2$. The group ${\bf Z}_4$
consists of rotations
$$\begin{array}{l}
s_1:\ (x,y,z)\mapsto(y,-x,z),\\
s_2:\ (x,y,z)\mapsto(-x,-y,z),\\
s_3:\ (x,y,z)\mapsto(-y,x,z)
\end{array}$$
and the identity $s_0=e$. ${\bf T}_x$ and ${\bf T}_y$
are the groups of translations in the $x$ and $y$ directions, respectively:
$$\begin{array}{l}
\gamma^x_\alpha:\ (x,y,z)\mapsto(x+\alpha,y,z),\\
\gamma^y_\alpha:\ (x,y,z)\mapsto(x,y+\alpha,z)
\end{array}$$
where $0\le\alpha<L$ ($\gamma^x_L=\gamma^y_L=e$).
${\bf T}_{xy}$ is the group of translations along the diagonal:
$$\gamma^{xy}_\alpha:\ (x,y,z)\mapsto(x+\alpha,y+\alpha,z).$$
The group ${\bf Z}_2$ is generated by reflections about the horizontal midplane:
$$r:\ (x,y,z)\mapsto(x,y,1-z).$$

If magnetic field is present, the group of symmetries of the system
is augmented by the symmetry reversing magnetic field
$$q:\ ({\bf v},\theta,{\bf b})\mapsto({\bf v},\theta,-{\bf b}).$$

The symmetry about a vertical axis, $s_2$, and parity invariance about a centre
(located on the midplane) which is a composition of $s_2$ and $r$, are of
special interest, since in such MHD states the global $\alpha-$effect is zero:
In the presence of the global $\alpha-$effect the system is inherently unstable
to large-scale perturbations, while in its absence the instability, if present,
develops on time scales of a higher order (see Zheligovsky, 2009b).
Although we have not studied spatio-temporal symmetries of non-steady attractors
in detail, we note that we have encountered periodic orbits possessing
the symmetry about a vertical axis or parity invariance with a time shift
(equal to a half of the temporal period). A spatio-temporal symmetry with a time
shift differs from the respective spatial symmetry in that it relates fields
at two time instances, separated by an interval whose length is equal
to the time shift; for instance, the symmetry $s_2$ about a vertical axis
through the point $(a_x,a_y,0)$ with a time shift $T$ of a vector field $\bf f$
is defined by the conditions
\renewcommand{\arraycolsep}{1pt}
$$\begin{array}{rcl}
f_x(a_x-x,a_y-y,z,t)&=&-f_x(a_x+x,a_y+y,z,t+T),\\
f_y(a_x-x,a_y-y,z,t)&=&-f_y(a_x+x,a_y+y,z,t+T),\\
f_z(a_x-x,a_y-y,z,t)&=&f_z(a_x+x,a_y+y,z,t+T).
\end{array}$$
Note that compositions $s_2\gamma^d_\alpha$ and $s_2r\gamma^d_\alpha$ are
the symmetry about a vertical axis and parity invariance, in which the axis
(the centre, respectively) of symmetry is displaced in the direction $d$ by
$-\alpha/2$; changing the order of factors in these compositions to the opposite
one results in the displacement of the axis and the centre, respectively, in the
reverse direction. $s_2$ is used as a generic notation for the symmetry about
a vertical axis with no specific axis indicated; consequently, we do not
distinguish notationally $s_2\gamma^d_\alpha$ from $s_2$ (except for in Table 5).
In what follows, for
construction of figures displaying vector fields which possess the symmetry
$s_2$ we employ a coordinate system with the origin lying on the axis of $s_2$.

\section{Hydrodynamic convective attractors}

\begin{figure}
\centerline{\psfig{file=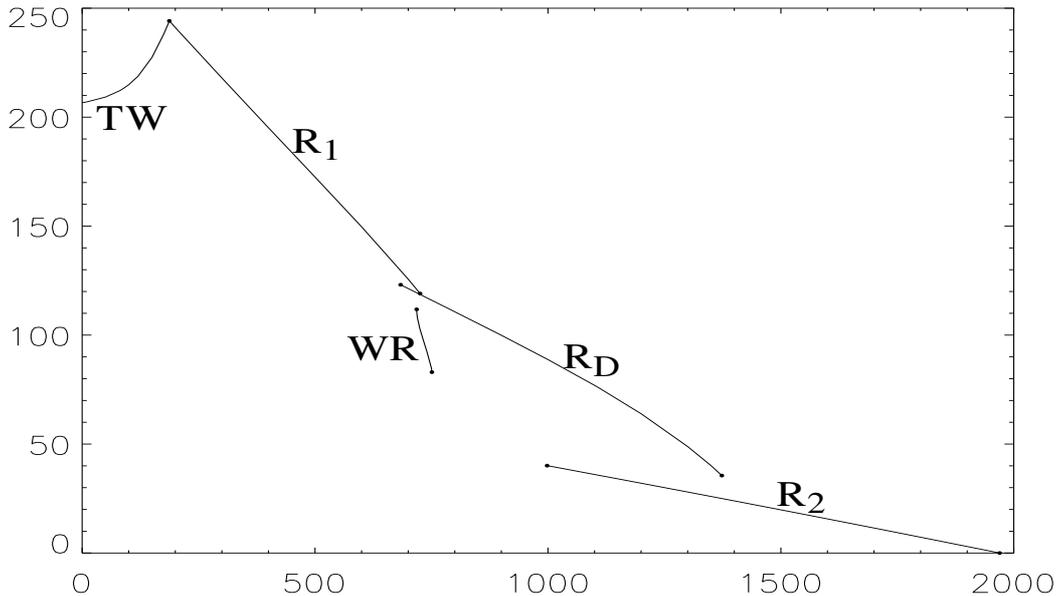,width=14cm,height=8cm}}
\caption{Kinetic energy (vertical axis) of hydrodynamic attractors for
$0<Ta\le 2000$ (horizontal axis). Labelling of attractors is explained
in Section 4 (see also Table~\ref{tab:conv_ev}). Note that R$_1$ does not
bifurcate from $\rm R_D$: energies of the two flows at the point, where R$_1$
becomes unstable, differ by about 0.5,
which is not seen in the scale of the figure.\label{fig:conv_ev}}
\end{figure}

For a small Rayleigh number, i.e. a small temperature difference between the
upper and lower boundaries, the fluid is not moving and heat is transported by
thermal diffusion only. When $R$ exceeds the critical value (which increases
with $Ta$), a fluid motion sets in. For $P=1$ considered here, the instability
is monotonic, the primary motion has the form of rolls. The wave number (in the
horizontal direction) of the most unstable mode monotonically increases with
$Ta$. If periodicity in horizontal directions is imposed, the horizontal
wave number must be compatible with the periodicity cell size and it can take
only discrete values, and for an increasing $Ta$ the wave numbers of the
emerging rolls constitute a piecewise constant growing function. At very high
rotation rates an array of very thin rolls emerges (Bassom and Zhang, 1994).

Five types of attractors of hydrodynamic convection (governed by the system
(1)--(3) for ${\bf b}=0$), that we have found in computations, are shown on
the bifurcation diagram \xrf{fig:conv_ev} (see also Table~\ref{tab:conv_ev}).
Since the
Rayleigh number is not far from the critical value for the onset of convection,
the flows are of a simple spatial structure (see \xrf{fig:conv_isolv}).

\begin{table}[t]
\caption{Attractors found numerically for hydrodynamic (${\bf b}=0$) convection.
Column 2 presents the interval, where existence of branches of attractors is
confirmed numerically, column 3 presents the symmetry group for which
an attractor is pointwise invariant, column 4 generators of the group (for
an appropriately chosen Cartesian coordinate system); if a group is a product
of several subgroups, generators of the subgroups are separated by semicolons.
Column 5 presents locations ($Ta$) and types of bifurcations in
which a steady state becomes unstable (S denotes a steady-state bifurcation,
H a Hopf bifurcation), column 6 the dimension of the respective center eigenspace,
column 7 the action of the symmetry group on the null eigenspace, and the last
column elements of the group which act trivially. $\bf 1$ denotes a trivial
group of symmetries.\label{tab:conv_ev}}
\begin{center}
\begin{tabular}{|c|c|c|c|c|c|c|c|}\hline
Type&Interval of&Symmetry&Generators&Bifurcation&D&Action&Kernel\\
&existence ($Ta$)&group&&&&&\\\hline
TW&[1,\,187]&${\bf Z}_2$&$r\gamma^x_{L/2}$&&&&\\\hline
R$_1$&[188,\,725]&${\bf D}_2\ltimes\bf T$&$r\gamma^x_{L/2}$, $s_2$;
$\gamma^y$&187.19, H&4&{\bf O}(2)&$r\gamma^x_{L/2}$\\
&&&&725.45, S&2&{\bf O}(2)&$r\gamma^x_{L/2}$\\\hline
WR&[719,\,750]&${\bf D}_2$&$r\gamma^x_{L/2}$, $s_2$&
718.16, S&1&$\bf 1$&$r\gamma^x_{L/2}$, $s_2$\\
&&&&750.82, S&1&${\bf Z}_2$&$s_2r\gamma^x_{L/2}$\\\hline
$\rm R_D$&[684,\,1373]&${\bf D}_2\ltimes\bf T$&
$r\gamma^x_{L/2}$, $s_2$; $\gamma^{xy}$&683.64, S&2&{\bf O}(2)&
$r\gamma^x_{L/2}$\\
&&&&1373.32, S&2&{\bf O}(2)&$r\gamma^{x,-y}_{L/4}$\\\hline
R$_2$&[999,\,1969]&${\bf D}_4\ltimes\bf T$&$r\gamma^x_{L/4}$, $s_2$;
$\gamma^y$&998.18, S&2&{\bf O}(2)&$r\gamma^{xy}_{L/4}$\\\hline
\end{tabular}\end{center}\end{table}

Three types of attractors are steady rolls of different
spatial periods. In agreement with the linear stability theory, their wave
numbers increase with $Ta$ (remaining the same within each branch).
The rolls obtained in computations are parallel to a side of the periodicity
cell with either the same period as the periodicity cell size, or a half of it;
or they are parallel to the diagonal of the cell. The branches are labelled
R$_1$, R$_2$ and $\rm R_D$, respectively.

In a non-rotating layer, bifurcations of rolls are well documented. However,
these results cannot be directly applied to convection in a rotating layer,
even if the rotation rate is small -- this can only be done after unfolding of
bifurcations of rolls is performed, i.e. after it is determined how
the bifurcations alter, if small terms breaking the reflection symmetry
are added. This is an interesting problem in its own right, but it is beyond
the scope of the paper.

On increasing $R$, in a non-rotating layer rolls can bifurcate to travelling
waves in a Hopf bifurcation for
$P\stackrel{\raisebox{-1mm}{$\textstyle <$}}{\raisebox{-1mm}{$\sim$}}1$ (Getling, 1998).
This bifurcation, with the {\bf O}(2) symmetry group, was studied in detail in
Golubitsky\al\ (1988). In it, branches of standing and travelling waves emerge;
if both branches bifurcate supercritically, one of them is stable. A travelling
wave, TW, is time-periodic in a coordinate frame at rest and it is steady
in a frame moving with the speed of the pattern. In a non-rotating convection
such bifurcation from R$_1$ to a stable travelling wave takes place
near $R\approx1755$ (Podvigina, 2006). The instability causes
a sinusoidal bending of rolls, the pattern travelling along the axis of a roll.

\begin{figure}
\centerline{
\psfig{file=fig2/tw-ta50.ps,width=45mm,clip=}
\psfig{file=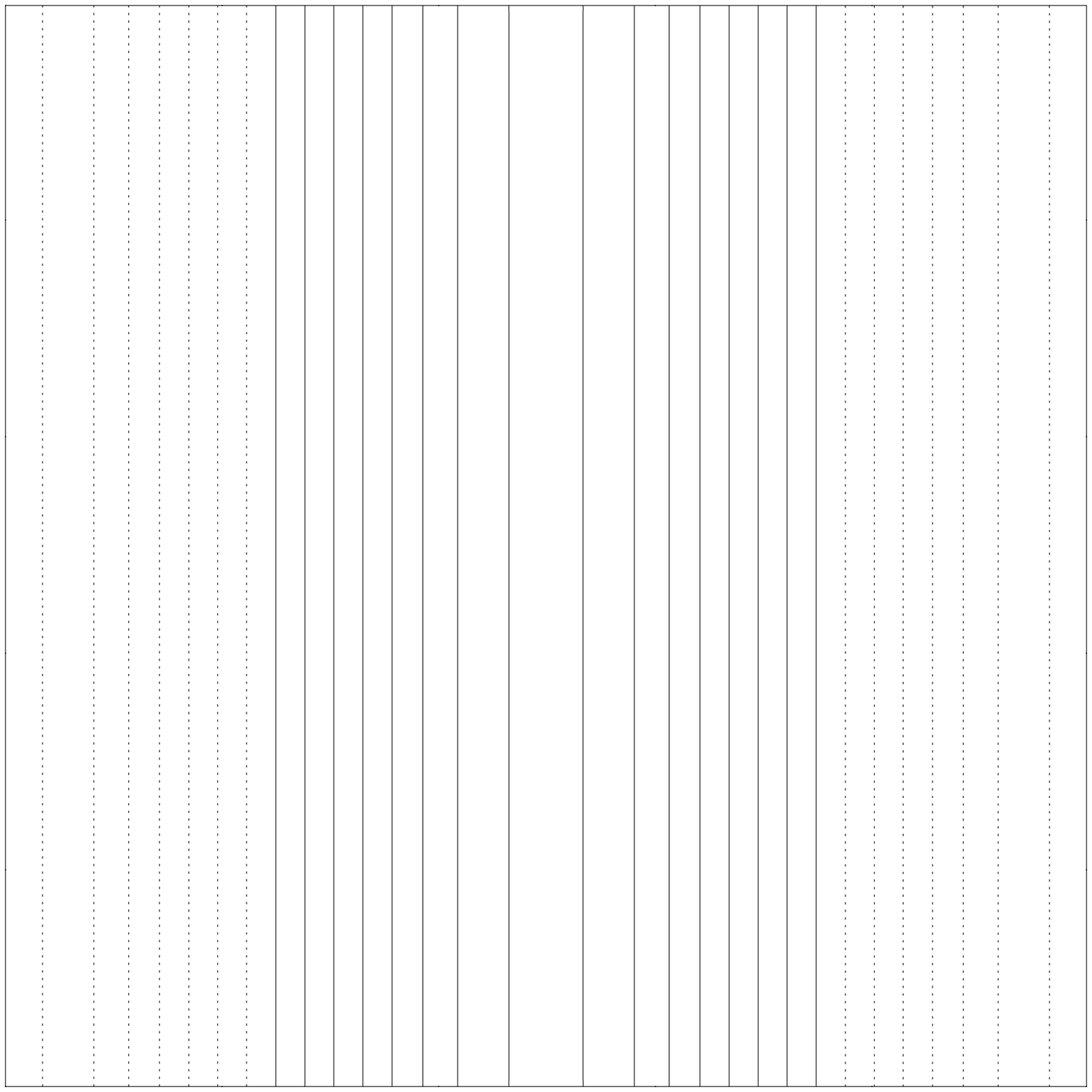,width=45mm,clip=}
\psfig{file=fig2/wr-ta720.ps,width=45mm,clip=}}

\vspace*{1mm}
\hspace{32mm}(a)\hspace{42mm}(b)\hspace{42mm}(c)

\vspace*{2mm}
\centerline{
\psfig{file=fig2/rd-ta685.ps,width=45mm,clip=}
\psfig{file=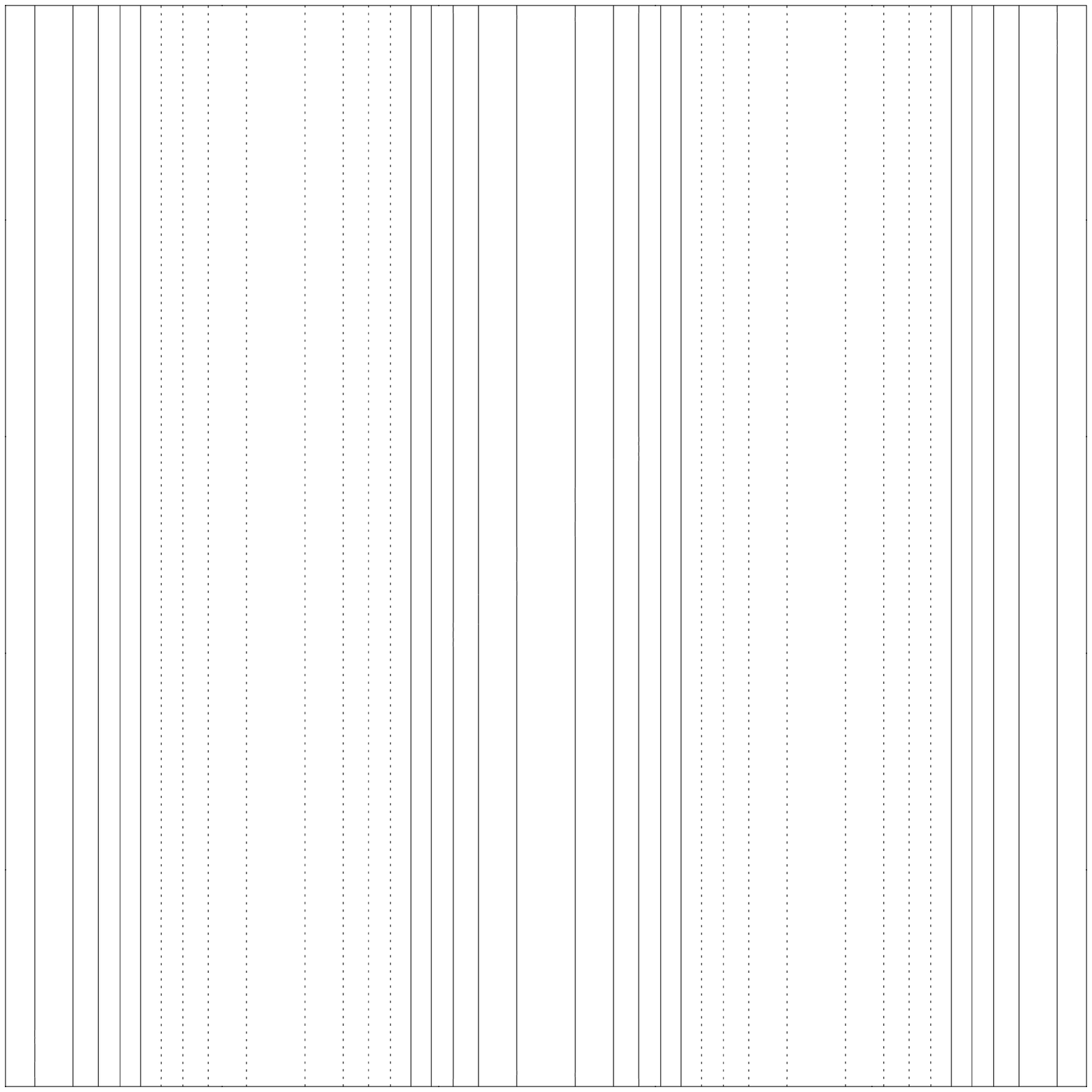,width=45mm,clip=}}

\vspace*{1mm}
\hspace{55mm}(d)\hspace{42mm}(e)
\caption{Isolines (step 2) of $v_z$ on the horizontal midplane $z=1/2$ for
hydrodynamic attractors TW, $Ta=50$ (a); R$_1$, $Ta=500$ (b); WR, $Ta=720$ (c);
$\rm R_D$, $Ta=685$ (d); R$_2$, $Ta=1100$ (e). Solid lines indicate non-negative,
dashed lines negative values. $x$: horizontal axis, $y$: vertical
axis.\label{fig:conv_isolv}}\end{figure}

\begin{figure}
\centerline{\psfig{file=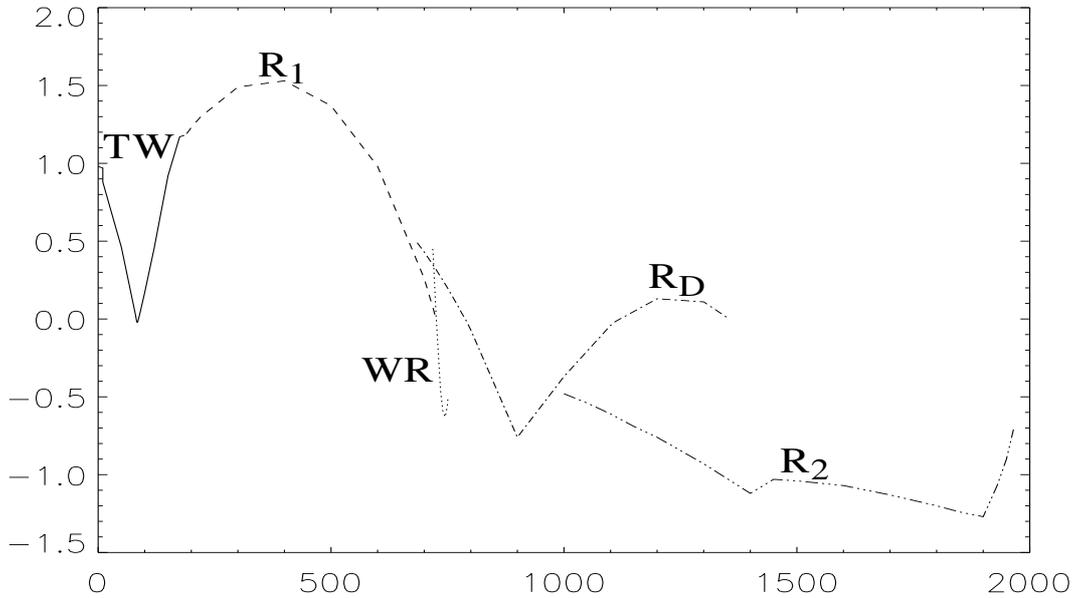,width=14cm,height=8cm}}
\caption{Growth rates (vertical axis) of magnetic field generated
kinematically by hydrodynamic attractors as a function of $Ta$
(horizontal axis).\label{fig:conv_lambda}}\end{figure}

We also observe the TW: On decreasing $Ta$, R$_1$ loses stability and the
stable travelling wave emerges in a similar supercritical Hopf bifurcation
with the {\bf O}(2) symmetry group. Due to rotation, no reflection symmetries
are present, enabling the travelling wave to drift in the direction
perpendicular to axes of rolls. All other bifurcations of rolls of the three
types are subcritical pitchfork bifurcations, with no stable branches emerging.
On increasing $Ta$, the branch R$_2$ terminates at $Ta=1969.67$ (this number is
in a good agreement with the value of the Taylor number, obtained from the
Chandrasekhar's (1961) formula for the critical value of the Rayleigh number
for the onset of convection in a layer with free boundaries) on the trivial
steady state; no non-trivial convective regimes exist for higher rotation rates.

Another attractor, distinct from rolls, is wavy rolls, WR. In a non-rotating
layer emergence of WR from R$_1$ is a consequence of the
$1:\sqrt{2}$ mode interaction (Podvigina and Ashwin, 2007). Similar arguments
show that in a rotating layer WR can also bifurcate from rolls.
On decreasing $Ta$, WR disappear in a saddle-node bifurcation -- the
branch turns back and becomes unstable. Apparently, it bifurcates from R$_1$
at $Ta=725.45$: WR and R$_1$ have similar shapes and close energy values,
and by the equivariant branching lemma (Golubitsky\al, 1988) the action of the
symmetry group of R$_1$ (see Table~\ref{tab:conv_ev}) implies that a branch
with the symmetries of WR bifurcates from R$_1$.

\section{Magnetic field generation}

We have investigated whether the flows that are convective hydrodynamic
attractors are capable of kinematic magnetic field generation. For steady
hydrodynamic states we have been computing (applying the algorithm
of Zheligovsky, 1993a,b) dominant eigenvalues of the magnetic induction
operator. For travelling waves, equation (2.b) in the co-moving reference
frame yields a similar eigenvalue problem (see Podvigina, 2006).

\begin{table}
\caption{Dominant kinematic magnetic modes of the hydrodynamic convective
attractors. Column 2 presents generators for symmetry groups of the
hydrodynamic attractors,
column 3 the interval of $Ta$ where a magnetic mode is dominant, column 4
dimension of the eigenspace associated with the dominant eigenvalue,
column 5 the action of the symmetry group on the eigenspace,
column 6 the symmetries which act trivially,
the last column the eigenvalue $\lambda$ of the magnetic induction operator
for which the maximal growth rate is attained in the interval of $Ta$
specified in column 3.\label{tab:conv_lambda}}
\begin{center}
\begin{tabular}{|c|c|c|c|c|c|c|}\hline
Flow&Generators&$Ta$&D&Action&Kernel&$\lambda$\\\hline
TW&$r\gamma^x_{L/2}$&[1,\,84]&2&${\bf Z}_2$&$qr\gamma^x_{L/2}$&$0.97\pm28{\rm i}$\\
&&[85,\,187]&2&${\bf Z}_2$&$r\gamma^x_{L/2}$&$1.18\pm19{\rm i}$\\\hline
R$_1$&$s_2$, $r\gamma^x_{L/2}$; $\gamma^y$&[188,\,725]&
2&{\bf O}(2)& $r\gamma^x_{L/2}$, $q\gamma^y_{L/2}$&1.53\\\hline
WR&$s_2$, $r\gamma^x_{L/2}$&[719,\,750]&1&${\bf Z}_2$&$qs_2$, $r\gamma^x_{L/2}$&0.45\\\hline
$\rm R_D$&$s_2$, $r\gamma^x_{L/2}$; $\gamma^{xy}$&[684,\,925]&
2&{\bf O}(2)&$rq\gamma^x_{L/2}$, $q\gamma^{xy}_{L/4}$&0.49\\
&&[930,\,1373]&
2&{\bf O}(2)&$rq\gamma^x_{L/2}$, $q\gamma^{xy}_{L/2}$&0.13\\\hline
R$_2$&$s_2$, $r\gamma^x_{L/4}$; $\gamma^y$&[999,\,1420]&
2&{\bf O}(2)&$r\gamma^x_{L/4}$, $q\gamma^y_{L/2}$&-0.48\\
&&[1430,\,1969]&
2&{\bf O}(2)&$rq\gamma^{xy}_{L/4}$, $q\gamma^y_{L/2}$&-0.62\\\hline
\end{tabular}\end{center}\end{table}

All the hydrodynamic attractors have non-trivial symmetry groups and
magnetic modes can be classified in the terms of the action of their symmetries.
Symmetry groups of rolls are continuous, they include shifts along the axis
of rolls, and bifurcating modes can have arbitrary periods in this direction.
We restrict our attention to magnetic modes that have the periodicity of
the hydrodynamic convective attractors. The computed growth rates are
shown on \xrf{fig:conv_lambda}, and symmetries of dominant modes are presented
in Table~\ref{tab:conv_lambda}. It turns out that all the attractors, except
for R$_2$, can generate magnetic field. Note, that since TW bifurcates
from R$_1$, the steady magnetic mode of R$_1$ becomes a time-periodic mode
of TW, whose frequency at the point of bifurcation coincides with the one of TW
and the eigenvalue becomes the real part of the TW magnetic eigenvalue.

\begin{figure}
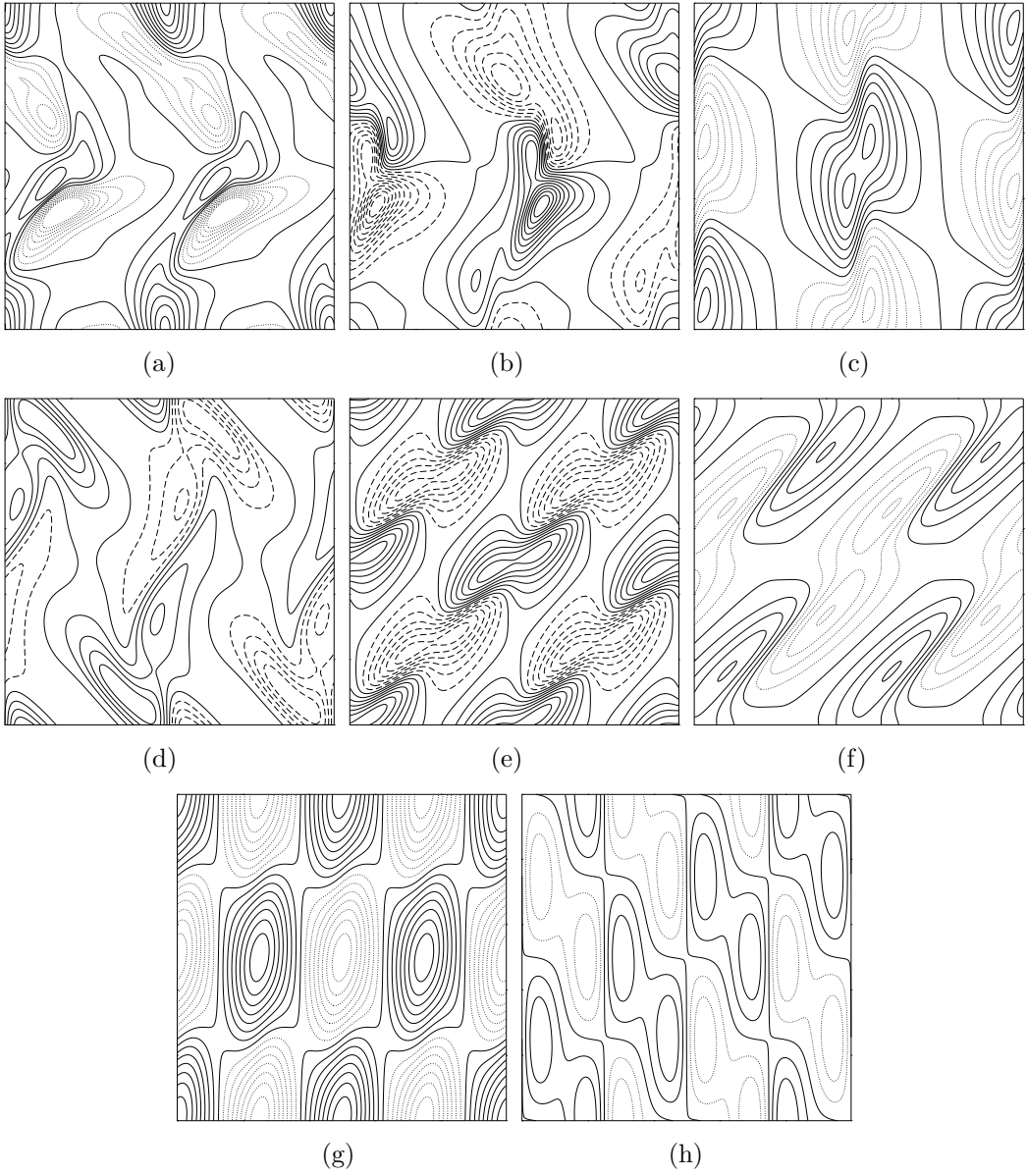

\centerline{
\psfig{file=fig4/tw-ta50.ps,width=45mm,clip=}
\psfig{file=fig4/tw-ta150.ps,width=45mm,clip=}
\psfig{file=fig4/r1-ta500.ps,width=45mm,clip=}}

\vspace*{1mm}
\hspace{33mm}(a)\hspace{42mm}(b)\hspace{42mm}(c)

\vspace*{2mm}
\centerline{
\psfig{file=fig4/wr-ta720.ps,width=45mm,clip=}
\psfig{file=fig4/rd-ta685.ps,width=45mm,clip=}
\psfig{file=fig4/rd-ta1200.ps,width=45mm,clip=}}

\vspace*{1mm}
\hspace{33mm}(d)\hspace{42mm}(e)\hspace{42mm}(f)

\vspace*{2mm}
\centerline{
\psfig{file=fig4/r2-ta1100.ps,width=45mm,clip=}
\psfig{file=fig4/r2-ta1700.ps,width=45mm,clip=}}

\vspace*{1mm}
\hspace{57mm}(g)\hspace{42mm}(h)
\caption{Isolines step 0.2 of $b_z$ on the horizontal midplane $z=1/2$ for
dominant magnetic modes: $Ta=50$, TW (a); $Ta=150$, TW (b); $Ta=500$, R$_1$ (c);
$Ta=720$, WR (d); $Ta=685$, $\rm R_D$ (e); $Ta=1200$, $\rm R_D$ (f);
$Ta=1100$, R$_2$ (g); $Ta=1700$, R$_2$ (h). Solid lines indicate non-negative,
dashed lines negative values. $x$: horizontal axis, $y$: vertical axis.
\label{fig:conv_isolb}}\end{figure}

\begin{figure}
\centerline{
\psfig{file=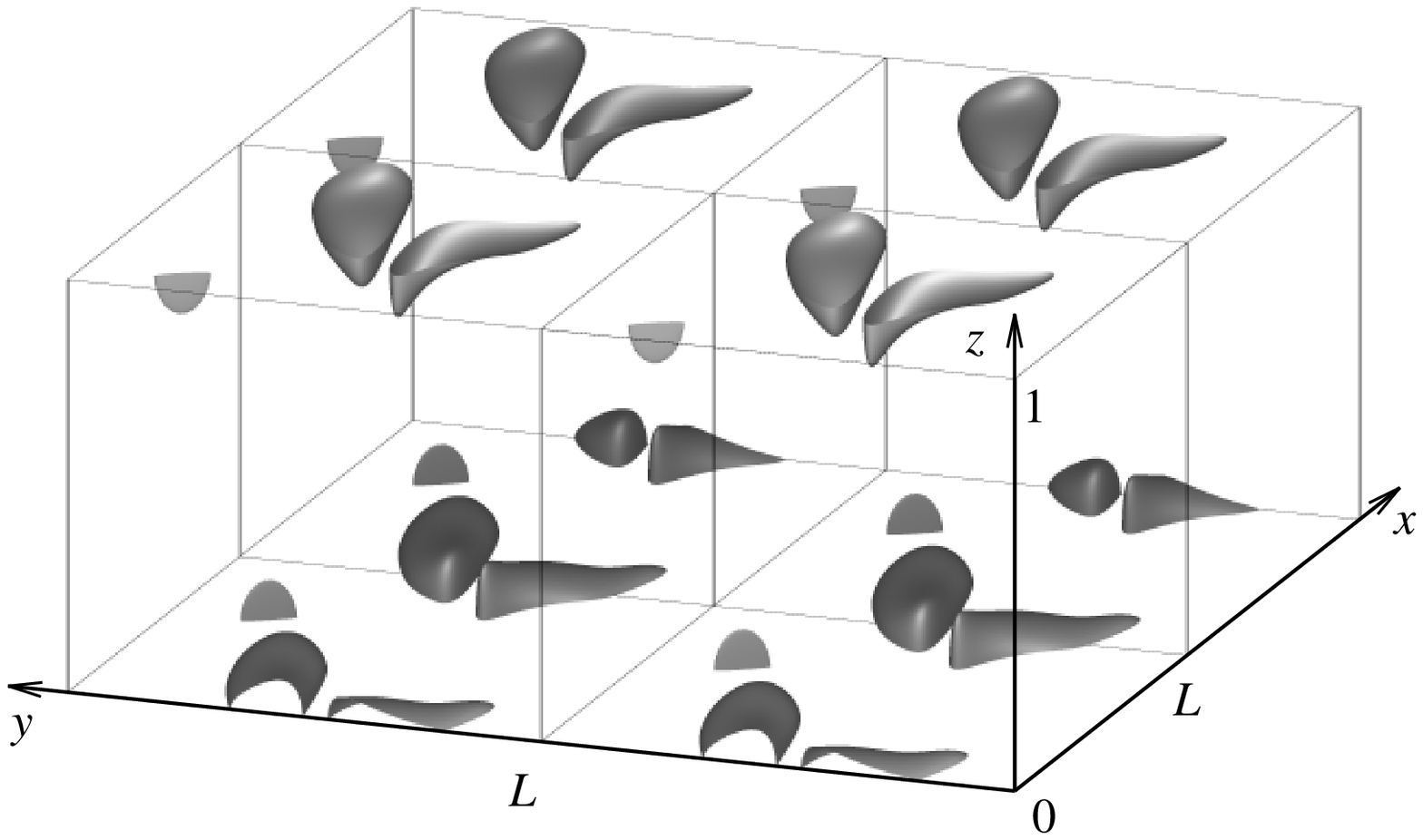,width=7cm,clip=}
\psfig{file=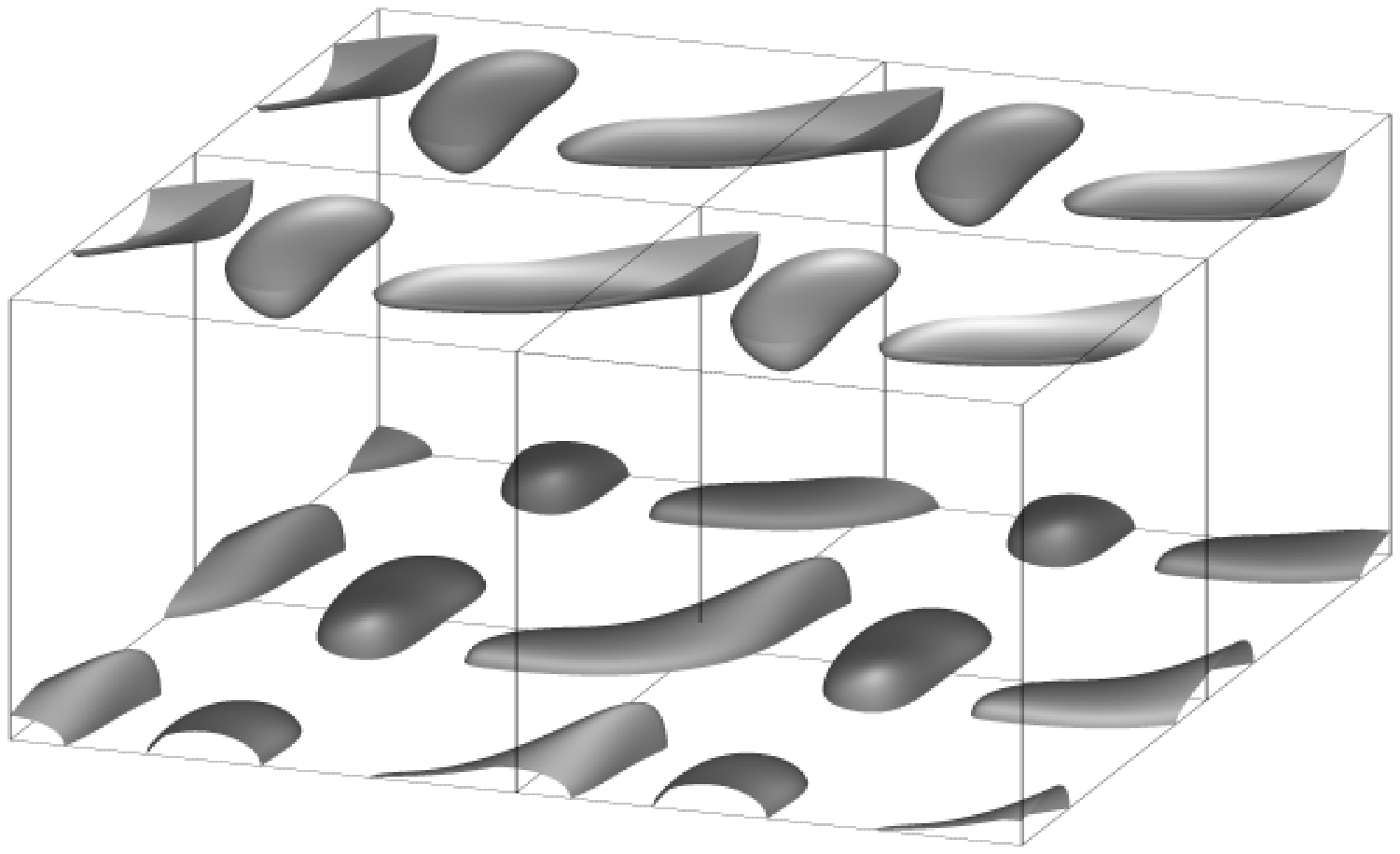,width=7cm,clip=}}
\hspace{42mm}(a)\hspace{66mm}(b)

\vspace*{2mm}
\centerline{
\psfig{file=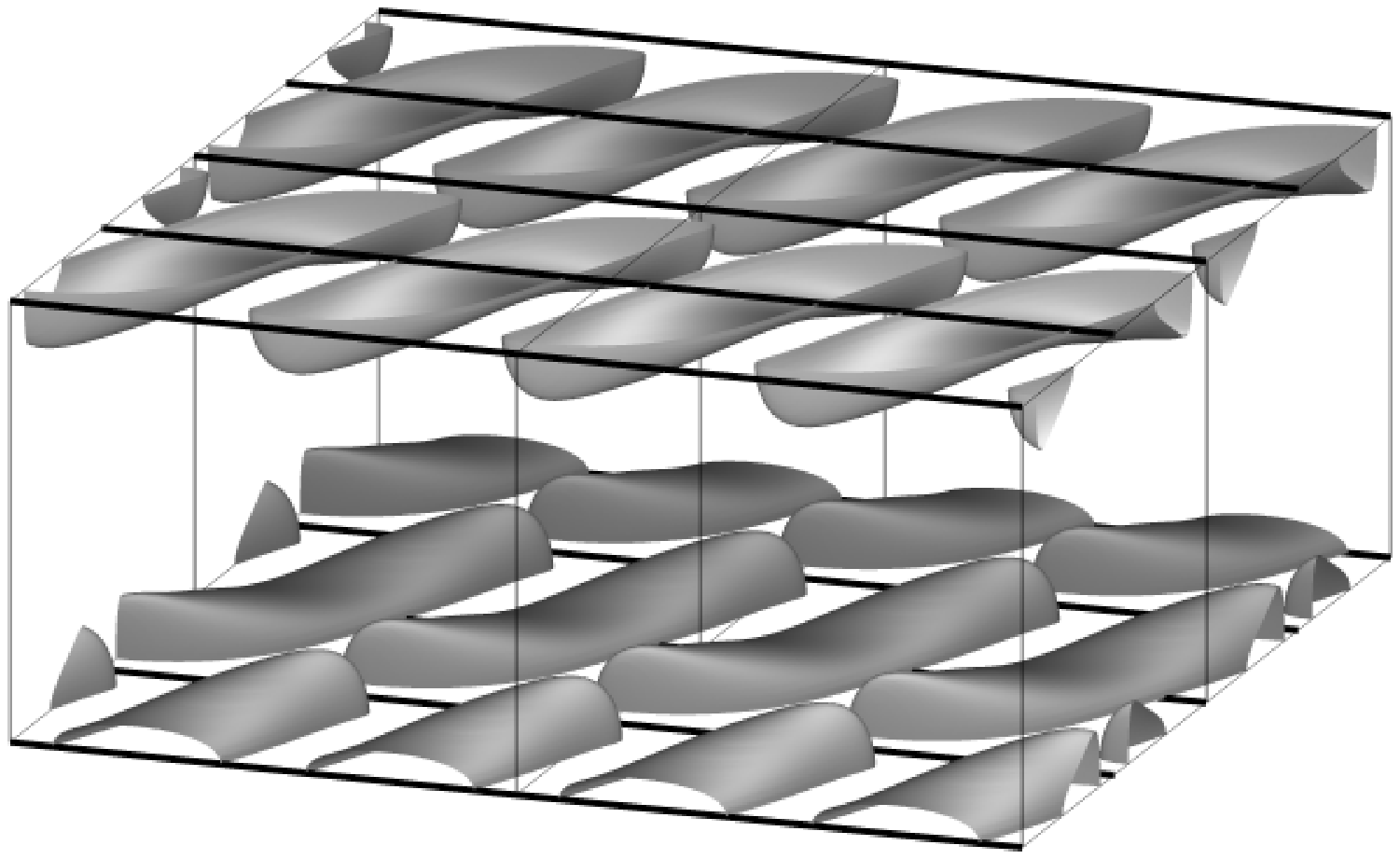,width=7cm,clip=}
\psfig{file=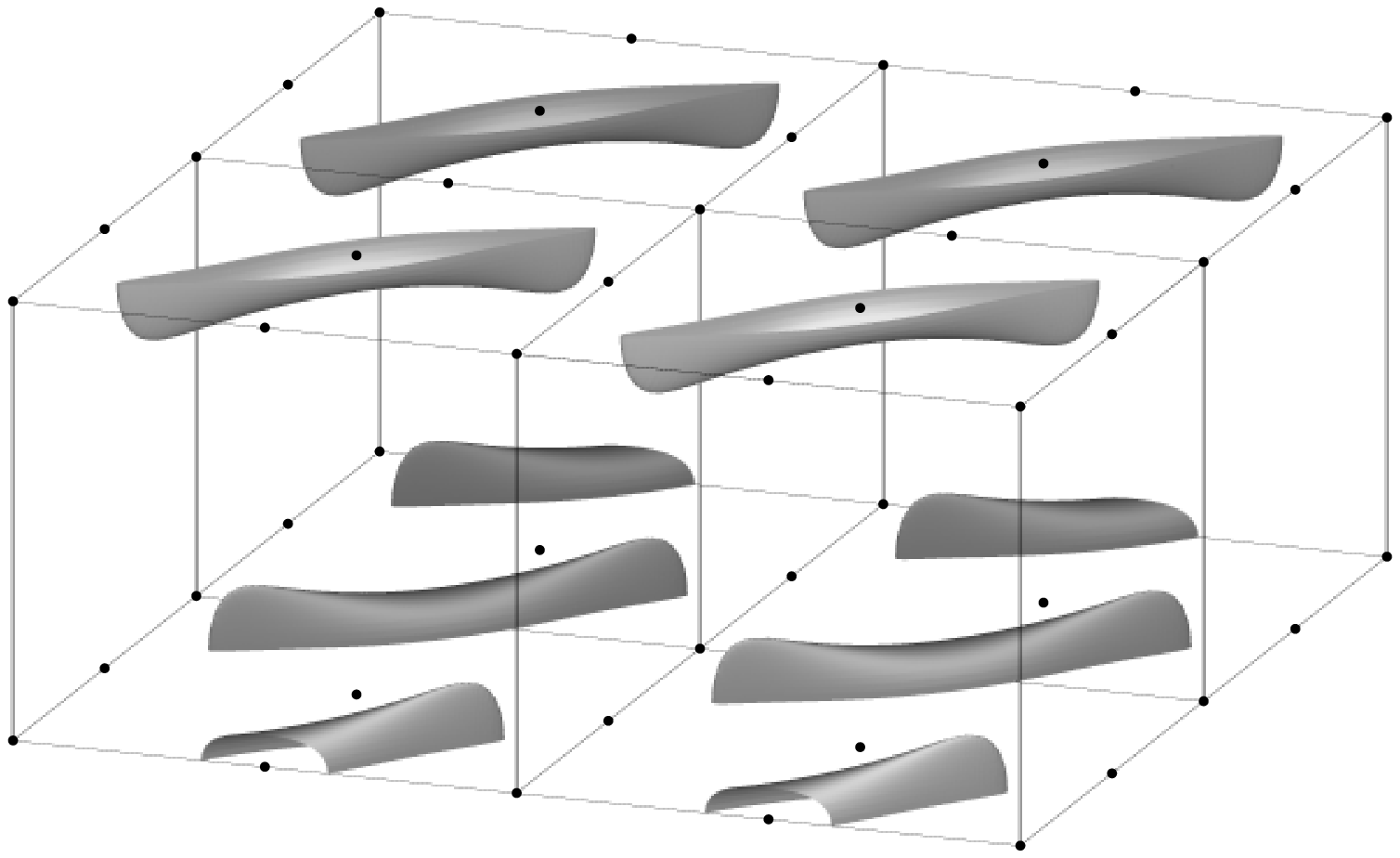,width=7cm,clip=}}
\hspace{42mm}(c)\hspace{66mm}(d)

\vspace*{2mm}
\centerline{
\psfig{file=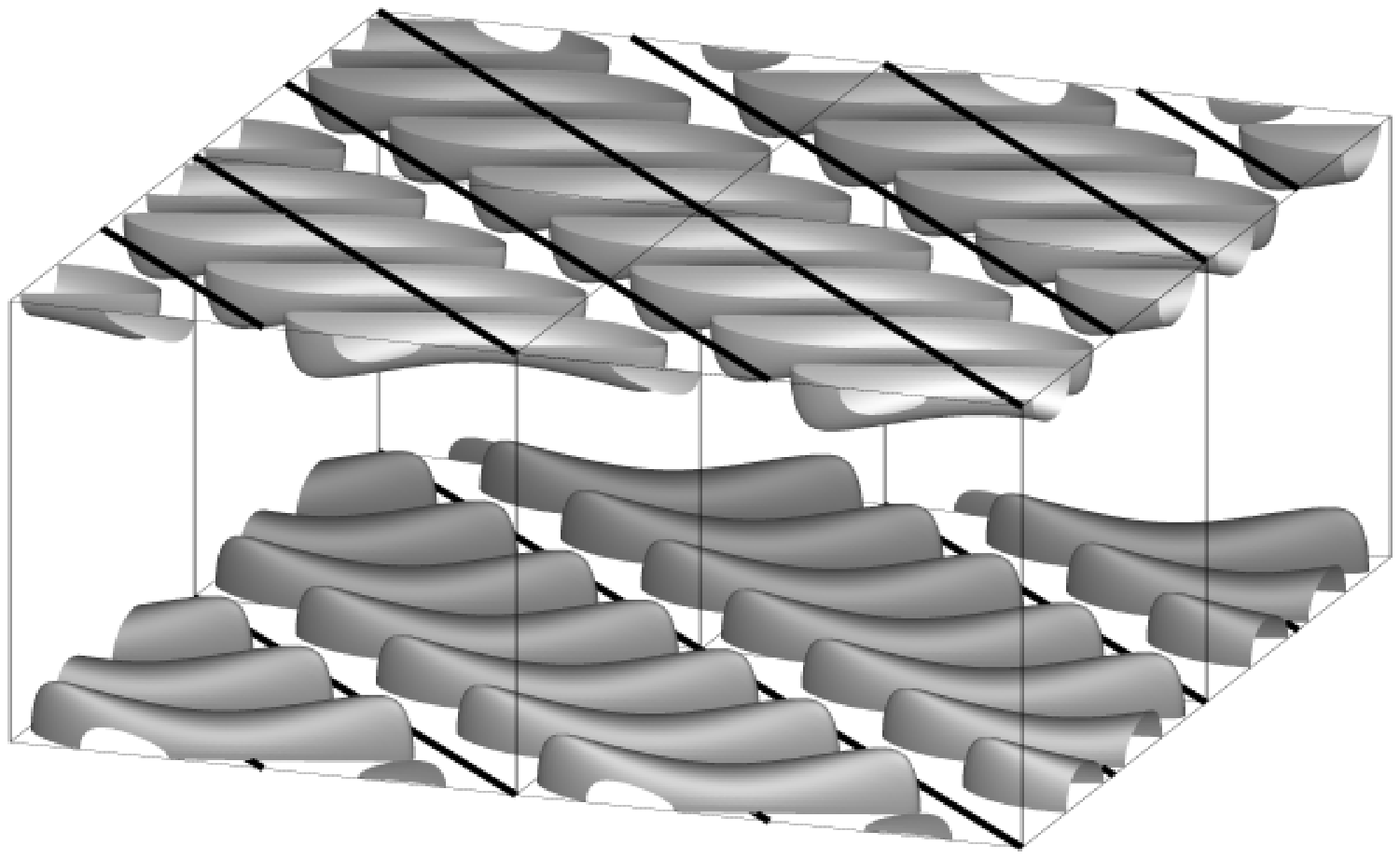,width=7cm,clip=}
\psfig{file=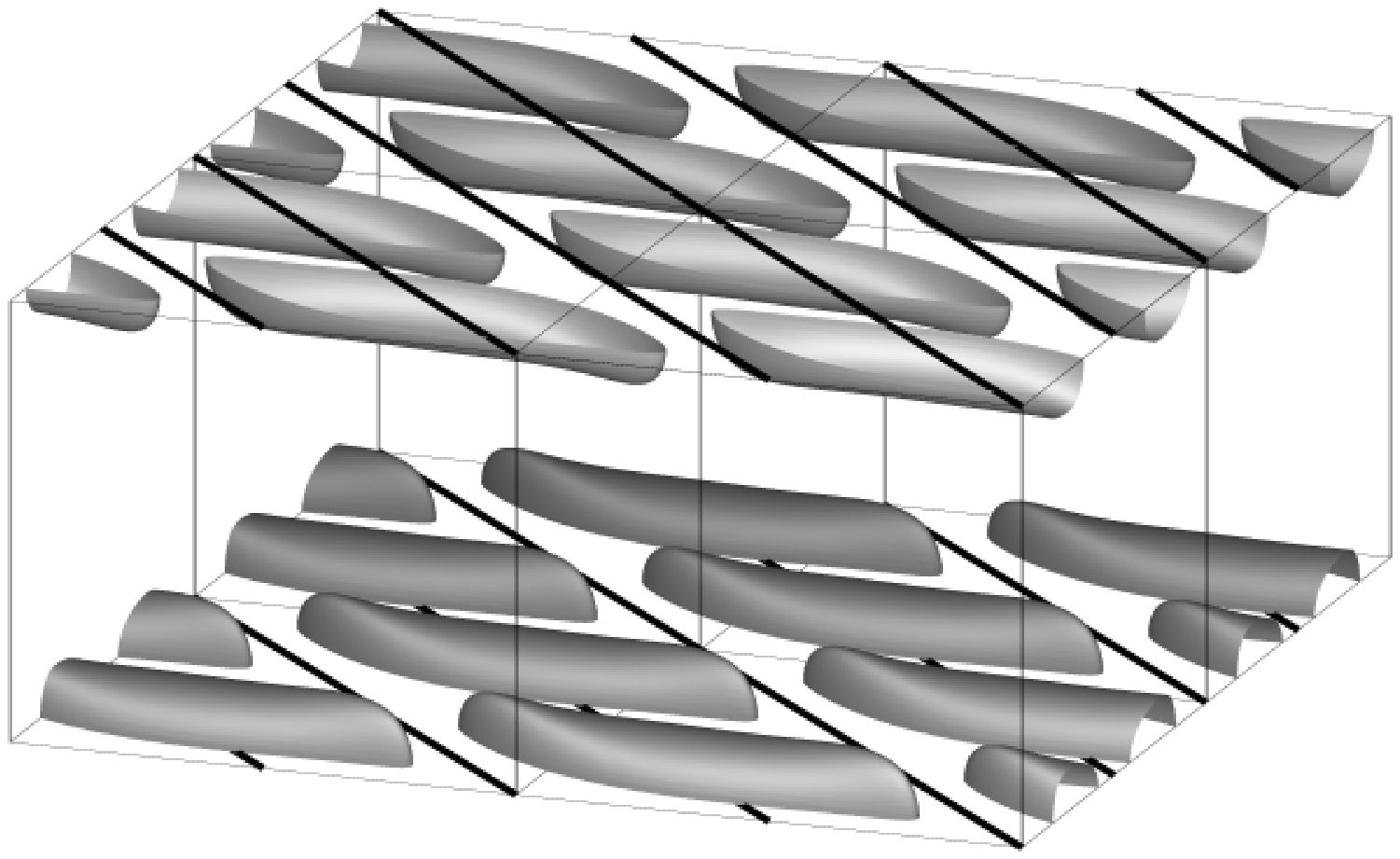,width=7cm,clip=}}
\hspace{42mm}(e)\hspace{66mm}(f)

\vspace*{2mm}
\centerline{
\psfig{file=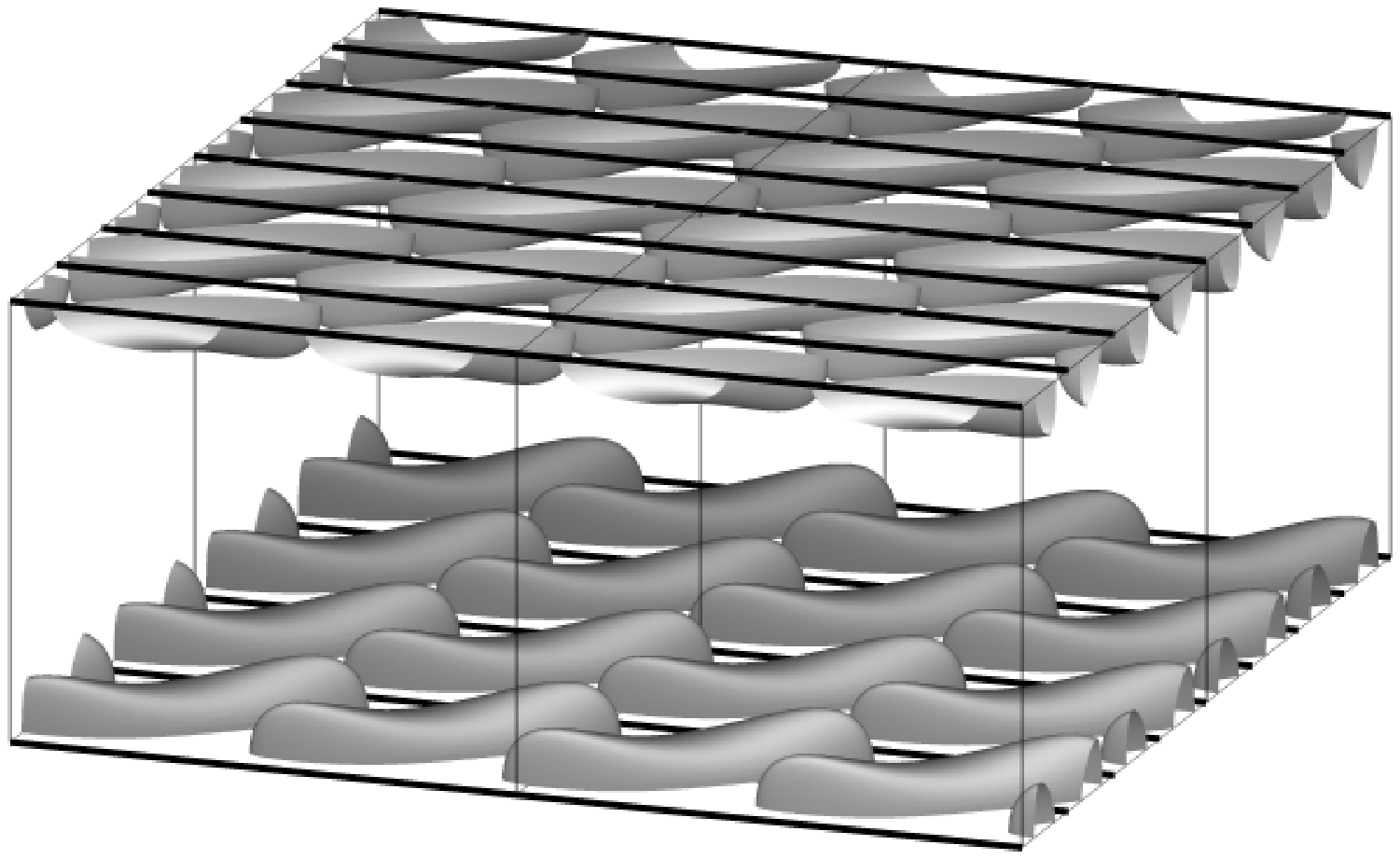,width=7cm,clip=}
\psfig{file=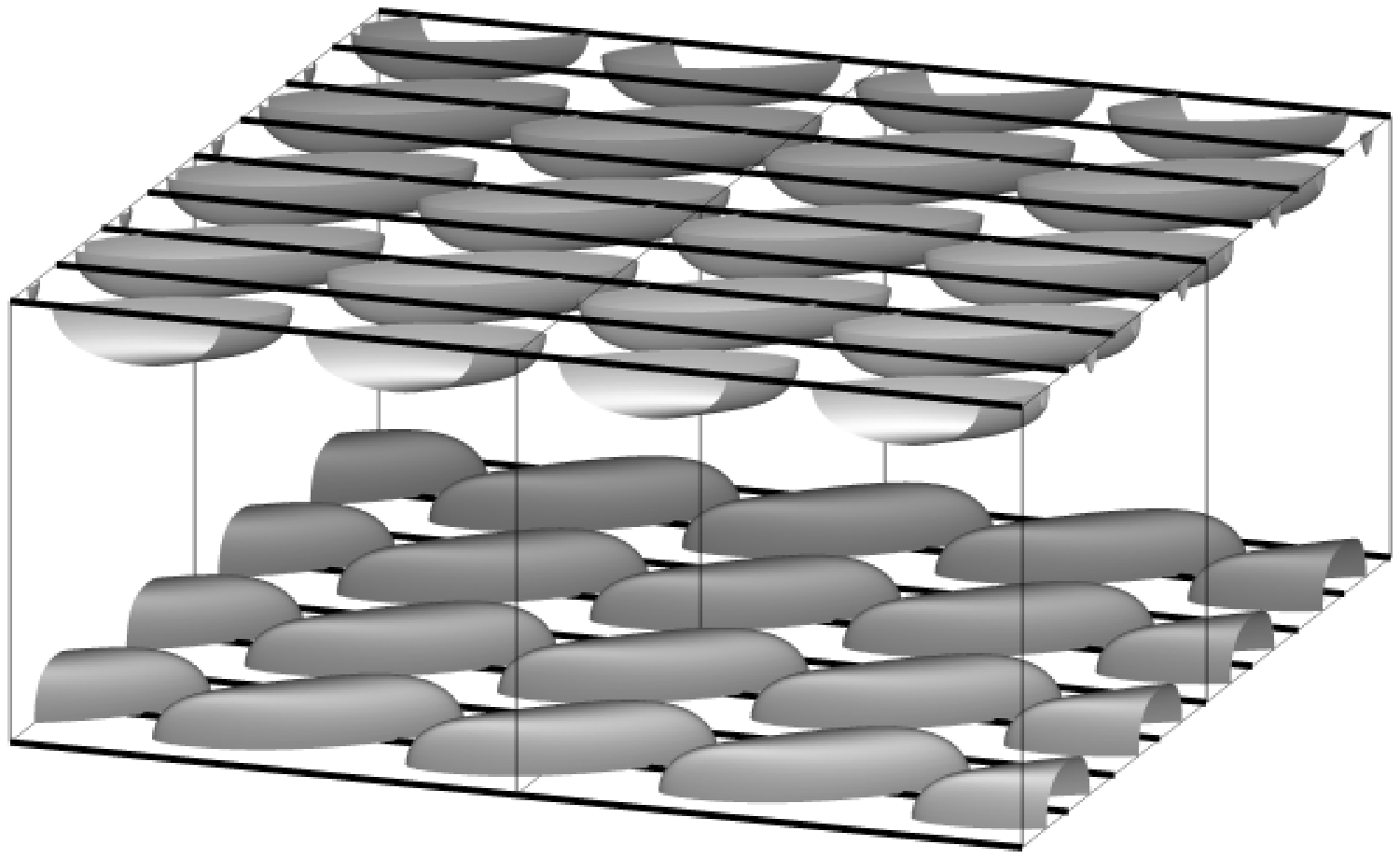,width=7cm,clip=}}
\hspace{42mm}(g)\hspace{66mm}(h)
\caption{Isosurfaces of magnetic energy density of the dominant magnetic modes,
at the level of a half of the maximum, for $Ta=50$, TW (a); $Ta=150$, TW (b);
$Ta=500$, R$_1$ (c); $Ta=720$, WR (d); $Ta=685$, $\rm R_D$ (e);
$Ta=1200$, $\rm R_D$ (f); $Ta=1100$, R$_2$ (g); $Ta=1700$, R$_2$ (h).
Isolated stagnation points and lines of stagnation points of the flow
on the horizontal boundaries are shown by dots and bold lines.
Four periodicity cells are displayed. On each panel, coordinate axes
are as shown on (a).\label{fig:conv_isosb}}
\end{figure}

\begin{figure}
\centerline{
\psfig{file=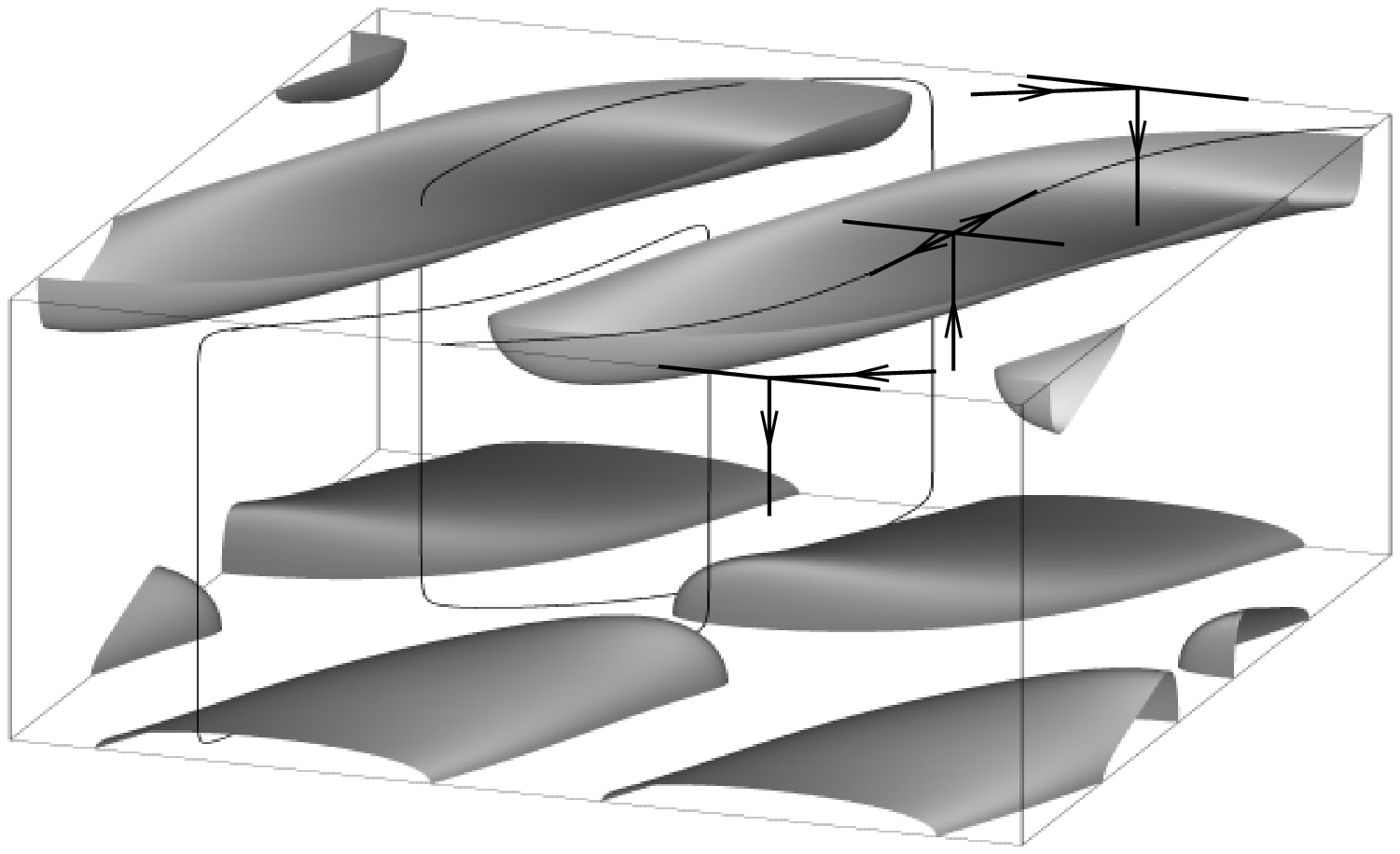,width=7cm,clip=}
\psfig{file=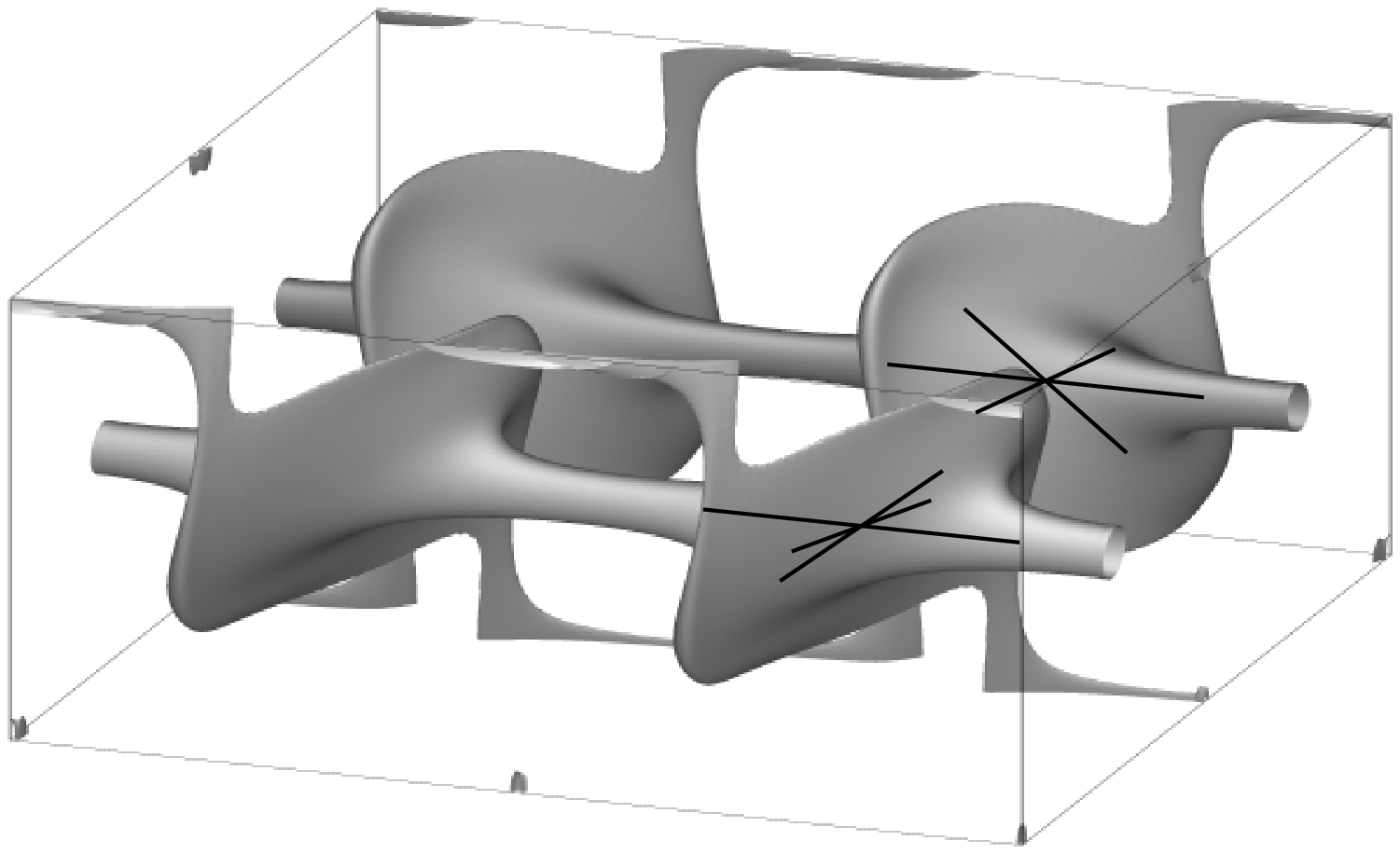,width=7cm,clip=}}
\hspace{42mm}(a)\hspace{66mm}(b)

\vspace*{2mm}
\centerline{
\psfig{file=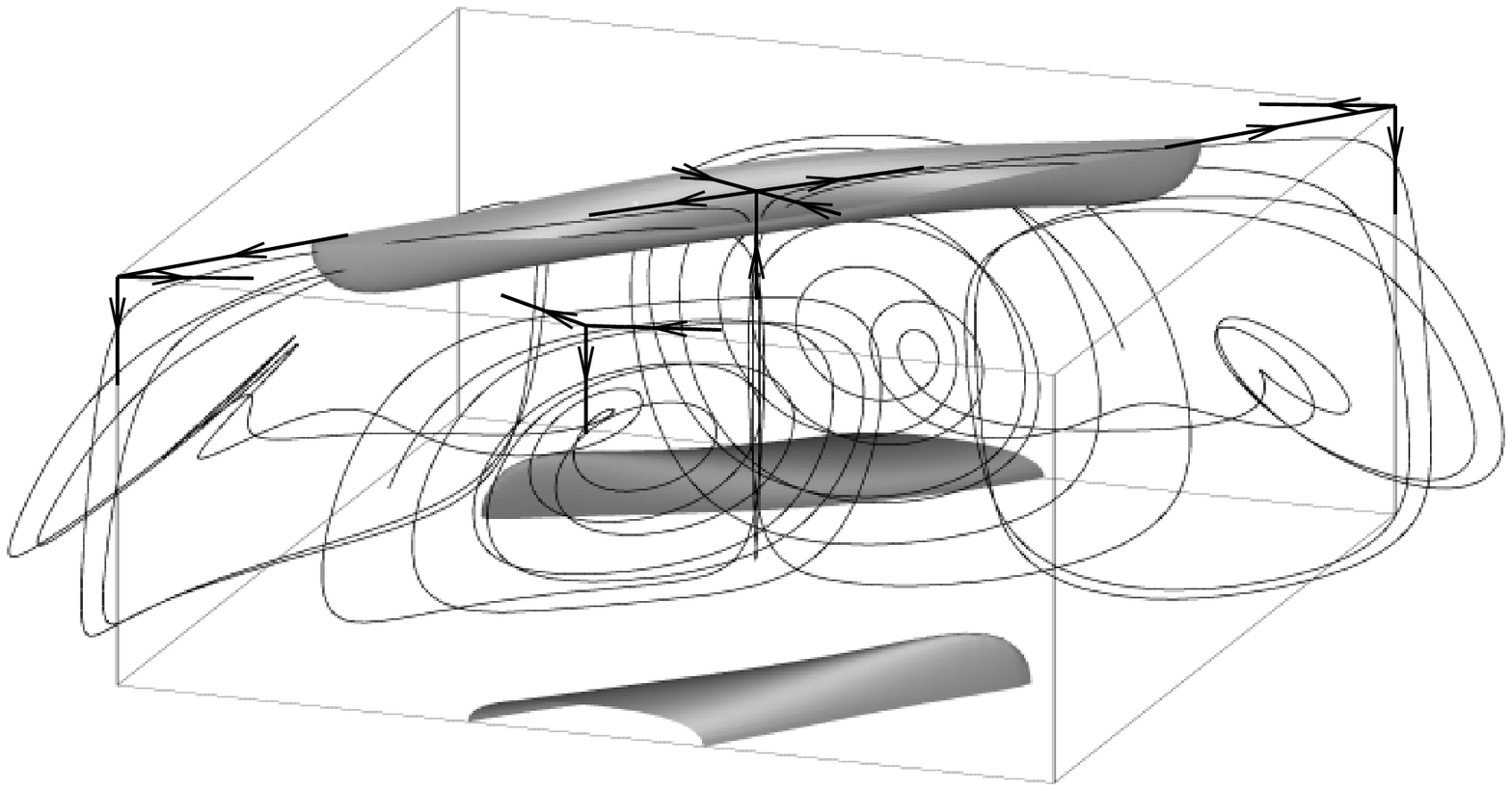,width=7cm,clip=}
\psfig{file=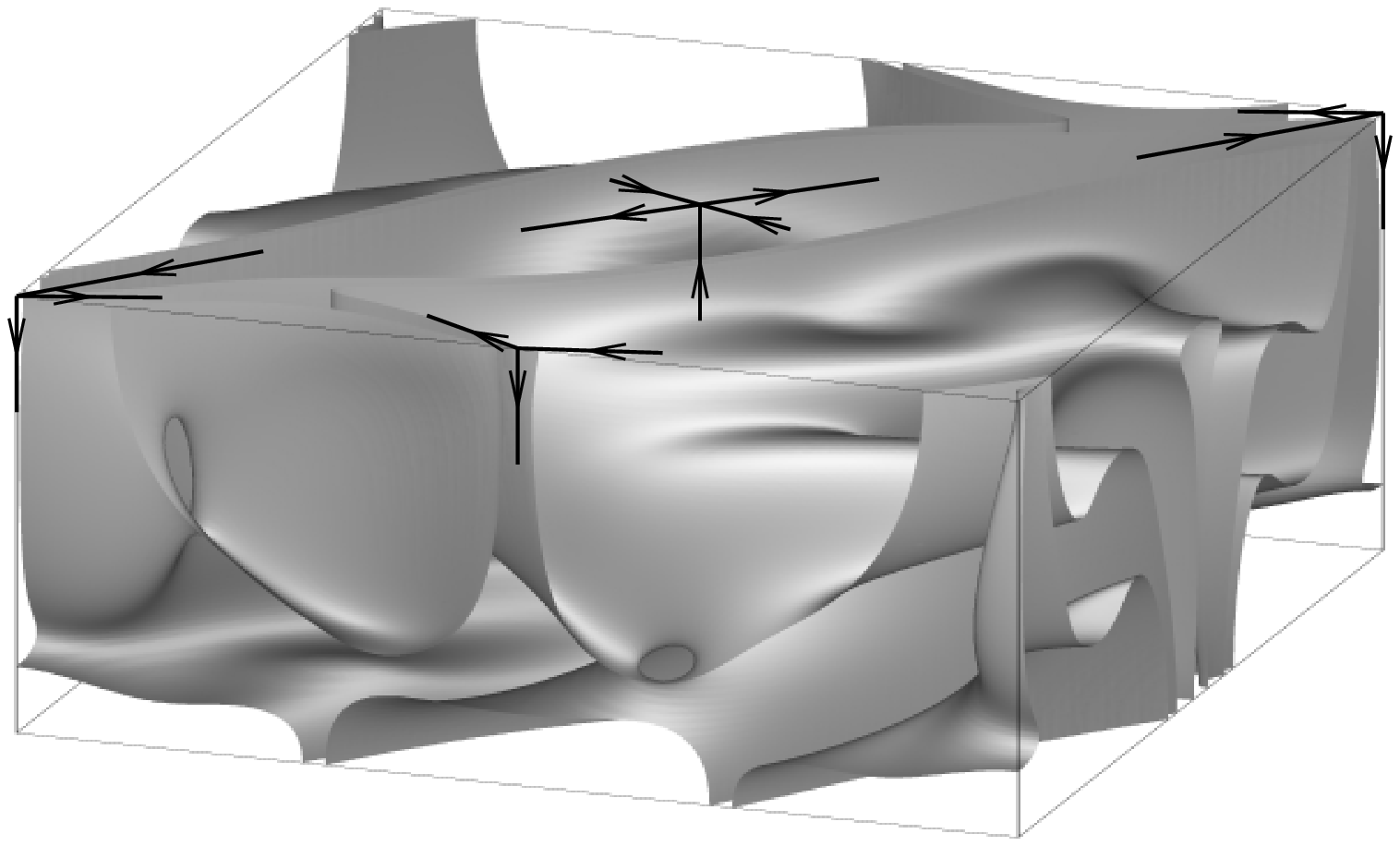,width=7cm,clip=}}
\hspace{42mm}(c)\hspace{66mm}(d)

\vspace*{2mm}
\centerline{
\psfig{file=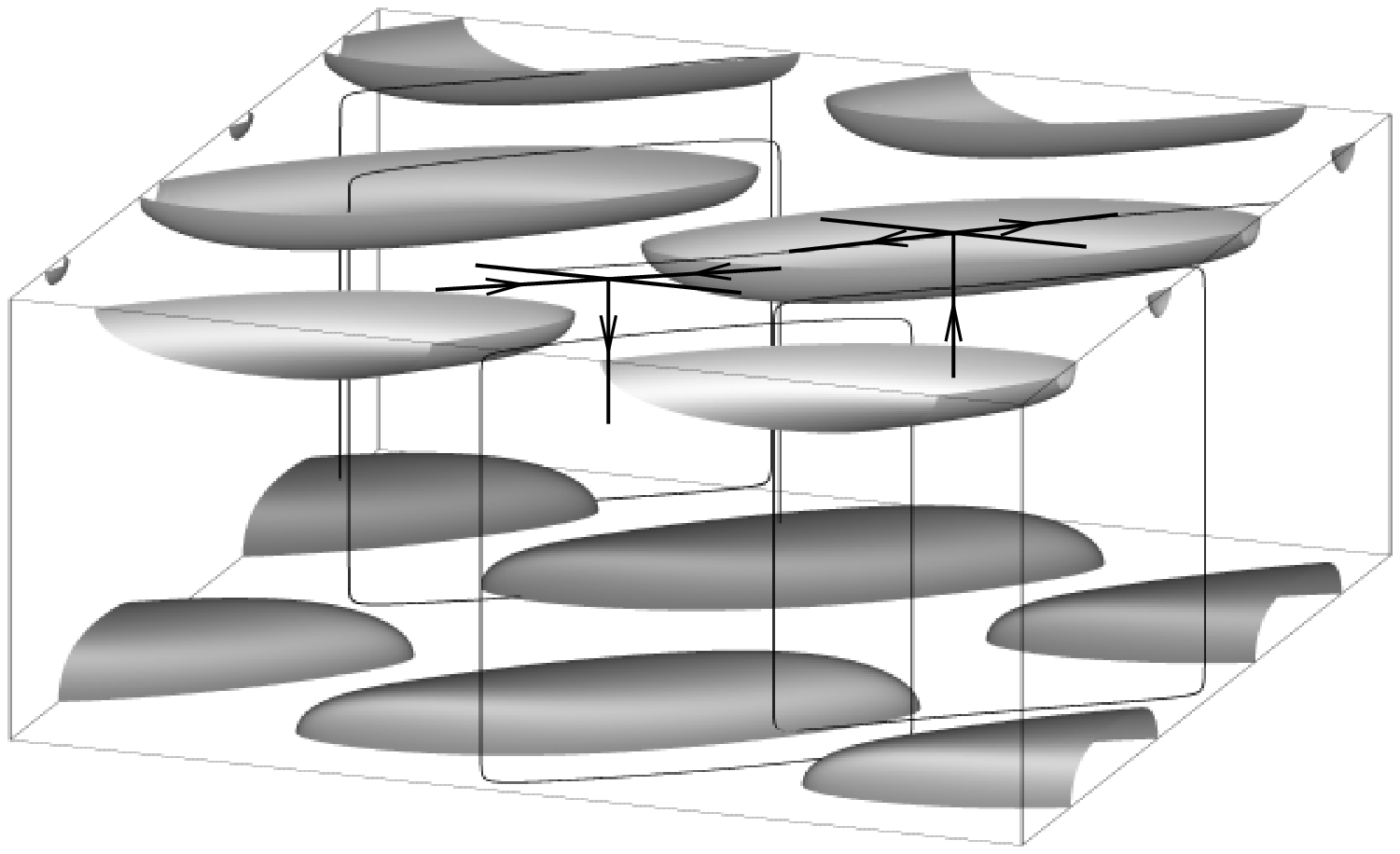,width=7cm,clip=}
\psfig{file=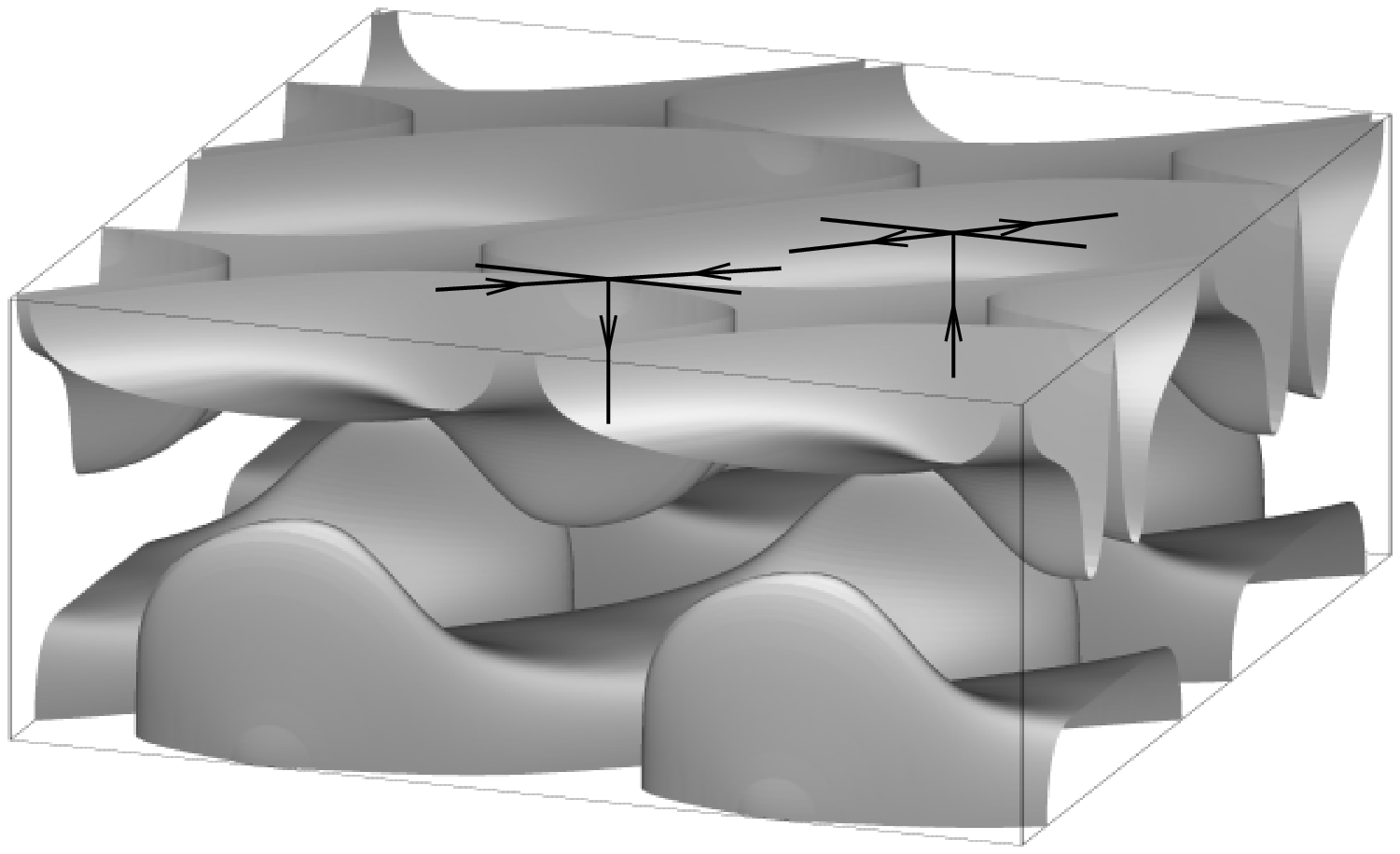,width=7cm,clip=}}
\hspace{42mm}(e)\hspace{66mm}(f)
\caption{Isosurfaces of magnetic energy density of the dominant kinematic
magnetic modes generated by R$_1$ for $Ta=500$ at the 50\% level of the
maximum (a) and at 3.5\% (b); by WR for $Ta=720$ at 50\% (c) and 10\%
(d); by R$_2$ at $Ta=1700$ at 50\% (e) and 12\% (f). Stable and unstable
directions of some stagnation points are shown by lines with arrows,
neutral directions associated with zero and imaginary eigenvalues by lines
without arrows. (See location of isolated stagnation points of the flow on the
horizontal boundaries and lines of stagnation points on \protect{\xrf{fig:conv_isosb}}
and note that axes of rolls R$_1$ and R$_2$ are also comprised of stagnation
points.) Sample trajectories of fluid particles on the upper boundary
and in the interior of the fluid layer are shown by thin lines. One periodicity
cell is displayed. On each panel, coordinate axes are as shown on
\protect{\xrf{fig:conv_isosb}} (a).\label{fig:nlin_stag}}
\end{figure}

There is no obvious relation between the structure of magnetic field generated
by TW and stagnation (in the co-moving reference frame) points of the flow
(Podvigina, 2006, found that the same was true for convective dynamos without
rotation). The spatial structure of growing magnetic modes is shown on
\xrf{fig:conv_isolb} and \ref{fig:conv_isosb}. For all convective flows
capable of magnetic field generation, in dominant magnetic modes the field
concentrates near horizontal boundaries in flattened half-ropes. A plausible
underlying physical mechanism for this kind of behaviour in the case of
perfectly conducting boundaries was proposed by St Pierre (1993).
For steady flows, in agreement with the kinematic dynamo theory (Galloway and
Zheligovsky, 1994), each half-rope is centered at a stagnation point
of the flow and oriented along the one-dimensional unstable manifold
of the flow at this point (see \xrf{fig:nlin_stag} (a), (c) and
(e)\,). Magnetic field is advected by the flow and, accordingly, the ropes are
stretched along the trajectories of fluid particles on the upper boundary;
however, advection is affected by magnetic diffusion, which must be responsible
for a (rather modest) deviation of the direction of half-ropes near their ends
from the direction of the trajectories (see \xrf{fig:nlin_stag} (a), (c)\,).
The half-ropes extend till they
begin to feel the influence of the adjacent stagnation points in the direction
of their stretching, which results in termination of the ropes. Magnetic field
is advected by the flow into the layer; it is redistributed to form vertical
two-dimensional magnetic flux sheets (shown on \xrf{fig:nlin_stag} (d) and
(f)\,) in the plane of the unstable/neutral directions of the stagnation points.

Sample trajectories of fluid particles inside the rolls, also shown on
\xrf{fig:nlin_stag} (a), (c), (e), attest that the motion is not planar
(and hence the Zeldovich, 1956, antidynamo theorem is unapplicable).
In R$_1$ and R$_2$ the fluid moves along
closed loops, because the flows are two-dimensional (i.e. independent of
a horizontal Cartesian coordinate, which in our notation is $y$) and possess
two symmetries, $s_2$ (more precisely, the symmetry about any vertical line
on the boundary between two adjacent rolls, rotating in opposite directions)
and $r\gamma^x_{L/2}$. (Due to solenoidality and two-dimensionality, $v_x$ and
$v_z$ are associated with a stream function, say, $\psi(x,z)$, and equations
of motion for the particles imply that their trajectories are
spirals residing on the surfaces $\psi(x,z)=\rm const$, topologically
equivalent to infinite cylinders. The two symmetries of the flow imply
$\psi(x,z)=\psi(-x,-z)$ and $v_y(x,z)=-v_y(-x,-z)$, in a coordinate system
with the origin on the axis of the roll; hence the helical trajectories
degenerate into closed loops.) The structure of trajectories in WR is
by far more complex (two trajectories in the interior of the volume of fluid
are shown on \xrf{fig:nlin_stag} (c)\,), however, visibly this does not affect
the complexity of the magnetic field -- neither at large magnetic energy
levels, nor at small ones.

\xrf{fig:nlin_stag} (b), (d), (f), showing isosurfaces of magnetic energy
density at low levels, expose two-dimensional structures of magnetic field
apparently related to stagnation points (for which eigenvectors of the stress
tensor $\|\partial v_{x,y,z}/\partial\{x,y,z\}\|$ are also shown) of the flows.
Formation of a magnetic flux
sheet tangent to the two-dimensional unstable manifold of a stagnation point
of a flow was investigated analytically by Childress (1979) and Childress and
Soward (1985). The structures that we observe are peculiar in that magnetic
energy density increases not in between the isosurfaces, as it does in magnetic
flux sheets, but, on the contrary, outside them, i.e. the isosurfaces reveal
two-dimensional ``magnetic flux sheet gaps''.

At first glance, emergence of such ``antistructures'' can be linked to the
fact that in most cases the respective stagnation points have neutral
eigendirections along lines of stagnation points (note that lines of
stagnation points reside not only on the horizontal boundaries, as shown on
\xrf{fig:conv_isosb}, but also constitute the axes of the rolls R$_1$ and R$_2$)
-- at such stagnation points the two-dimensional antistructures are associated
with the unstable and neutral eigendirection. In the presence of a neutral
direction the theory of Childress (1979) and Childress and Soward (1985) may be
unapplicable. This argument, however, breaks for isolated stagnation
points of WR in the middle of upper edges parallel to the $y$ axis (see
\xrf{fig:nlin_stag} (d)\,), which do not possess neutral eigendirections,
but have genuine two-dimensional unstable manifolds. By contrast, formation of
``bells of trombones'', the antistructures on \xrf{fig:nlin_stag} (b), is not
related to any unstable direction -- they are oriented along the eigenplane
associated with two imaginary eigenvalues of the stress tensor. Dynamics of
fluid in this case is significantly different from that considered {\it ibid.}:
the trajectories are closed oval loops centered at the axis of the roll.
\xrf{fig:conv_isolb} (c) reveals, that the ``bells of trombones'' constitute
boundaries between blobs of magnetic field of the opposite orientation.

\begin{figure}
\centerline{\psfig{file=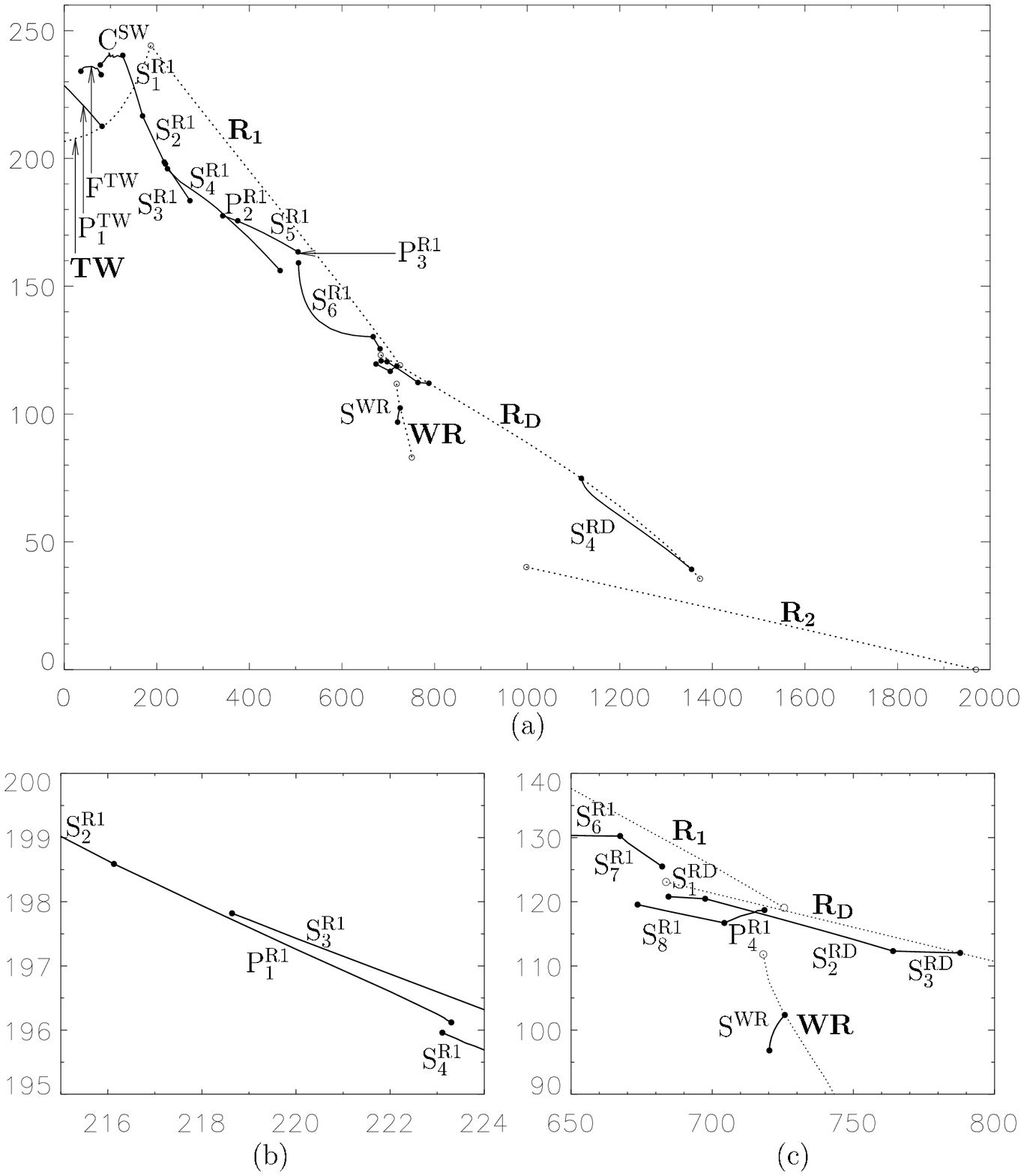,width=16cm}}
\caption{Kinetic $E_k$ energy (vertical axis) of convective MHD attractors
(solid line) and of hydrodynamic attractors (dashed line) for $0<Ta\le 2000$
(horizontal axis). Labelling of attractors is explained in Sections 4 and 6
(see also Table~\ref{tab:conv_lambda}).\label{fig:nlin_ev}}
\end{figure}

\begin{figure}
\centerline{\psfig{file=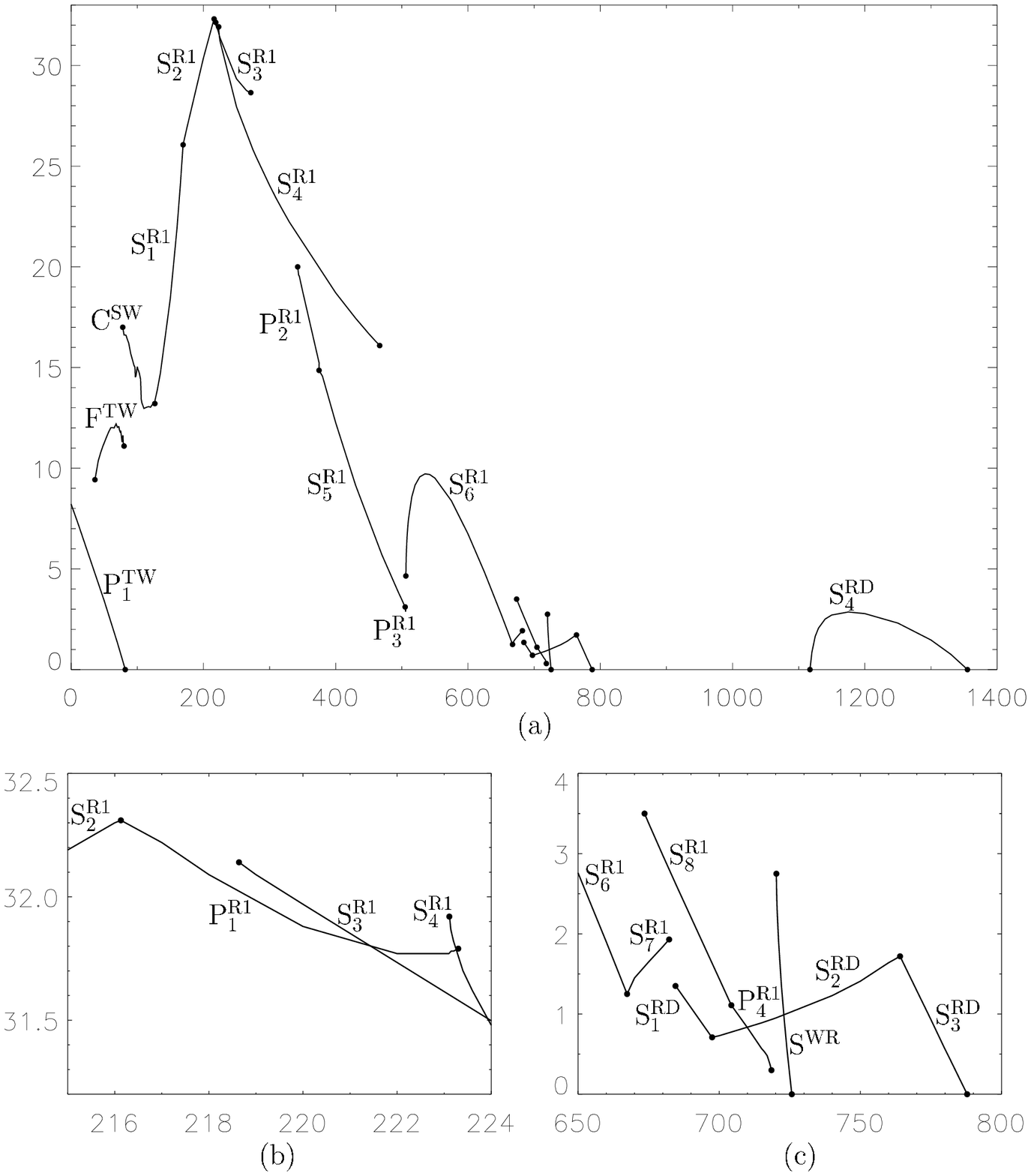,width=16cm}}
\caption{Magnetic $E_m$ energy (vertical axis) of convective MHD attractors
for $0<Ta\le 2000$ (horizontal axis). Labelling of attractors is explained
in Section 6 (see also Table~\ref{tab:conv_lambda}).\label{fig:nlin_eb}}
\end{figure}

\begin{figure}
\centerline{\psfig{file=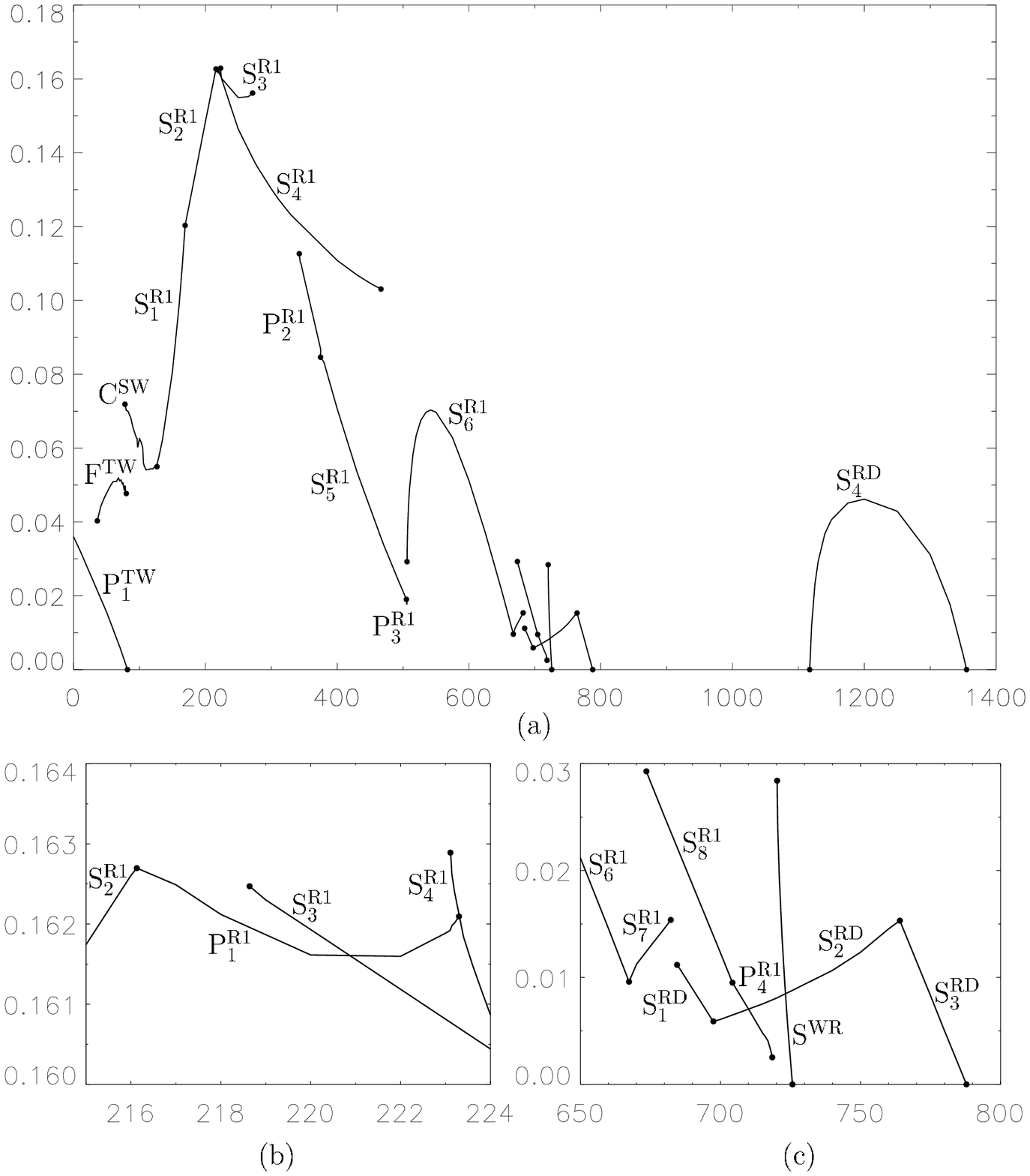,width=16cm}}
\caption{The ratio of energies $E_m/E_k$ (vertical axis) of MHD attractors
for $0<Ta\le 2000$ (horizontal axis).\label{fig:nlin_emek}}
\end{figure}

\begin{table}[p]
\caption{MHD attractors found in computations. Column 3 presents the
symmetry group for which an attractor is pointwise invariant, column 4
generators of the group (for appropriately chosen location of the origin
of the coordinate system); if a group is a product of several subgroups,
generators of the subgroups are separated by semicolons. Columns 5 and 6 present
time-averaged kinetic and magnetic energies, respectively.\label{tab:nlin_en}}
\begin{center}\begin{tabular}[t]{|c|c|c|c|c|c|}\hline
Type&Interval of&Group of&Generators&$E_k$&$E_m$\\
&existence ($Ta$)&symmetries&&&\\
&(frequency)&&&&\\\hline
P$^{\rm TW}_1$&[1,\,81]&${\bf Z}_2$&$rq\gamma^x_{L/2}$&228.3--212.5&8.1--0\\
&$f=4.29-4.19$&&&&\\\hline
F$^{\rm TW}$&[36,\,80]&${\bf Z}_2$&$r\gamma^x_{L/2}$&234.1--232.7&9.4--11.3\\\hline
C$^{\rm SW}$&[78,\,126]&${\bf Z}_2$&$r\gamma^x_{L/2}$&236.6--239.9&17.0--13.3\\\hline
S$^{\rm R1}_1$&[127,\,169]&${\bf D}_2$&$r\gamma^x_{L/2}$, $qs_2$&239.8--217.0&13.4--26.0\\\hline
S$^{\rm R1}_2$&[170,\,216]&${\bf D}_2\ltimes{\bf Z}_2$&$r\gamma^x_{L/2}$, $s_2$; $q\gamma^y_{L/2}$&216.3--198.7&26.2--32.3\\\hline
P$^{\rm R1}_1$&[217,\,223.3]&${\bf D}_2$&$q\gamma^y_{L/2}$, $rs_2$&198.3--196.1&32.2--31.8\\
&$f=0.056-0.041$&&&&\\\hline
S$^{\rm R1}_3$&[219,\,271]&${\bf D}_2$&$q\gamma^y_{L/2}$, $s_2$&197.7--183.6&32.1--28.6\\\hline
S$^{\rm R1}_4$&[223.11,\,466]&${\bf D}_2$&$q\gamma^y_{L/2}$, $rs_2$&196.0--156.2&31.9--16.1\\\hline
P$^{\rm R1}_2$&[343,\,377]&${\bf D}_2$&$rq\gamma^{xy}_{L/2}$, $qs_2$&177.6--175.4&19.9--15.2\\
&$f=0.44-0.45$&&&&\\\hline
S$^{\rm R1}_5$&[378,\,505]&${\bf D}_2$&$rq\gamma^{xy}_{L/2}$, $qs_2$&175.5--163.5&15.0--3.1\\\hline
P$^{\rm R1}_3$& [505.1,\,506.0]&${\bf Z}_2$&$rq\gamma^{xy}_{L/2}$&163.5--163.0&3.1--3.2\\
&$f=0.05-0.004$&&&&\\\hline
S$^{\rm R1}_6$&[506.1,\,667]&${\bf D}_4\ltimes{\bf Z}_2$&$q\gamma^y_{L/4}$, $s_2$; $rq\gamma^x_{L/2}$&159.2--130.3&4.7--1.4\\\hline
S$^{\rm R1}_7$&[668,\,682]&${\bf D}_2$&$rq\gamma^{xy}_{L/2}$, $s_2$&129.9--125.7&1.3--1.9\\\hline
S$^{\rm R1}_8$&[674,\,704]&${\bf D}_2$&$r\gamma^x_{L/2}$, $s_2$&119.5--116.8&3.4--1.1\\\hline
P$^{\rm R1}_4$&[705,\,718]&${\bf D}_2$&$r\gamma^x_{L/2}$, $s_2$&116.5--118.7&0.9--0.4\\
&$f=0.06-0.01$&&&&\\\hline
S$^{\rm WR}$&[721,\,725]&${\bf D}_2$&$r\gamma^x_{L/2}$, $qs_2$&98.9--102.0&1.9--0.2\\\hline
S$^{\rm RD}_1$&[685,\,697]&${\bf D}_6\ltimes{\bf Z}_2$&$q\gamma^{xy}_{L/6}$, $s_2$; $r\gamma^x_{L/2}$&120.8--120.5&1.3--0.7\\\hline
S$^{\rm RD}_2$&[698,\,764]&${\bf D}_2$&$rq\gamma^y_{L/2}$, $s_2$&120.4--112.7&0.7--1.7\\\hline
S$^{\rm RD}_3$&[765,\,787]&${\bf D}_4\ltimes{\bf Z}_2$&$q\gamma^{xy}_{L/4}$, $s_2$; $rq\gamma^x_{L/2}$&112.6--112.0&1.6--0.\\\hline
S$^{\rm RD}_4$&[1118,\,1355]&${\bf D}_2\ltimes{\bf Z}_2$&$r\gamma^y_{L/2}$, $s_2$; $rq\gamma^x_{L/2}$&74.3--39.2&0.--2.9\\\hline
\end{tabular}\end{center}\end{table}

\begin{table}[p]
\caption{Bifurcations of MHD steady states.
Column 2 presents the symmetry group of a steady state, column 3 the generators
of the symmetry group, column 4 the critical $Ta$, column 5 the type of a bifurcation
(S.P. denotes subcritical pitchfork, S.H. subcritical Hopf and S.
saddle-node bifurcations), column 6 the dimension of the respective center eigenspace,
column 7 the action of the system symmetry group on the eigenspace, and
the last column elements of the group which act trivially. Note that $s_2$
mentioned in columns 3 and 8 have the {\it same} axis.\label{tab:bifur}}
\begin{center}
\begin{tabular}{|c|c|c|c|c|c|c|c|}\hline
Type&Symmetry&Generators&$Ta$&B&D&Action&Kernel\\
&group&&&&&&\\\hline
S$^{\rm R1}_1$&${\bf D}_2$&$r\gamma^x_{L/2}$, $qs_2$&126.55&to P$^{\rm SW}$&2&${\bf Z}_2$&$r\gamma^x_{L/2}$\\
&&&169.27&from S$^{\rm R1}_2$&&&\\\hline
S$^{\rm R1}_2$&${\bf D}_2\ltimes{\bf Z}_2$&$r\gamma^x_{L/2}$, $s_2$; $q\gamma^y_{L/2}$&169.27&to S$^{\rm R1}_1$&1&${\bf Z}_2$&$r\gamma^x_{L/2}$, $qs_2\gamma^y_{L/2}$\\
&&&216.13&to P$^{\rm R1}_1$&2&${\bf Z}_2$&$rs_2\gamma^x_{L/2}$, $q\gamma^y_{L/2}$\\\hline
S$^{\rm R1}_3$&${\bf D}_2$&$q\gamma^y_{L/2}$, $s_2$&218.64&S.H.&2&${\bf Z}_2$&$q\gamma^y_{L/2}$\\
&&&271.66&S.P.&1&${\bf Z}_2$&$q\gamma^y_{L/2}$\\\hline
S$^{\rm R1}_4$&${\bf D}_2$&$q\gamma^y_{L/2}$, $rs_2$&223.10&S.&1&{\bf 1}&$q\gamma^y_{L/2}$, $rs_2$\\
&&&466.50&S.P.&1&${\bf Z}_2$&$qrs_2\gamma^y_{L/2}$\\\hline
S$^{\rm R1}_5$&${\bf D}_2$&$rq\gamma^{xy}_{L/2}$, $qs_2$&377.89&to P$^{\rm R1}_2$&2&{\bf 1}&$rq\gamma^{xy}_{L/2}$, $qs_2$\\
&&&505.05&to P$^{\rm R1}_3$&2&${\bf Z}_2$&$rq\gamma^{xy}_{L/2}$\\\hline
S$^{\rm R1}_6$&${\bf D}_4\ltimes{\bf Z}_2$&$q\gamma^y_{L/4}$, $s_2$; $rq\gamma^x_{L/2}$&506.07&S.&1&{\bf 1}&$q\gamma^y_{L/4}$, $s_2$; $rq\gamma^x_{L/2}$\\
&&&667.37&to S$^{\rm R1}_7$&2&${\bf D}_4$&$rq\gamma^{xy}_{L/2}$\\\hline
S$^{\rm R1}_7$&${\bf D}_2$&$rq\gamma^{xy}_{L/2}$, $s_2$&667.37&from S$^{\rm R1}_6$&&&\\
&&&682.26&S.P.&1&${\bf Z}_2$&$s_2$\\\hline
S$^{\rm R1}_8$&${\bf D}_2$&$r\gamma^x_{L/2}$, $s_2$&673.55&S.P.&1&${\bf Z}_2$&$s_2$\\
&&&704.29&to P$^{\rm R1}_4$&2&{\bf 1}&$r\gamma^x_{L/2}$, $s_2$\\\hline
S$^{\rm WR}$&${\bf D}_2$&$r\gamma^x_{L/2}$, $qs_2$&720.25&S.&1&{\bf 1}&$r\gamma^x_{L/2}$, $qs_2$\\
&&&725.71&from WR&&&\\\hline
S$^{\rm RD}_1$&${\bf D}_6\ltimes{\bf Z}_2$&$q\gamma^{xy}_{L/6}$, $s_2$; $r\gamma^x_{L/2}$&684.51&S.P.&2&${\bf D}_6$&$r\gamma^x_{L/2}$\\
&&&697.49&to S$^{\rm RD}_2$&2&${\bf D}_6$&$rq\gamma^y_{L/2}$\\\hline
S$^{\rm RD}_2$&${\bf D}_2$&$s_2$, $rq\gamma^y_{L/2}$&697.49& from S$^{\rm RD}_1$&&&\\
&&&764.05& from S$^{\rm RD}_3$&&&\\\hline
S$^{\rm RD}_3$&${\bf D}_4\ltimes{\bf Z}_2$&$q\gamma^{xy}_{L/4}$, $s_2$; $rq\gamma^x_{L/2}$&764.05&to S$^{\rm RD}_2$&2&${\bf D}_4$&$rq\gamma^y_{L/2}$\\
&&&787.82&from RD&&&\\\hline
S$^{\rm RD}_4$&${\bf D}_2\ltimes{\bf Z}_2$&$r\gamma^y_{L/2}$, $s_2$; $rq\gamma^x_{L/2}$&1117.47&from RD&&&\\
&&&1355.20&from RD&&&\\\hline
\end{tabular}\end{center}\end{table}

\section{Nonlinear magnetic field generation}

\begin{figure}[t]
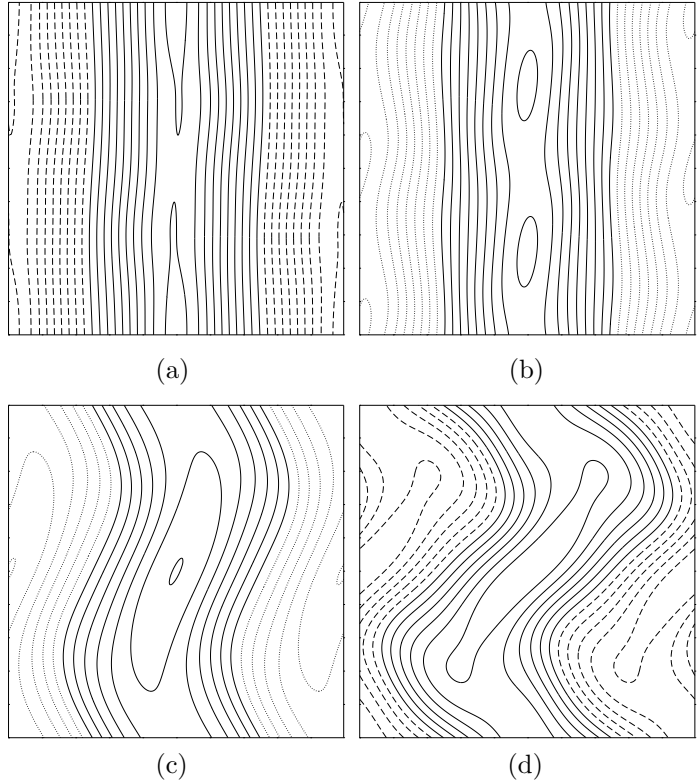

\centerline{
\psfig{file=fig9/sr1-1-ta150.ps,width=45mm,clip=}
\psfig{file=fig9/sr1-4-ta250.ps,width=45mm,clip=}}

\vspace*{1mm}
\hspace{57mm}(a)\hspace{42mm}(b)

\vspace*{2mm}
\centerline{
\psfig{file=fig9/swr-1-ta680.ps,width=45mm,clip=}
\psfig{file=fig9/swr-2-ta721.ps,width=45mm,clip=}}
\hspace{57mm}(c)\hspace{42mm}(d)
\caption{Isolines (step 2) of $v_z$ on the horizontal midplane $z=1/2$ for
some steady convective MHD attractors: S$^{\rm R1}_1$, $Ta=150$ (a);
S$^{\rm R1}_3$, $Ta=250$ (b); S$^{\rm R1}_8$, $Ta=680$ (c); S$^{\rm WR}$,
$Ta=721$ (d). Solid lines indicate non-negative, dashed lines negative values.
$x$: horizontal axis, $y$: vertical axis.\label{fig:nlin_isolv}}
\end{figure}

Kinematic magnetic growth rates are not large (see \xrf{fig:conv_lambda}).
In agreement with this, for all convective MHD attractors that we have computed,
magnetic energy, $E_m$, is smaller (at least seven times) than the kinetic one,
$E_k$ (see Figs.~\ref{fig:nlin_ev}--\ref{fig:nlin_emek}). Consequently, all
the MHD attractors can be regarded as perturbations of hydrodynamic attractors
(note that the plot of the ratio $E_m/E_k$ is quite similar in shape to the plot
of magnetic energy, cf. Figs.~\ref{fig:nlin_emek} and \ref{fig:nlin_eb},
also suggesting that we are close to the onset of both convective motions and
of magnetic field generation) and the spatial structure of the flows is similar
to that in the absence of magnetic field (cf. Figs.~\ref{fig:conv_isolv} and
\ref{fig:nlin_isolv}). Hence, we label branches of MHD attractors as
follows: The main label denotes the type of an attractor: S for a steady state,
P for periodic, Q for quasiperiodic (in assigning these labels we disregard
temporal periodicities due to drifts along the $x$ and $y$ axes). The superscript
denotes the hydrodynamic attractor, the MHD attractor is genetically related to.
The subscript is the consecutive number within the collection of attractors
of the specified morphology (e.g., there are {\bf eight} different MHD
steady states with spatial structure similar to that of R$_1$). The numbering
of branches is in the order of increasing $Ta$.

Despite hydrodynamic attractors of just four types generate magnetic field
in the kinematic regime and magnetic field growth rates are not large, we have
observed a large variety of MHD attractors (see Table~\ref{tab:nlin_en}
and bifurcation diagrams Figs.~\ref{fig:nlin_ev} and \ref{fig:nlin_eb}).
In the description of branches of convective MHD attractors and bifurcations
bounding them we follow the ordering of hydrodynamic attractors in Table~\ref{tab:conv_ev},
and for each of them we start with unstable magnetic modes.
When on increasing $P_m$ a magnetic mode eigenvalue crosses the imaginary axis,
an MHD steady state or periodic orbit appears; for supercritical bifurcations
these objects are stable. We refer to them as primary magnetic attractors.
For $P_m=8$, for which the problem is solved here, these objects are not
necessarily stable; for a varying $Ta$ they constitute a branch, and
somewhere along the branch can gain stability. Because of emergence of
the branches of unstable convective MHD states, in the present problem
identification of hydrodynamic attractors with neutral kinematic magnetic
modes, giving rise to the branches of primary convective MHD attractors, is not
straightforward. We perform this identification by comparing the
spatial structure of the fields in the attractor with the respective
hydrodynamic attractor and the magnetic mode, including their symmetry groups.

To identify bifurcations of steady states we calculate eigenvalues and
the associated eigenspaces of the operator of linearisation of the system
(1)--(3). Hence for each bifurcation of a steady state we know the dimension
and the action of the steady state symmetry group on the eigenspace (see
Table~\ref{tab:bifur}), which is a necessary information for application of
the general theory of bifurcations for symmetric systems (Golubitsky\al, 1988).

For each growing magnetic mode there exists a primary MHD attractor.
At points of bifurcations from rolls, dimension of the kernel of the magnetic
induction operator is two due to the presence of translation symmetries (see
Table~\ref{tab:conv_lambda}). The action of the symmetry group on the eigenspace
is {\bf O}(2) generated by $s_2$ and translations. Along axes of rolls
eigenmodes can have periods $\ell/n$ with an integer $n>0$, where for rolls
parallel to coordinate axes $\ell=L$ and for rolls parallel to a diagonal
$\ell=\sqrt{2}L$. If a magnetic mode has a period $\ell/n$, the shift
by $\ell/(2n)$ acts as $-I$, hence the superposition of the shift by $\ell/(2n)$
and $q$ maps the mode into itself. Bifurcations from rolls are pitchfork
with the symmetry group {\bf O}(2). Thus a continuum of steady states emerges,
the symmetry group of each of them is a product of ${\bf Z}_2$ generated by
$s_2$ and the subgroup which acts trivially. Bifurcations from TW are Hopf ones,
and for WR it is pitchfork.

The spatial structure of magnetic fields is shown on \xrf{fig:nlin_isosb}.
A common dominant feature in the majority of nonlinear convective hydromagnetic
regimes is concentration of magnetic field flux in half-ropes located near
the horizontal boundaries. In the primary attractors this structure is
inherited from the respective kinematic dynamo modes. Concentration of
magnetic field near boundaries of the layer was observed by St Pierre (1993)
in simulations of subcritical magnetic field generation by thermal convection
of rapidly rotating fluid (for the same boundary conditions, as employed here).
Such behaviour of magnetic field was observed in many computations (see
a discussion and references in Zheligovsky, 2010) and it is usually expected
for electrically perfectly conducting boundaries (although Zheligovsky, 2010,
found for these boundary conditions an example of a time-periodic nonlinear
convective magnetic dynamo, in which magnetic field always remains concentrated
inside the layer of fluid).

In what follows we overview the MHD attractors found in computations and their
bifurcations.

\begin{figure}
\centerline{
\psfig{file=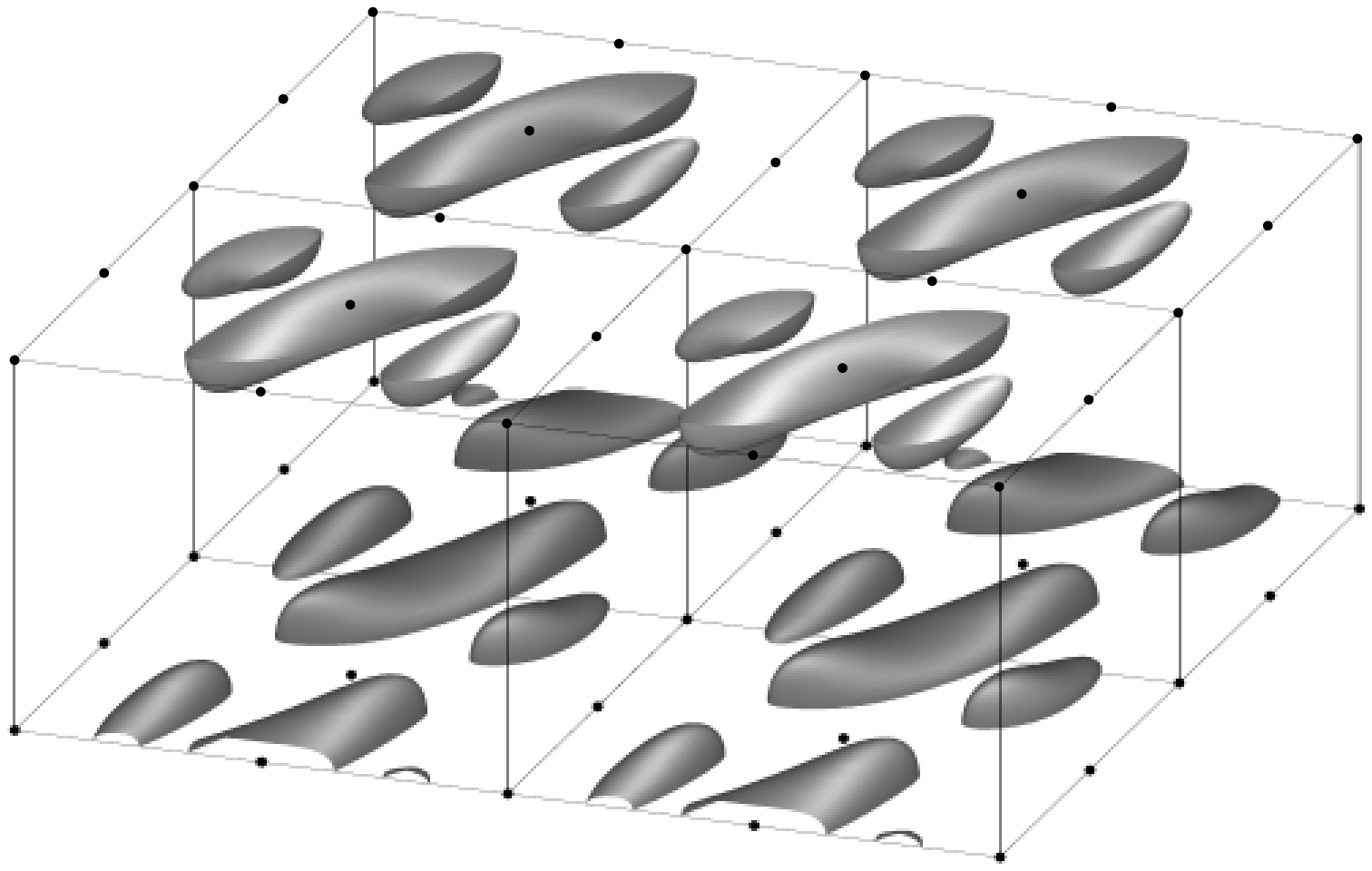,width=7cm,clip=}
\psfig{file=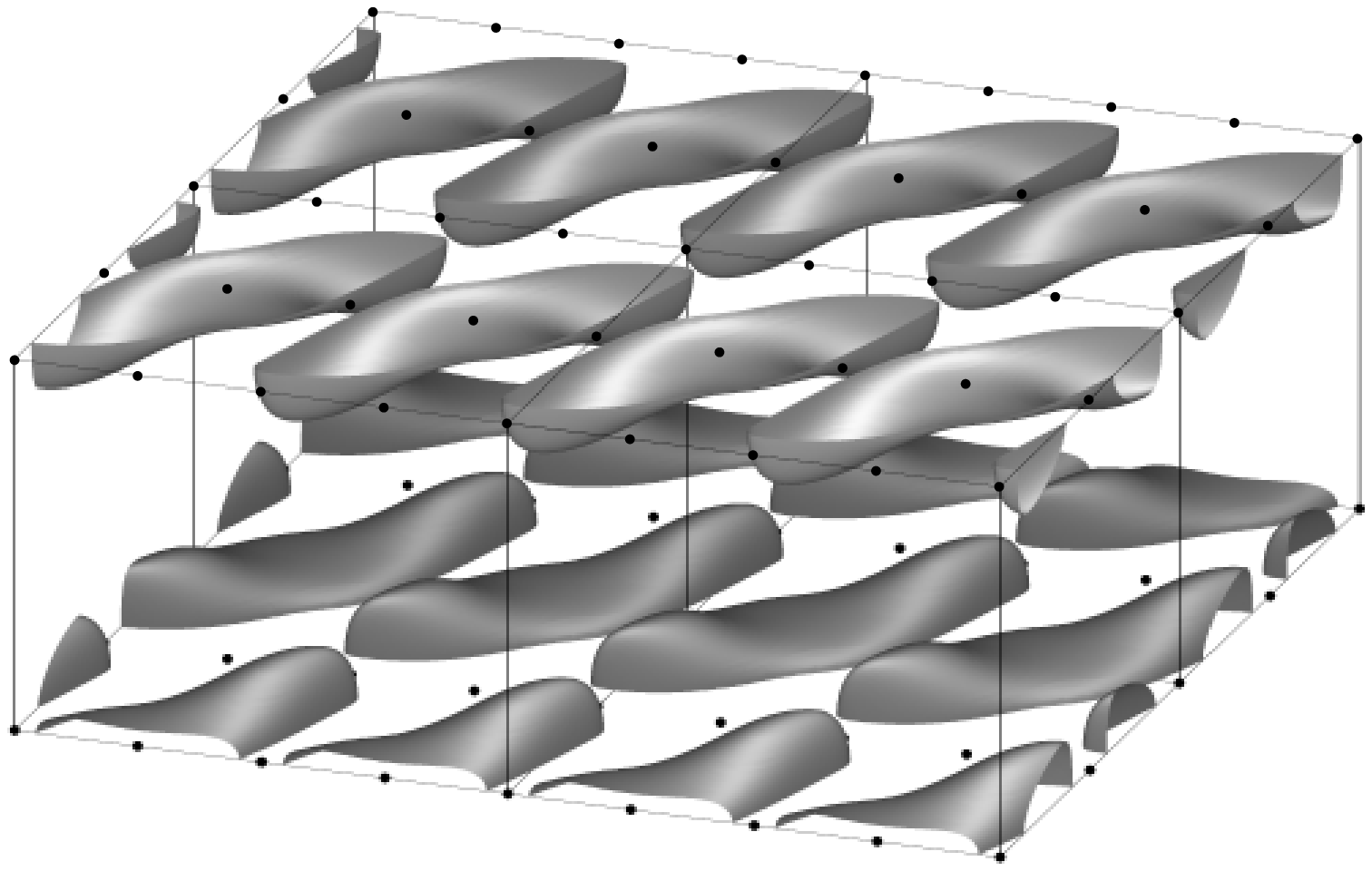,width=7cm,clip=}}
\hspace{42mm}(a)\hspace{66mm}(b)

\vspace*{2mm}
\centerline{
\psfig{file=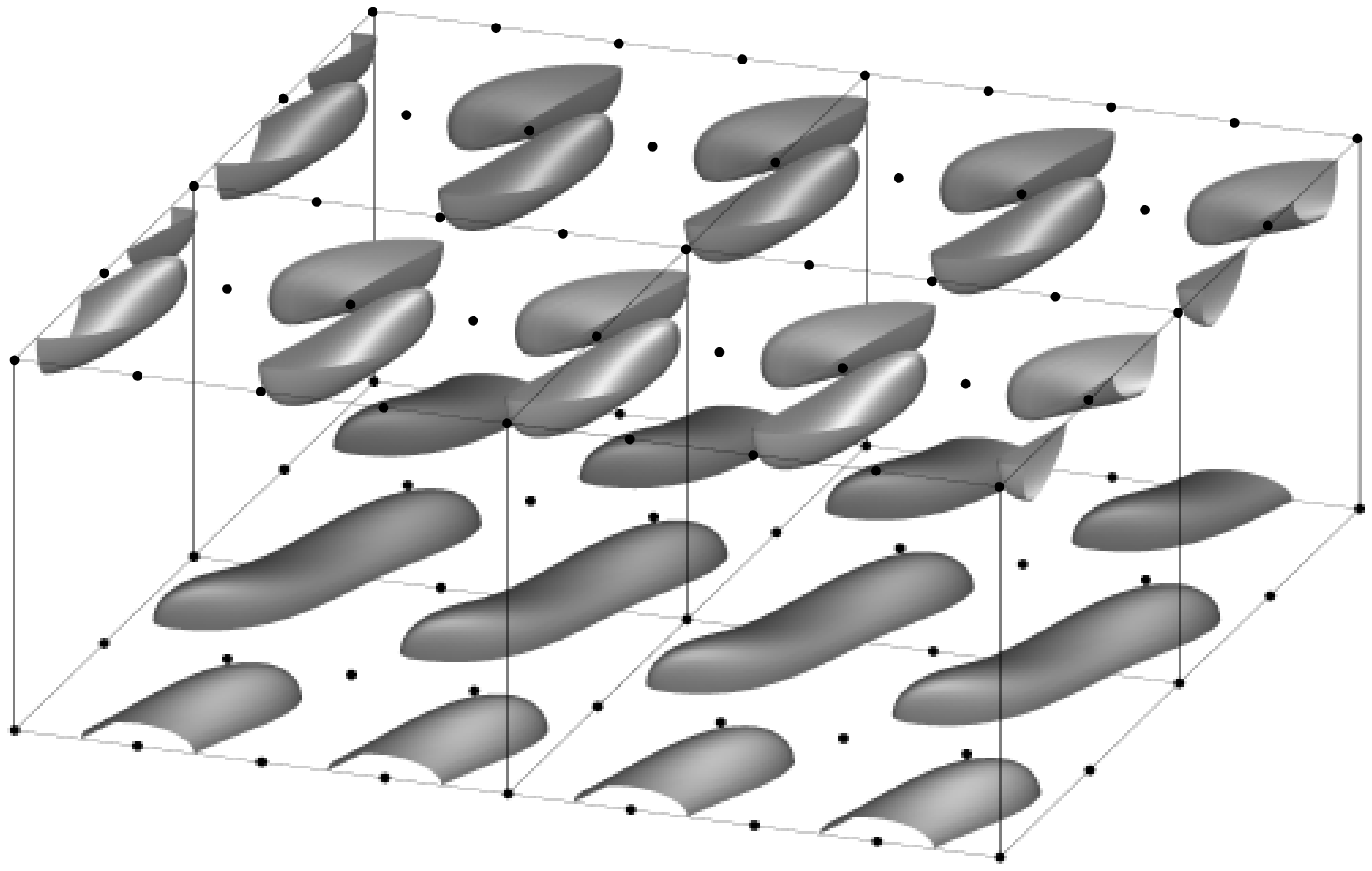,width=7cm,clip=}
\psfig{file=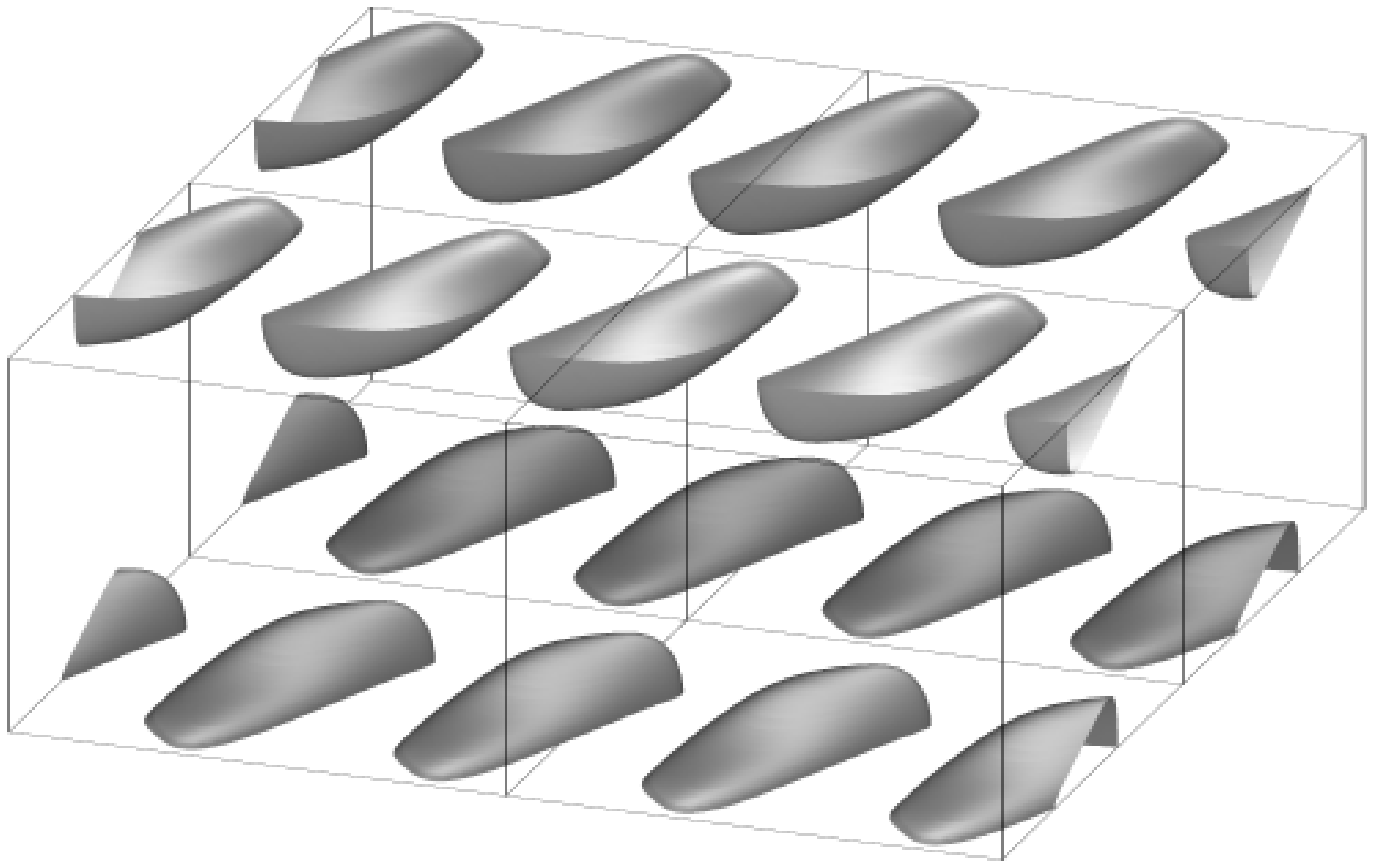,width=7cm,clip=}}
\hspace{42mm}(c)\hspace{66mm}(d)

\vspace*{2mm}
\centerline{
\psfig{file=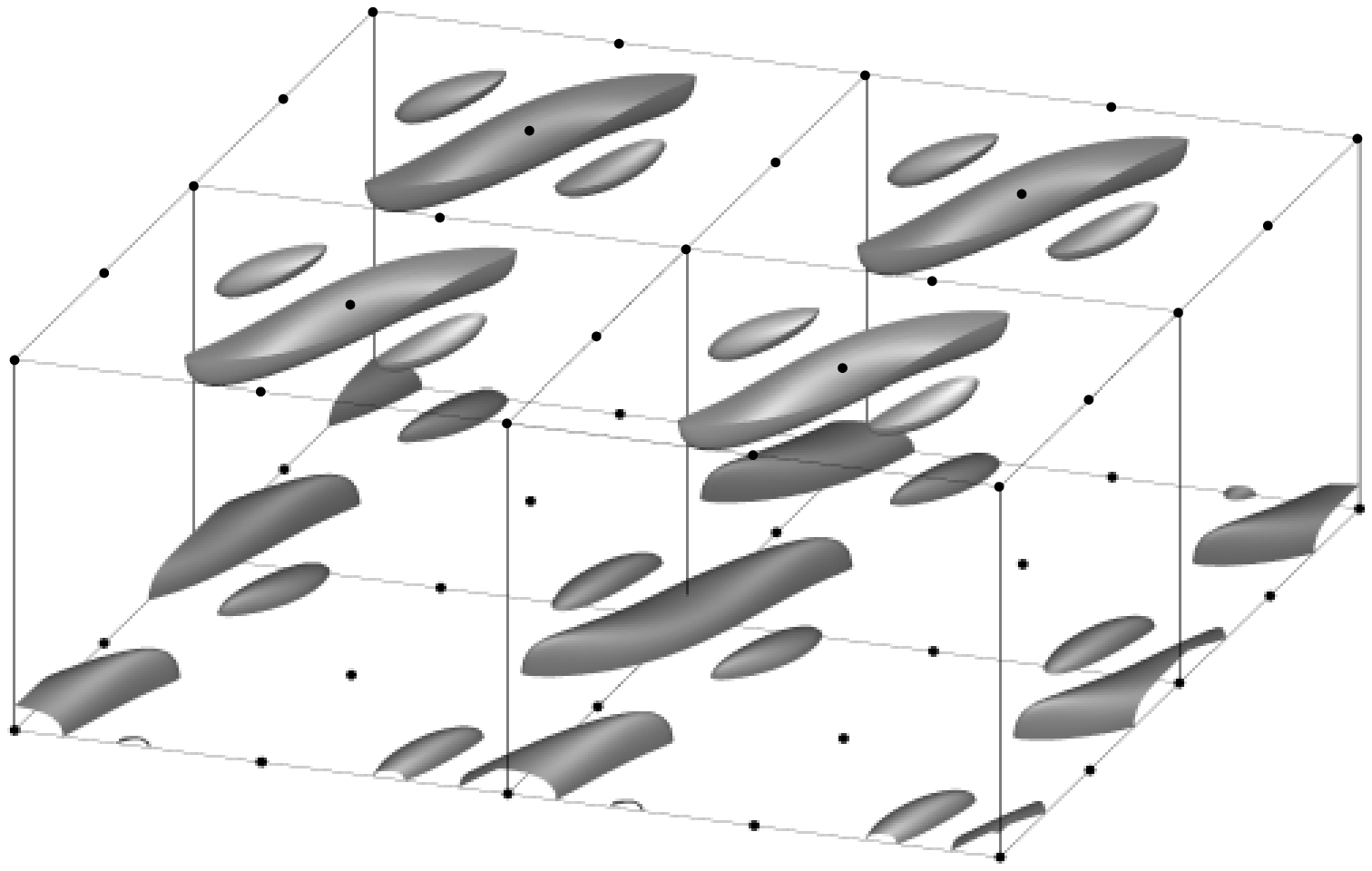,width=7cm,clip=}
\psfig{file=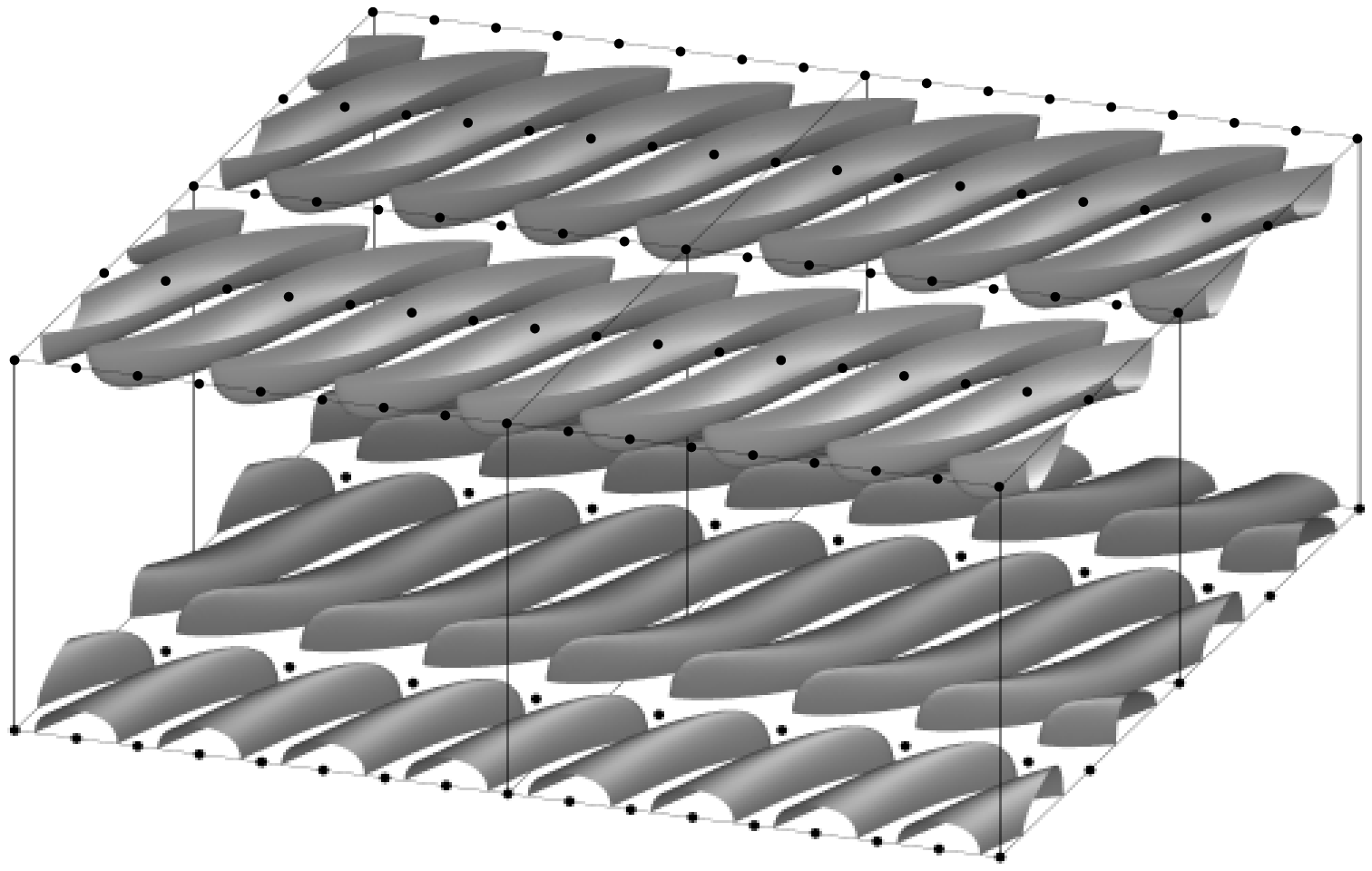,width=7cm,clip=}}
\hspace{42mm}(e)\hspace{66mm}(f)

\vspace*{2mm}
\centerline{
\psfig{file=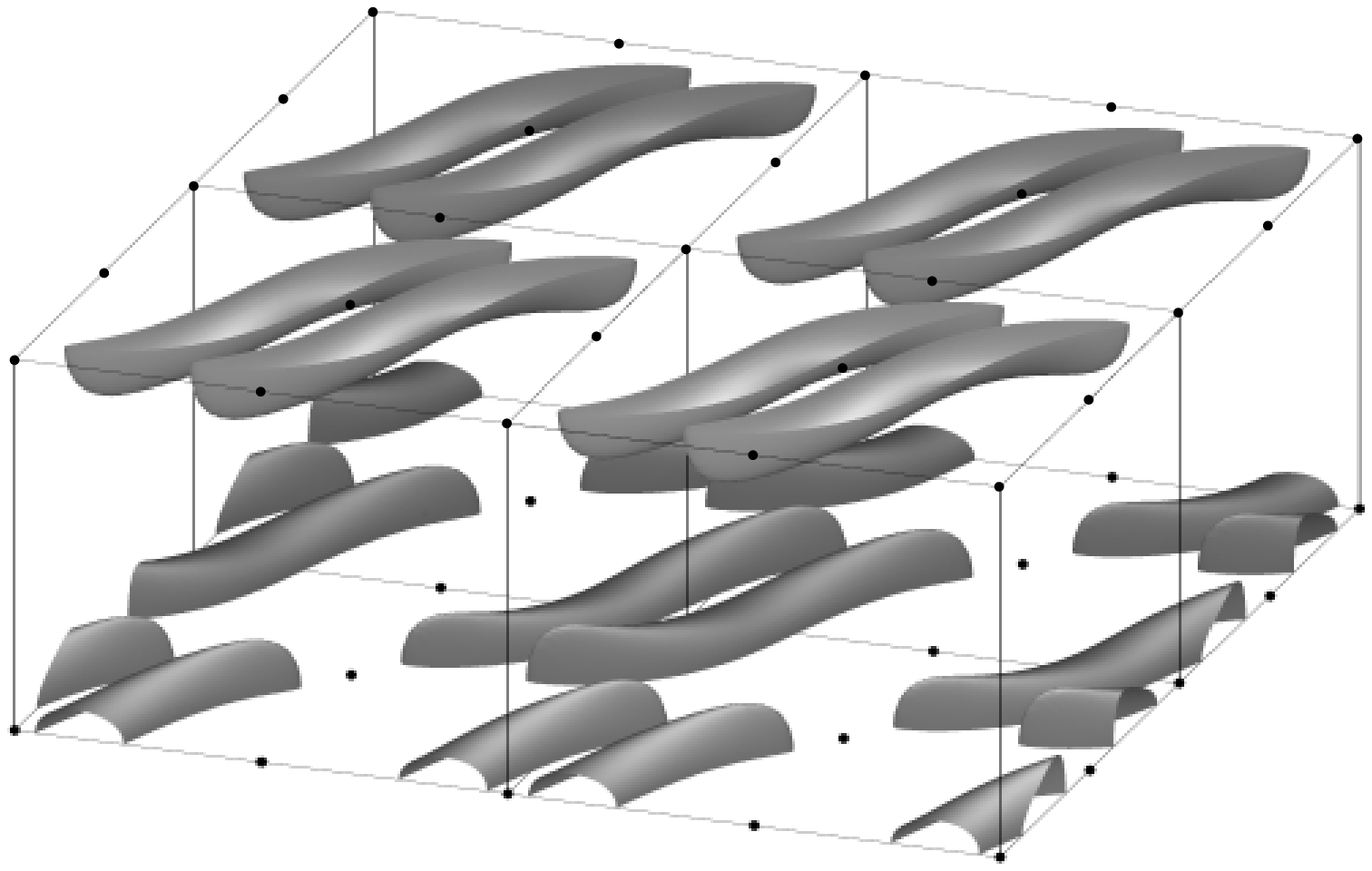,width=7cm,clip=}
\psfig{file=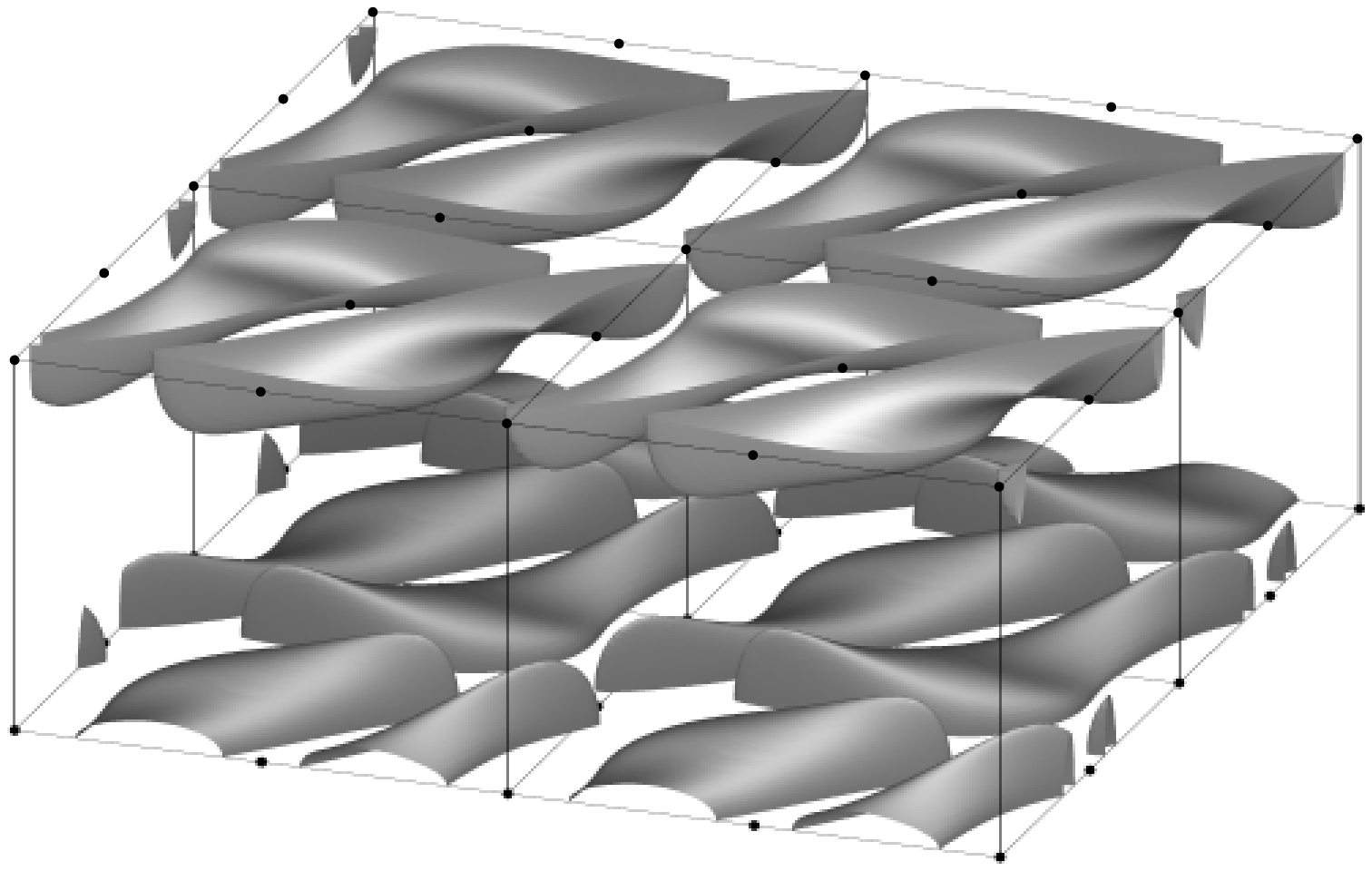,width=7cm,clip=}}
\hspace{42mm}(g)\hspace{66mm}(h)
\caption{Isosurfaces of magnetic energy density of magnetic fields,
at the level of a half of the maximum, in steady convective MHD attractors:
$Ta=150$, S$^{\rm R1}_1$ (a); $Ta=200$, S$^{\rm R1}_2$ (b);
$Ta=250$, S$^{\rm R1}_3$ (c); $Ta=300$, S$^{\rm R1}_4$ (d);
$Ta=430$, S$^{\rm R1}_5$ (e); $Ta=600$, S$^{\rm R1}_6$ (f);
$Ta=675$, S$^{\rm R1}_7$ (g); $Ta=680$, S$^{\rm R1}_8$ (h);
$Ta=721$, S$^{\rm WR}$ (i); $Ta=690$, S$^{\rm RD}_1$ (k);
$Ta=740$, S$^{\rm RD}_2$ (l); $Ta=775$, S$^{\rm RD}_3$ (m);
$Ta=1175$, S$^{\rm RD}_4$ (n).
Stagnation points of the flow on the horizontal boundaries are shown
by dots. Four periodicity cells are displayed.\label{fig:nlin_isosb}}
\end{figure}

\begin{figure}
\centerline{
\psfig{file=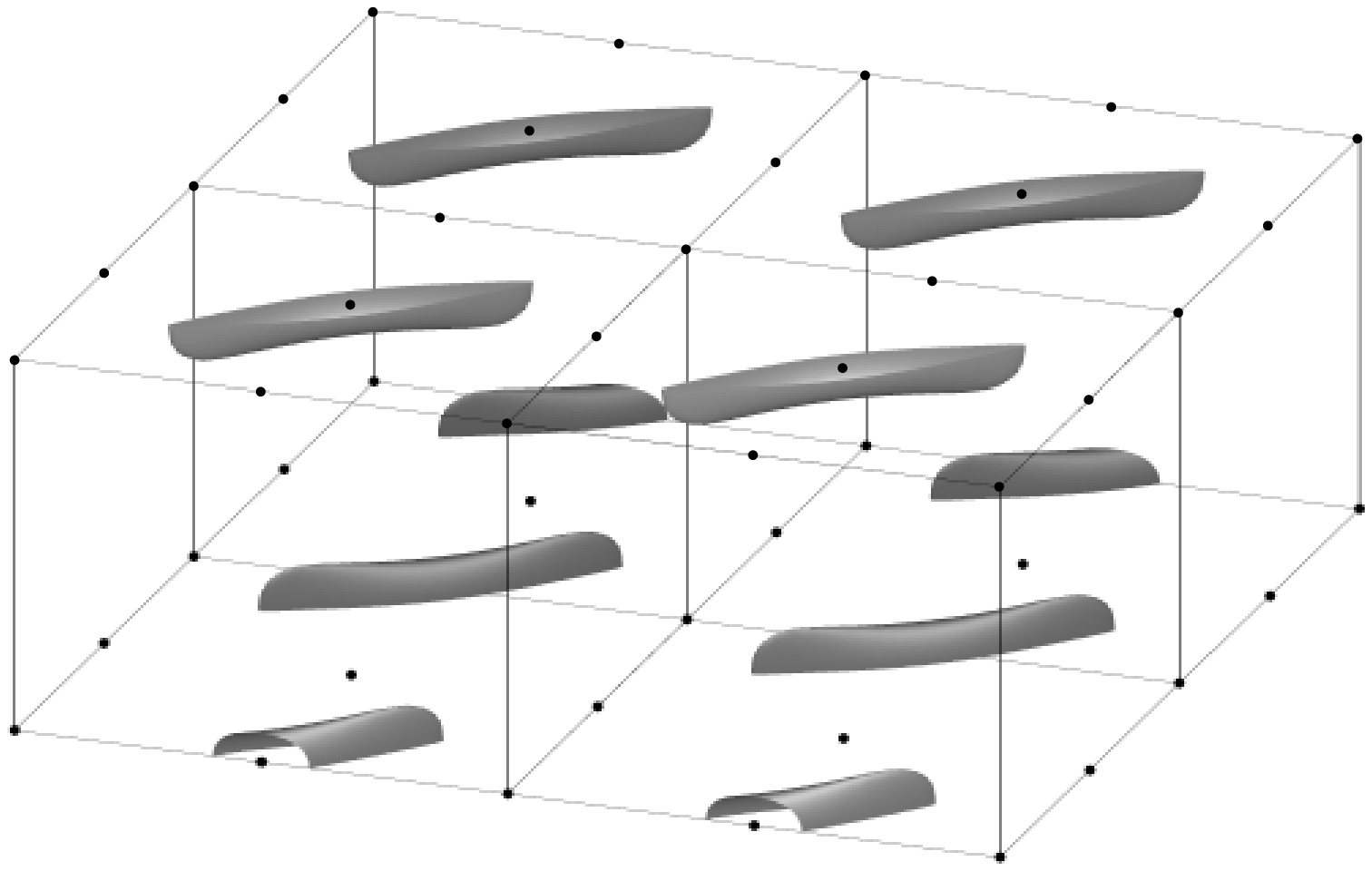,width=7cm,clip=}
\psfig{file=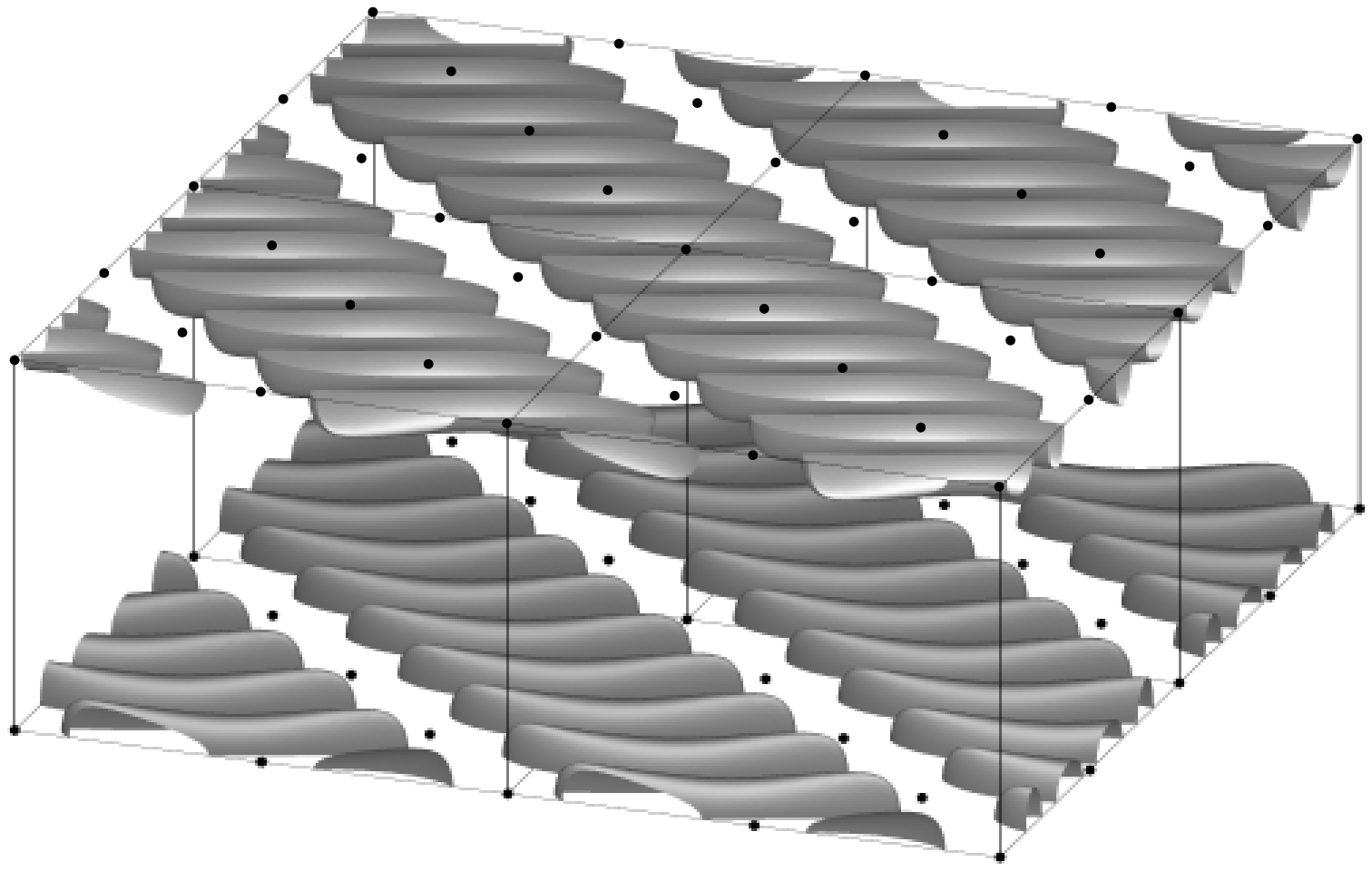,width=7cm,clip=}}
\hspace{42mm}(i)\hspace{66mm}(k)

\vspace*{2mm}
\centerline{
\psfig{file=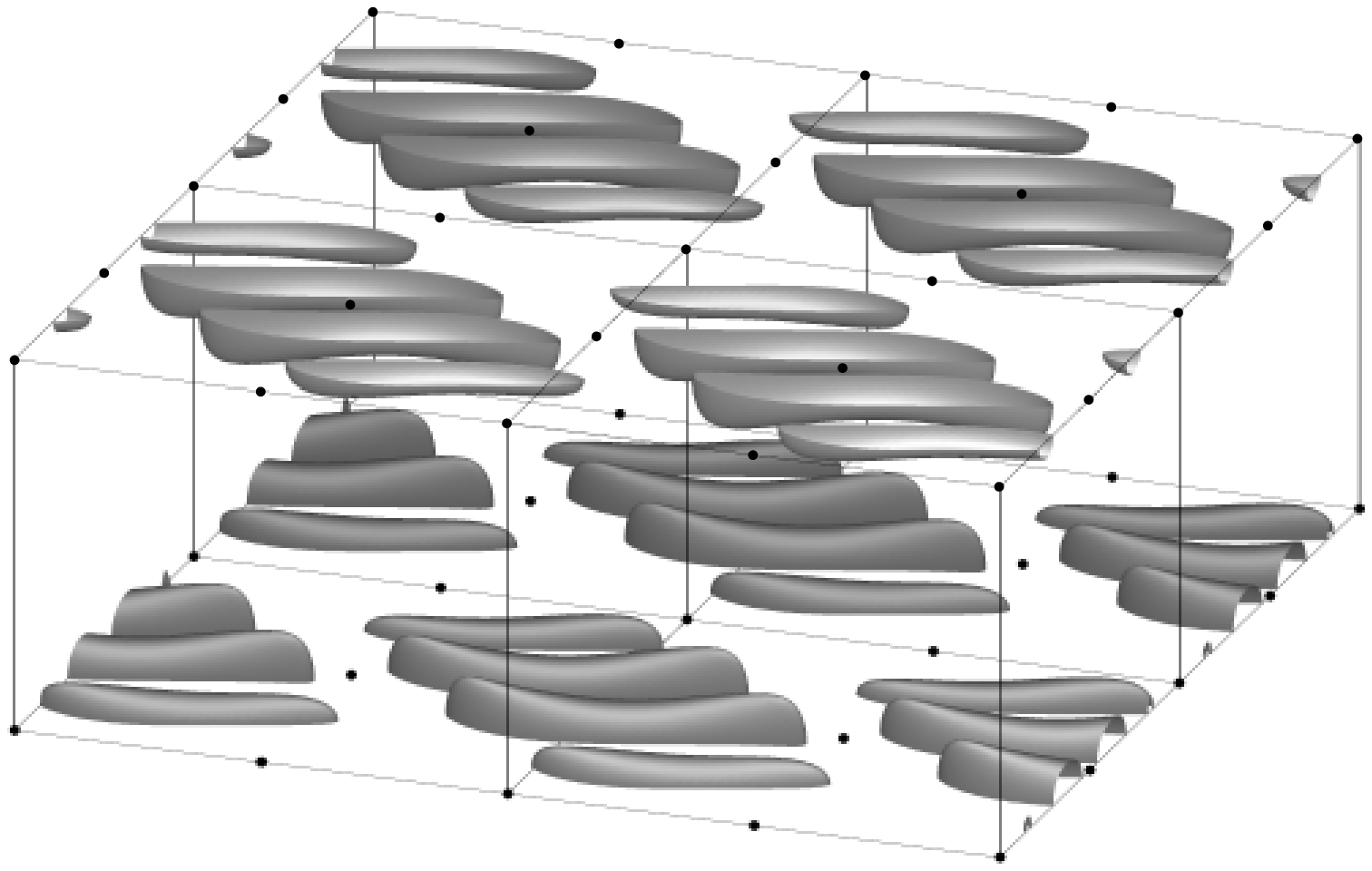,width=7cm,clip=}
\psfig{file=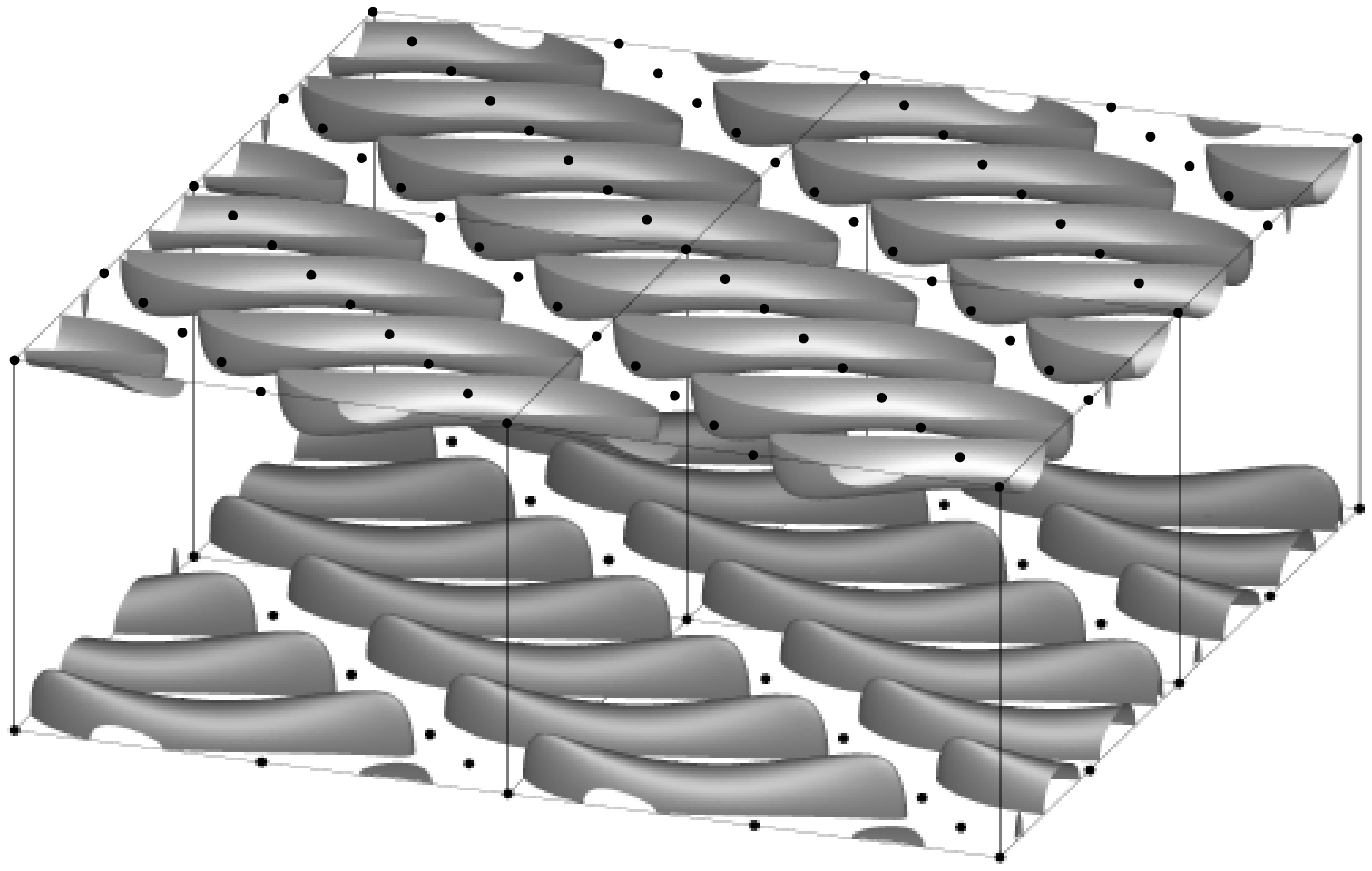,width=7cm,clip=}}
\hspace{42mm}(l)\hspace{66mm}(m)

\vspace*{2mm}
\centerline{
\psfig{file=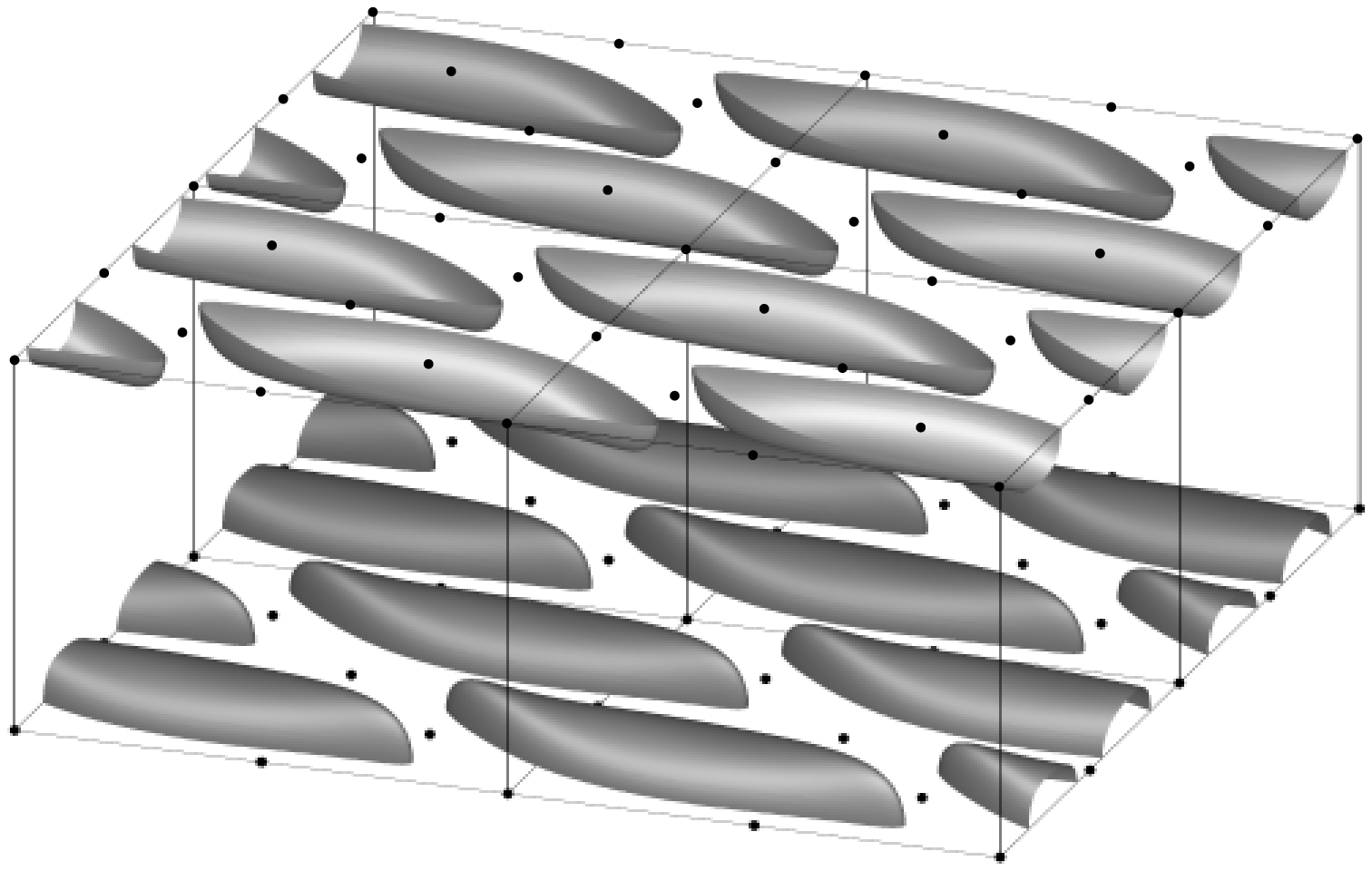,width=7cm,clip=}}
\centerline{(n)}

\vspace*{5mm}
\centerline{\small\protect{\xrf{fig:nlin_isosb}}: continuation.}
\end{figure}

\subsection{MHD attractors emerging from {\rm TW}, $0<Ta\le 81$}

The bifurcation diagram of MHD regimes at the interval $0<Ta<200$
is shown on \xrf{diagr}.

The interval of $Ta$, where TW exists, consists of two subintervals of kinematic
dynamo action, separated by a window $82\le Ta\le86$ of non-generating TW
regimes. Growing magnetic modes generated in the two subintervals differ, for
instance, in their symmetries (see Table~\ref{tab:conv_lambda}). This results
in emergence of MHD attractors of different types. At the lower subinterval of
$Ta$ the primary MHD attractor is P$^{\rm TW}_1$, similarly to TW drifting along
both horizontal axes, $x$ and $y$, and periodic in the co-moving reference frame.
At the right end the MHD branch P$^{\rm TW}_1$ terminates on TW at $Ta=81.80$,
where the respective magnetic eigenvalue of TW crosses the imaginary axis;
to the left it continues to $Ta=0$.

{\tabcolsep1.5mm
\begin{table}[p]
\caption{Families of attractors F$^{\rm TW}$ and C$^{\rm SW}$. Column 2
presents the type of attractors (disregarding drift frequencies): P periodic,
Q quasiperiodic, C chaotic, column 3 basic frequencies of periodic and
quasiperiodic regimes, columns 5 and 6 time-averaged kinetic and magnetic
energies, respectively, the last column the list of individual runs as pairs
$Ta$(integration time).\label{tab:qtw}}
\begin{center}\begin{tabular}{|c|c|c|c|c|c|c|c|}\hline
Fa-&Type&\multicolumn{2}{c|}{Basic}&Interval of&$E_k$&$E_m$&Individual\\
mily&&\multicolumn{2}{c|}{frequencies}&existence&&&runs\\\hline
F$^{\rm TW}$&P$^{\rm TW}_2$&2.86--2.87&&[36,\,57] &234.10--235.93&9.43--11.82
&36(351),41(22),\\
&&&&&&&45(29),48(59),\\
&&&&&&&54(251),57(1274)\\\cline{2-8}
&Q&2.87--2.94&0.209--0.151&[60,\,68]&235.95--235.11&12.01--12.20&60(337),63(370),\\
&&&&&&&65(2378),68(352) \\\cline{2-8}
&Q&2.93&0.071&70       &235.24        &12.03       &70(1646)\\\cline{2-8}
&Q&2.93&0.034&71.5     &235.20        &12.05       &71.5(2022)\\\cline{2-8}
&Q&2.93&0.016&72       &235.16        &12.08       &72(1696)\\\cline{2-8}
&C&    & &[73,\,74]&234.76--234.60&11.91--11.81&73(1032),74(2620)\\\cline{2-8}
&Q&2.92-2.94&0.035--0.030&[75,\,77]&234.05--233.31&11.86--11.30&75(457),77(2413)\\\cline{2-8}
&Q&2.91&0.015&[78,\,78.5]&233.90--233.62&11.62--11.60&78(2602),78.5(1427)\\\cline{2-8}
&Q&2.91&0.069--0.061&[79,\,80] &232.62--232.74&11.04--11.25&79(223),80(1784)\\\hline
C$^{\rm SW}$&C&&      &78       &236.55        &17.00 &78(2796)\\\cline{2-8}
&Q&2.86&0.048--0.051&[80,\,81]&236.76--236.86&16.63--16.62&80(1681),81(1379)\\\cline{2-8}
&C&&&[82,\,84] &237.35--237.91&16.68--16.49&82(2558),83(3174),\\
&&&&&&&84(1950)\\\cline{2-8}
&Q&2.90&0.017&85       &237.83        &16.39       &85(4052)\\\cline{2-8}
&C&&&[87,\,96.5]&238.27--240.01&16.22--15.01&87(1812),90(2851),\\
&&&&&&&95(3582),96(953)\\
&&&&&&&96.5(1379)\\\cline{2-8}
&Q&2.93&0.016--0.021&[97,\,98]  &241.13--240.94&14.52--14.58&97(928),98(2511)\\\cline{2-8}
&C&&&[99,100]    &239.86--239.85&15.02--15.05   &99(1682),100(1999)\\\cline{2-8}
&Q&2.88&0.026&[101,\,102]&239.98--240.05&14.91--14.83&101(979),102(3607)\\\cline{2-8}
&Q&2.88&0.052--0.056&[103,\,105]&240.12--240.38&14.73--14.41&103(1339),105(3847)\\\cline{2-8}
&P&2.90-2.89&&[106,\,119]&239.59--239.93&13.42--13.04&106(1177),107(199),\\
&&&&&&&110(511),115(571),\\
&&&&&&&117.5(3768),119(1181)\\\cline{2-8}
&P$^{\rm SW}$&2.89-2.88&&[120,\,126]&239.82--239.90&13.03--13.31&120(675),121.5(517),\\
&&&&&&&123(411),126(1002)\\\hline
\end{tabular}\end{center}\end{table}}

\subsection{MHD attractors emerging from {\rm TW}, mode interaction}

As noted in Section 3, TW bifurcates supercritically in a Hopf bifurcation
from R$_1$ as $Ta$ decreases below 188. For the employed value of $P_m$,
at the point of bifurcation R$_1$ generates magnetic field kinematically.
The respective magnetic eigenvalue is real and not large, hence attractors
observed for $Ta$ not far from 188 can be related to interaction of two
instability modes of R$_1$, hydrodynamic and magnetic ones.

The symmetry group of R$_1$ is generated by $s_2$, $r\gamma^x_{L/2}$,
$\gamma^y$ and $q$. In the hydrodynamic eigenspace the symmetries $s_2$ and
$\gamma^y$ act non-trivially, and in the magnetic eigenspace the action of
the symmetries $s_2$, $q$ and $\gamma^y$ is non-trivial. Hence, the results of
Golubitsky\al\ (1988) (Section XX \S 2) on Hopf / steady-state mode interaction
with the {\bf O}(2) symmetry group are applicable. (The symmetry $q$ plays no
r\^ole, since, in the notation {\it ibid}., $q=(\pi,\pi)$ is a superposition of
two shifts, in $\phi=\pi$ and $\theta=\pi$, which are elements of {\bf SO}(2)
and ${\bf S}^1$, respectively. In our problem, the action of {\bf SO}(2) is
generated by the shifts $\gamma^y$ along the direction of the rolls.)
According to Golubitsky\al\ (1988), the trivial steady state (R$_1$ in our case)
bifurcates with emergence of an $s_2$-symmetric steady state, when a real
eigenvalue becomes positive, and with a simultaneous emergence of rotating
(i.e. travelling, in our parlance), TW, and standing waves, SW, when a complex pair
of eigenvalues crosses the imaginary axis. The rotating wave can further
bifurcate to a modulated rotating wave (which in fact is a 2-torus).

\begin{figure}
\centerline{\psfig{file=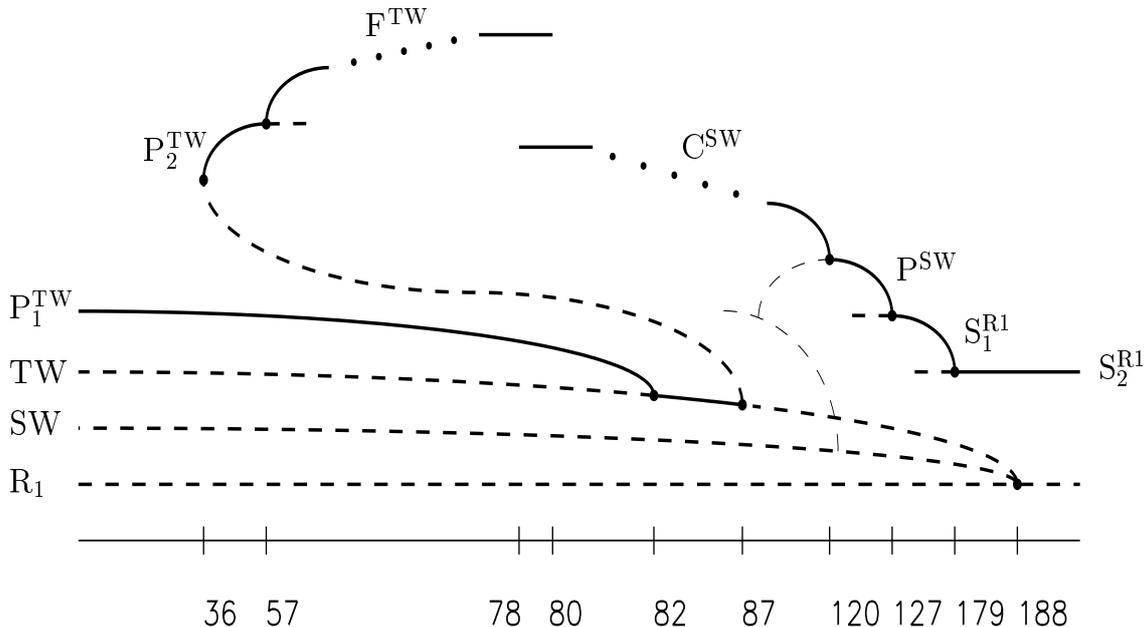,width=15cm}}
\caption{Bifurcation diagram of the MHD system for $0<Ta<200$. Solid lines
denote stable branches, dashed unstable, thin dashed conjectured.}
\label{diagr}\end{figure}

In our system, in agreement with this general theory, R$_1$ bifurcates to TW
(in the hydrodynamic subspace) and to S$^{\rm R1}_2$. TW further bifurcates
to a modulated travelling wave, P$^{\rm TW}_2$ (which we classify as a periodic
orbit, since we omit drift frequencies in the description of attractors).
At $Ta=87$ the bifurcation where P$^{\rm TW}_2$ emerges is subcritical,
and close to the point of bifurcation the periodic orbit is unstable.
It gains stability as it turns back in a saddle-node
bifurcation at $Ta=36$. We classify the stable periodic orbit observed in
computations as the modulated rotating wave predicted by the theory,
because 1) of the similarity of the spatial structure of TW and the flow in
P$^{\rm TW}_2$, as well as of the dominant magnetic mode of TW and magnetic field
in P$^{\rm TW}_2$; 2) the symmetry group of P$^{\rm TW}_2$ is the one expected
on the theoretical grounds for the branch emerging at $Ta=87$ from TW due to the
mode interaction (note that it includes a spatio-temporal symmetry: the symmetry
about a vertical axis with a shift in time by a half of a period); and 3)
the temporal frequency of P$^{\rm TW}_2$ is close to the Hopf frequency of TW.

The steady state S$^{\rm R1}_2$ bifurcates as $Ta$ is decreased to a less
symmetric steady state S$^{\rm R1}_1$, which is outside the Hopf / steady-state
mode interaction center manifold. As $Ta$ is decreased further, S$^{\rm R1}_1$
becomes unstable in a Hopf bifurcation with a stable periodic
orbit emerging. We label it P$^{\rm SW}$, because its spatial structure and
temporal frequency are similar to those of the unstable standing wave SW emerging
from R$_1$ simultaneously with the travelling wave TW, and its symmetry group
is a subgroup of the one of SW. The general theory of Hopf / steady-state mode
interaction predicts two types of periodic orbits bifurcating from SW,
but judging by its symmetries none of them is our P$^{\rm SW}$. The conjectured
relation of P$^{\rm SW}$ with SW is shown by thin dashed lines on \xrf{diagr}.

\subsection{MHD attractors emerging from {\rm TW}, $36<Ta\le 127$}

Both periodic orbits, P$^{\rm TW}_2$ and P$^{\rm SW}$, give rise to complex
families of attractors, F$^{\rm TW}$ and C$^{\rm SW}$, discussed in this
subsection (see Table~\ref{tab:qtw}). Like P$^{\rm TW}_1$, all attractors
constituting F$^{\rm TW}$ drift along both horizontal axes and thus have two
drift frequencies. The family F$^{\rm TW}$ starts from the primary
periodic MHD attractor P$^{\rm TW}_2$ at $Ta=36$.

As $Ta$ is increased beyond $Ta\approx 60$, the second frequency, $f_2$,
appears in a Hopf bifurcation, i.e. a stable torus emerges. Afterwards,
a sequence of bifurcations of halving of $f_2$ takes place (often such
a sequence is called a cascade of period doubling bifurcations for a torus;
we prefer to abstain from the use of this terminology, since the concept
of a period applied to a torus is not all too transparent). This sequence of
bifurcations is analogous to the Feigenbaum (1978) scenario of period doublings
for a logistic map, and we label this family of attractors by F.

Logistic map is defined by the recurrent relation $x_{n+1}=rx_n(1-x_n)$,
$0<x<1$. When $r$ is increased over 3, a stable 2-cycle is created in a flip
bifurcation. Subsequent flip bifurcations result in emergence of cycles of
lengths 4, 8, $\ldots$. The bifurcations accumulate at $r=3.57$\,. For larger
$r$, chaotic behaviour sets in, alternating with windows of periodic orbits of
periods $(2m+1)2^k$. For $r>4$, $x_n$ escapes from the interval $0<x<1$ and
diverges.

\begin{figure}
\centerline{
\psfig{file=fig14/eb_ta65.ps,width=7cm,height=45mm,clip=}
\psfig{file=fig14/eb_ta70.ps,width=7cm,height=45mm,clip=}}
\hspace{46mm}(a)\hspace{65mm}(b)

\vspace*{2mm}
\centerline{
\psfig{file=fig14/eb_ta71.5.ps,width=7cm,height=45mm,clip=}
\psfig{file=fig14/eb_ta72.ps,width=7cm,height=45mm,clip=}}
\hspace{46mm}(c)\hspace{65mm}(d)

\vspace*{2mm}
\centerline{
\psfig{file=fig14/eb_ta74.ps,width=7cm,height=45mm,clip=}
\psfig{file=fig14/eb_ta77.ps,width=7cm,height=45mm,clip=}}
\hspace{46mm}(e)\hspace{65mm}(f)

\vspace*{2mm}
\centerline{
\psfig{file=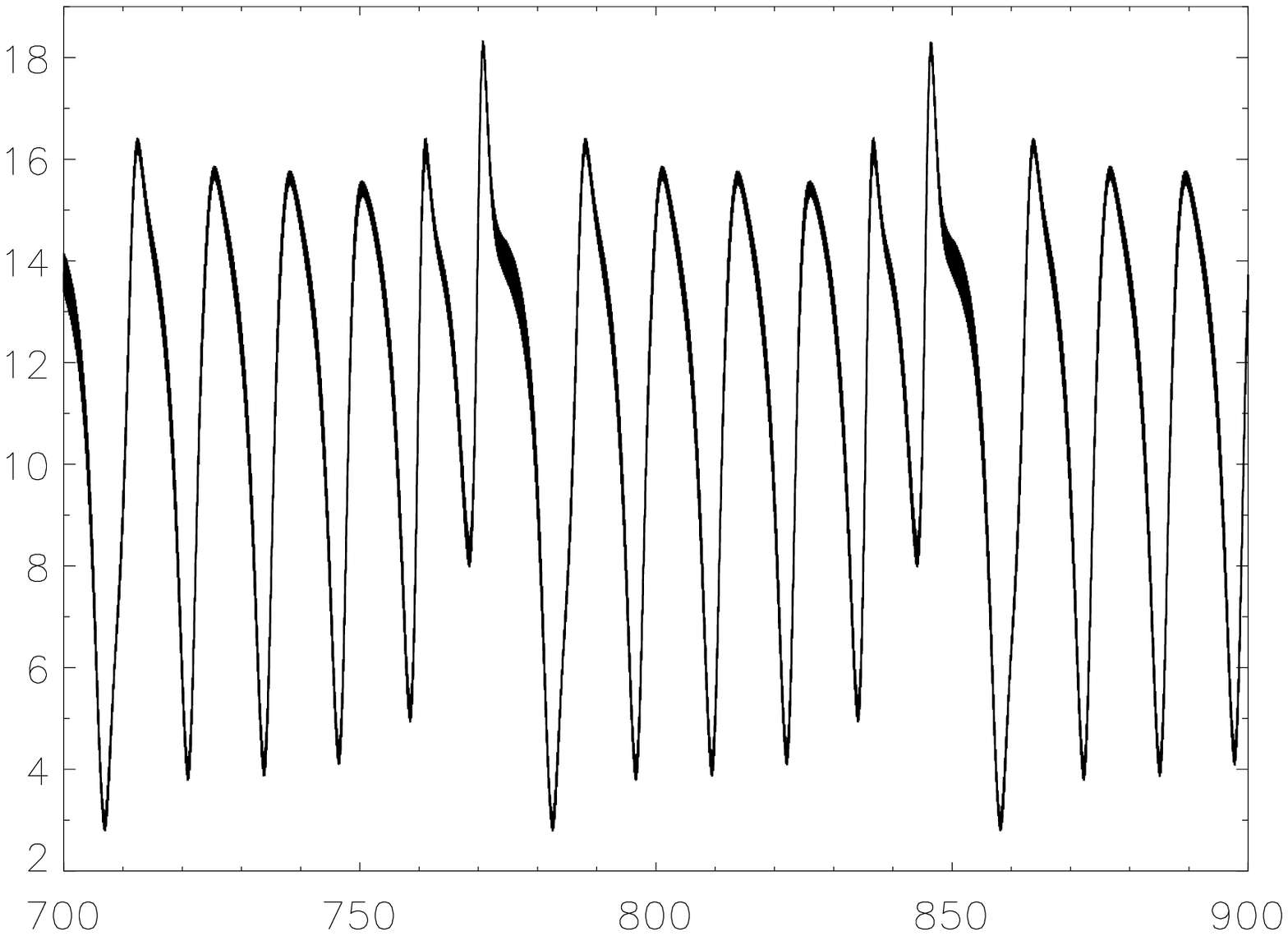,width=7cm,height=45mm,clip=}
\psfig{file=fig14/eb_ta80.ps,width=7cm,height=45mm,clip=}}
\hspace{46mm}(g)\hspace{65mm}(h)
\caption{Magnetic energy (vertical axis) versus time (horizontal axis)
for F$^{\rm TW}$: $Ta=65$ (a), $Ta=70$ (b), $Ta=71.5$ (c), $Ta=72$ (d)
$Ta=74$ (e), $Ta=77$ (f), $Ta=78.5$ (g), $Ta=80$ (h).\label{fig:qtw1_en}}
\end{figure}

In the convective hydromagnetic system, on increasing $Ta$ we have detected
three halvings of the frequency $f_2\approx0.15$, followed by a chaotic
behaviour. For larger $Ta$, lower basic frequencies set in, $f_2/3$ for
$75\le Ta\le 77$ and $f_2/6$ for $78\le Ta\le 78.5$; their appearance agrees
with the general theory. No more halvings have been observed.
This is consistent with the fact that a complete sequence of period doubling
bifurcations of tori is a structurally unstable scenario (see Coullet, 1984), and
generically it is interrupted by the onset of chaos. (We could also miss some
halvings just having considered not enough values of $Ta$.) For still larger
$Ta$, a frequency $\tilde f_2\approx f_2/2$ is observed. On the one hand,
emergence of this quasiperiodic regime is clearly outside the framework of
the Feigenbaum period doubling scenario (cf.~\xrf{fig:qtw1_en} (b) and (h)\,),
suggesting that attractors for $79\le Ta\le 80$ do not belong to our Feigenbaum
family. On the other, the standard theory takes into account only two first
terms in the Taylor expansion of the dynamical system on the central manifold,
and terms of higher order may be responsible for the birth of this new regime.

\begin{figure}
\centerline{
\psfig{file=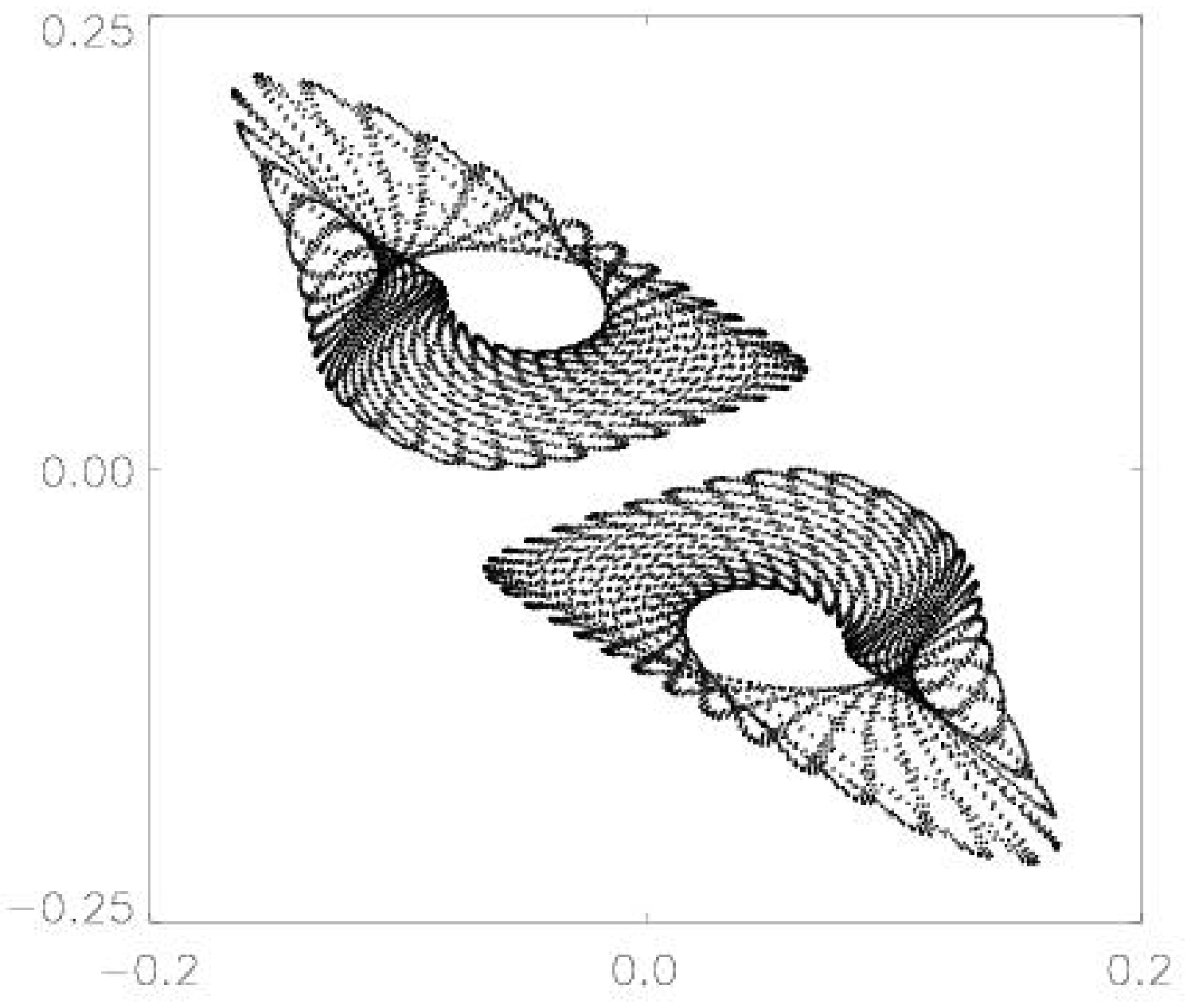,width=55mm,clip=}
\psfig{file=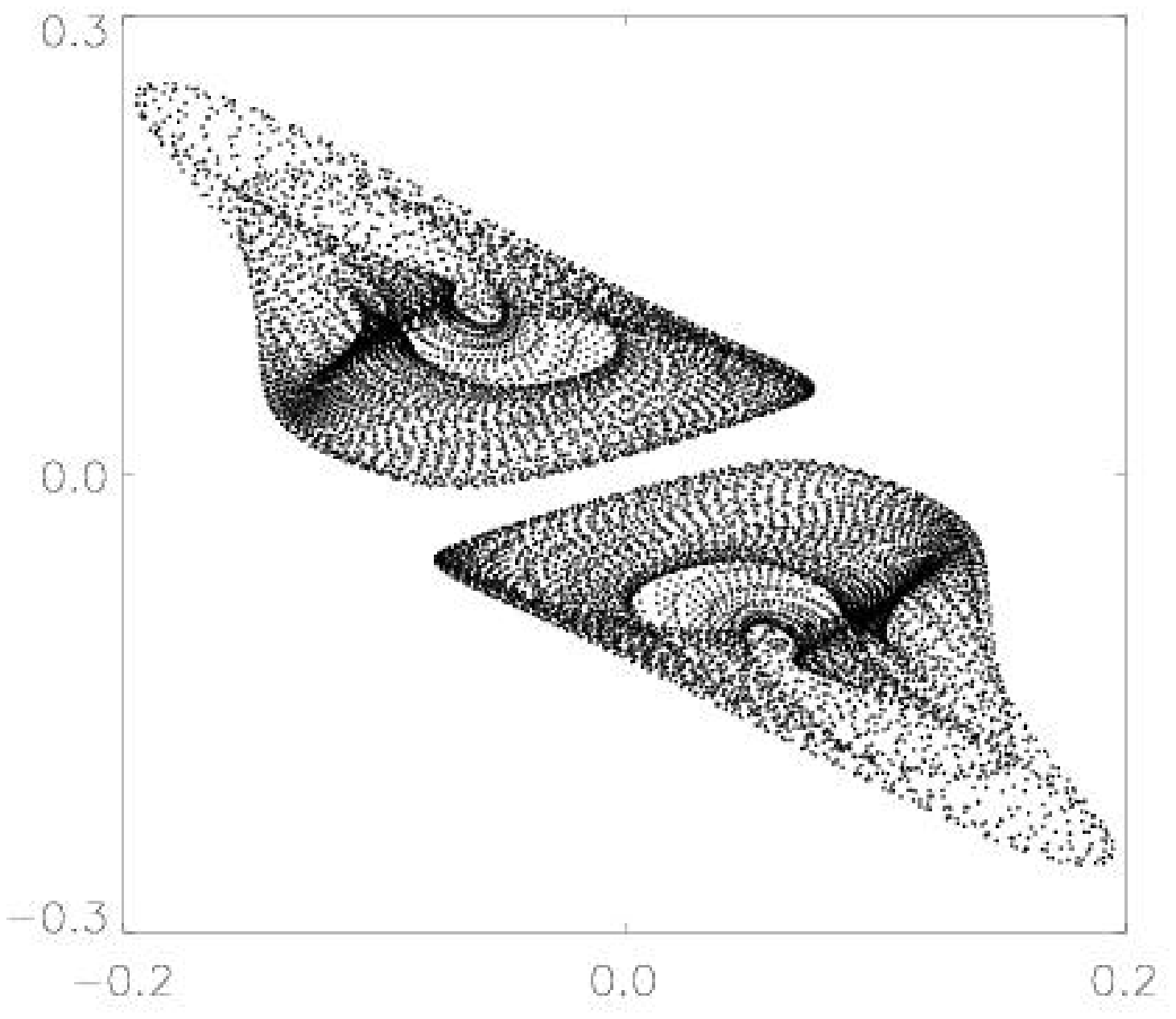,width=55mm,clip=}
\psfig{file=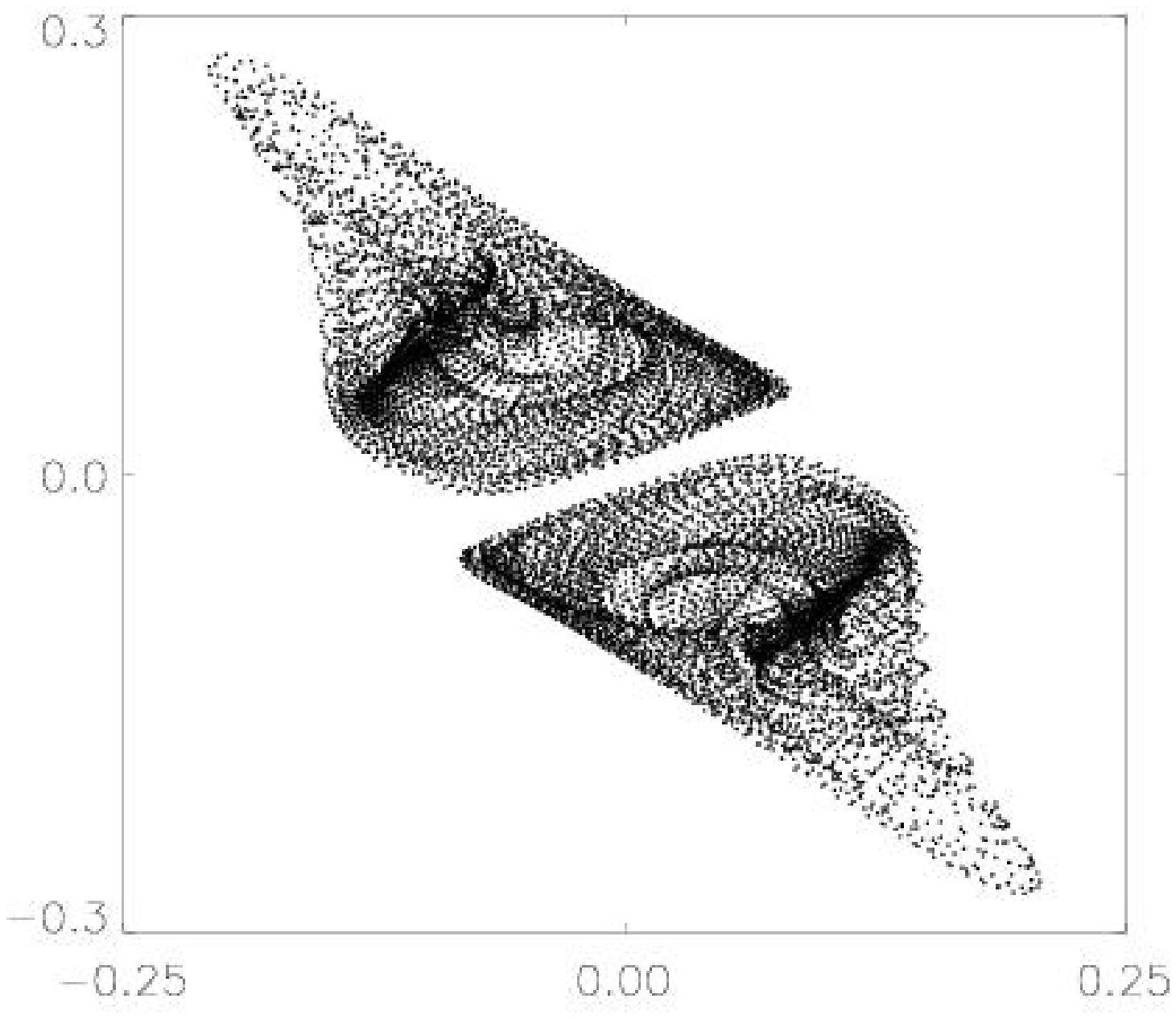,width=55mm,clip=}}
\hspace{27mm}(a)\hspace{51mm}(b)\hspace{51mm}(c)

\vspace*{2mm}
\centerline{
\psfig{file=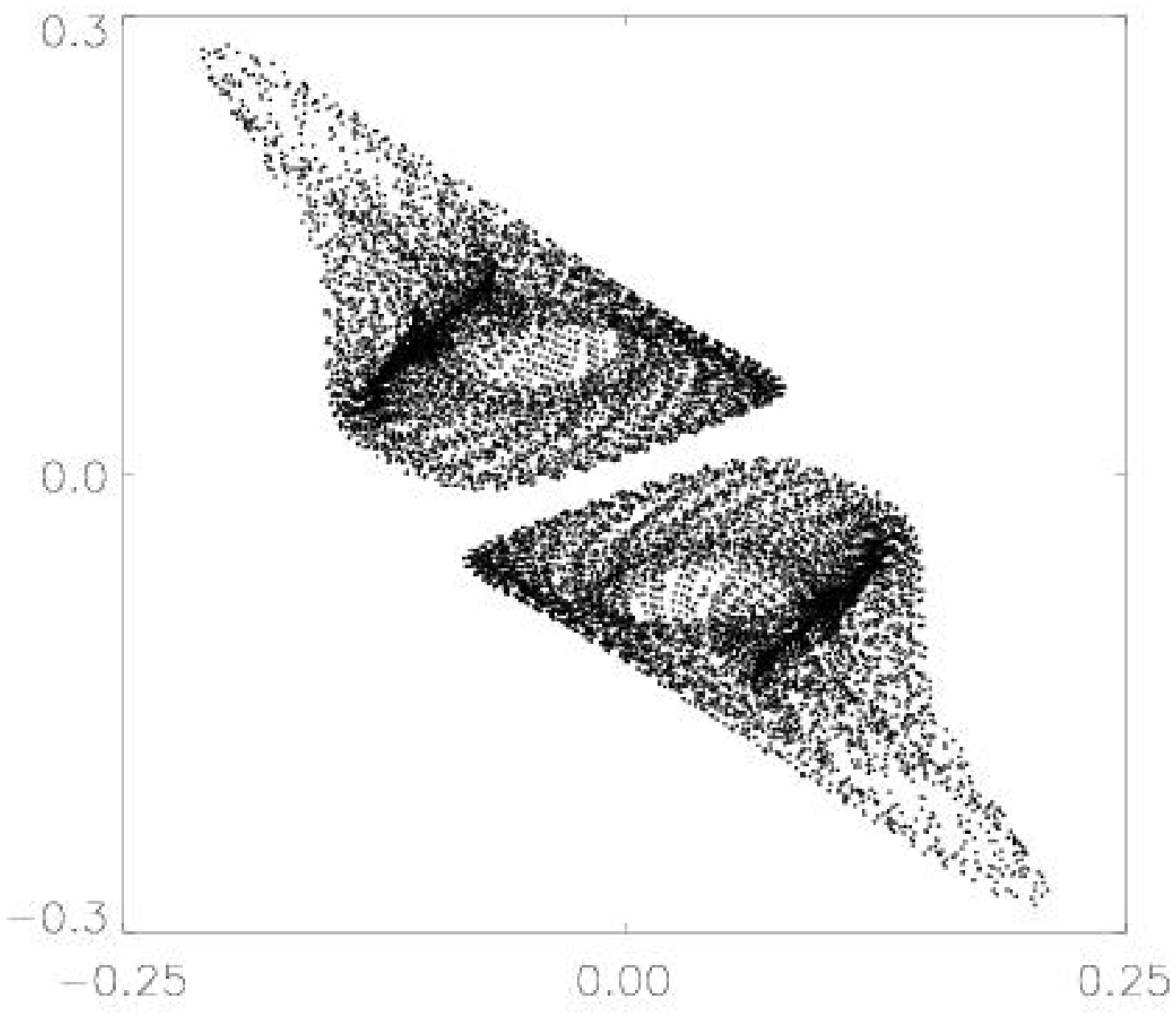,width=55mm,clip=}
\psfig{file=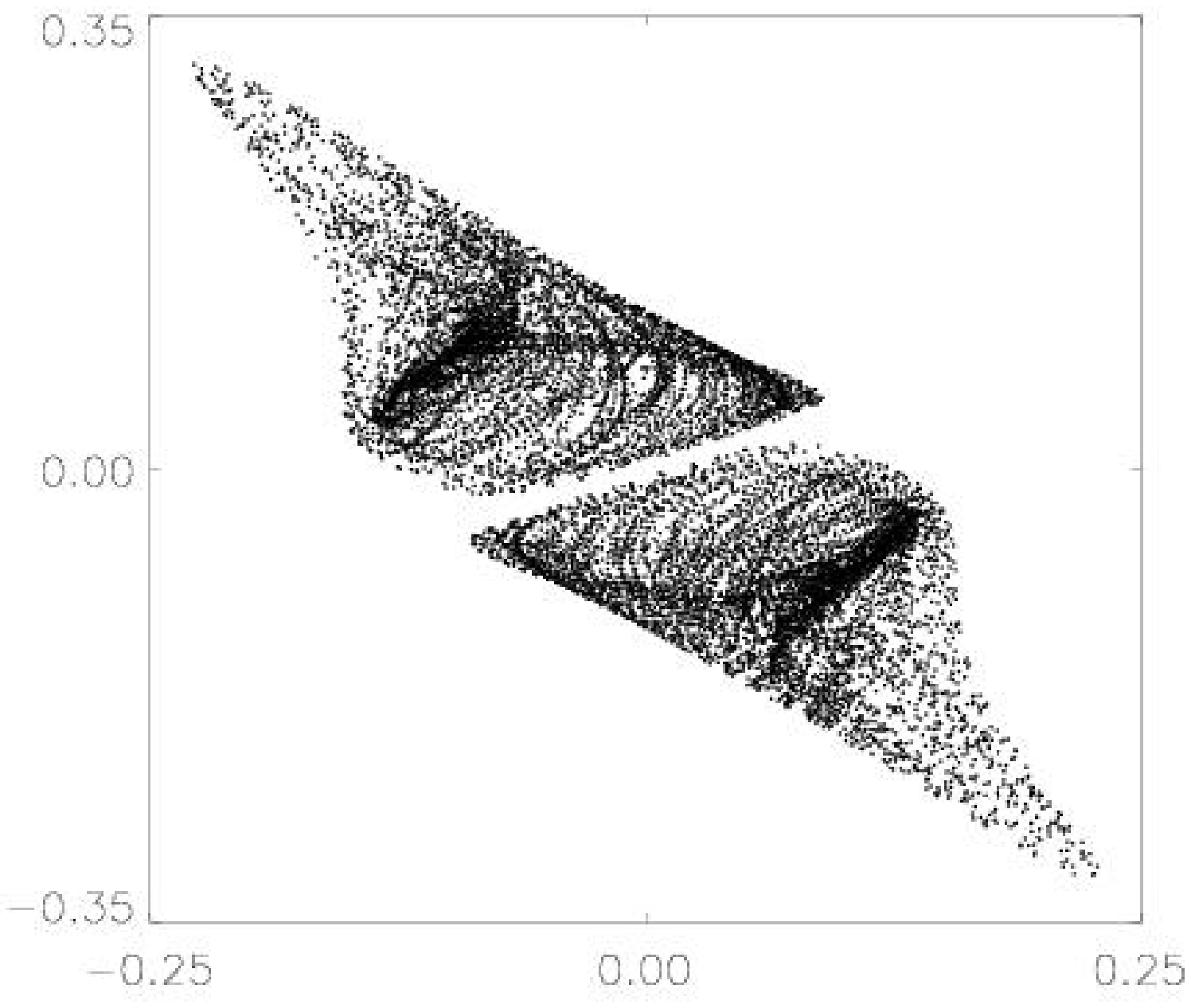,width=55mm,clip=}
\psfig{file=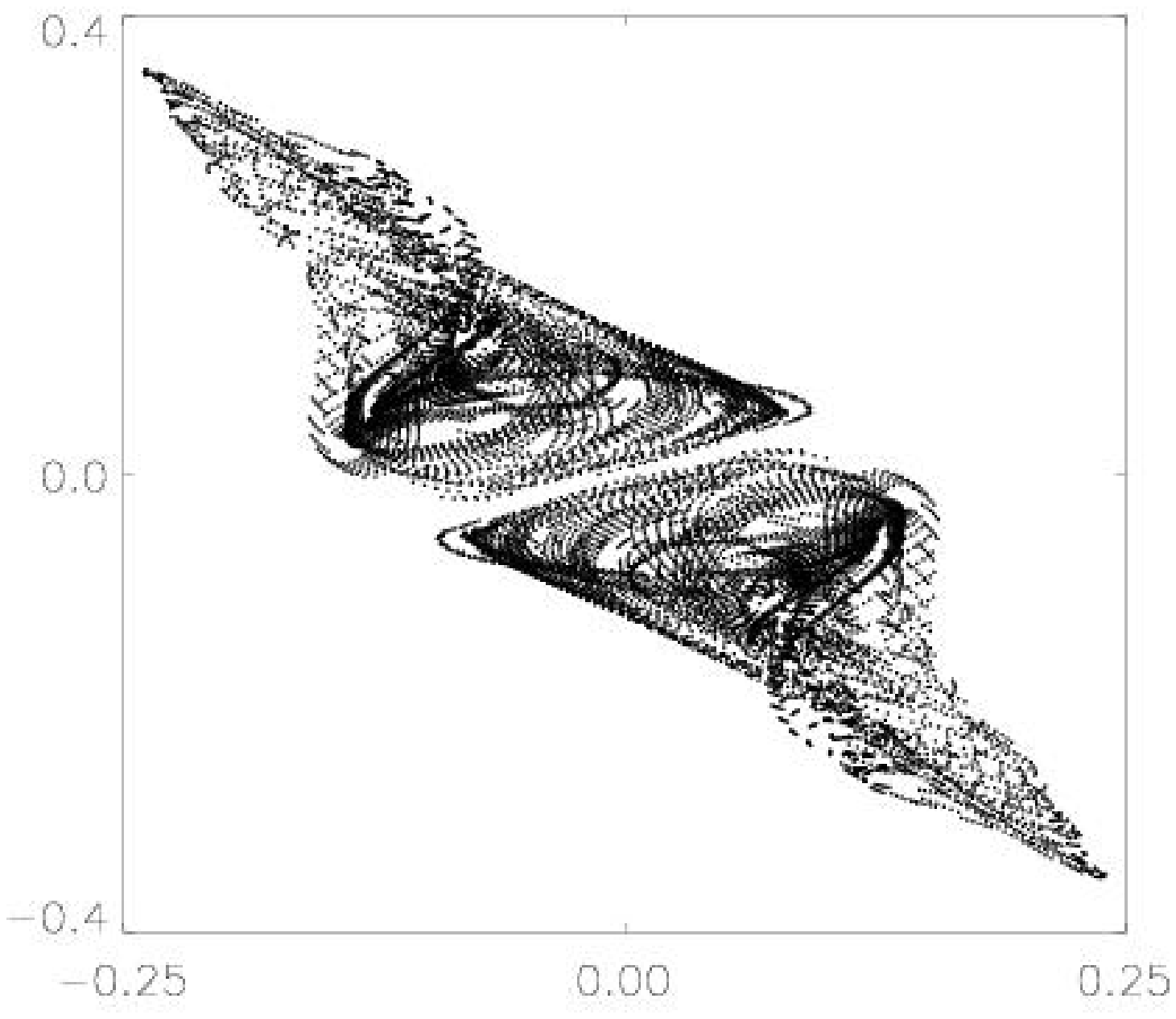,width=55mm,clip=}}
\hspace{27mm}(d)\hspace{51mm}(e)\hspace{51mm}(f)

\vspace*{2mm}
\centerline{
\psfig{file=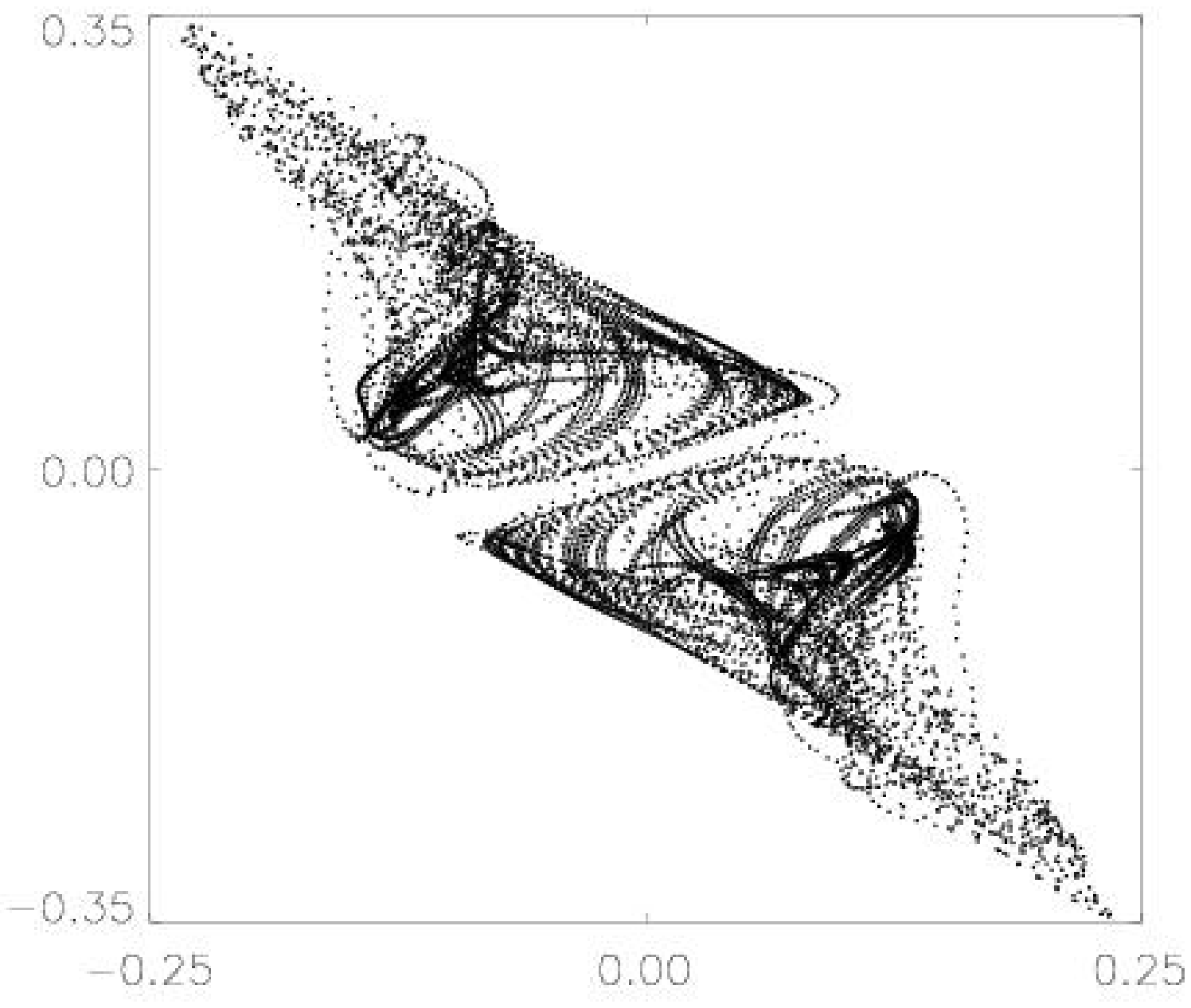,width=55mm,clip=}
\psfig{file=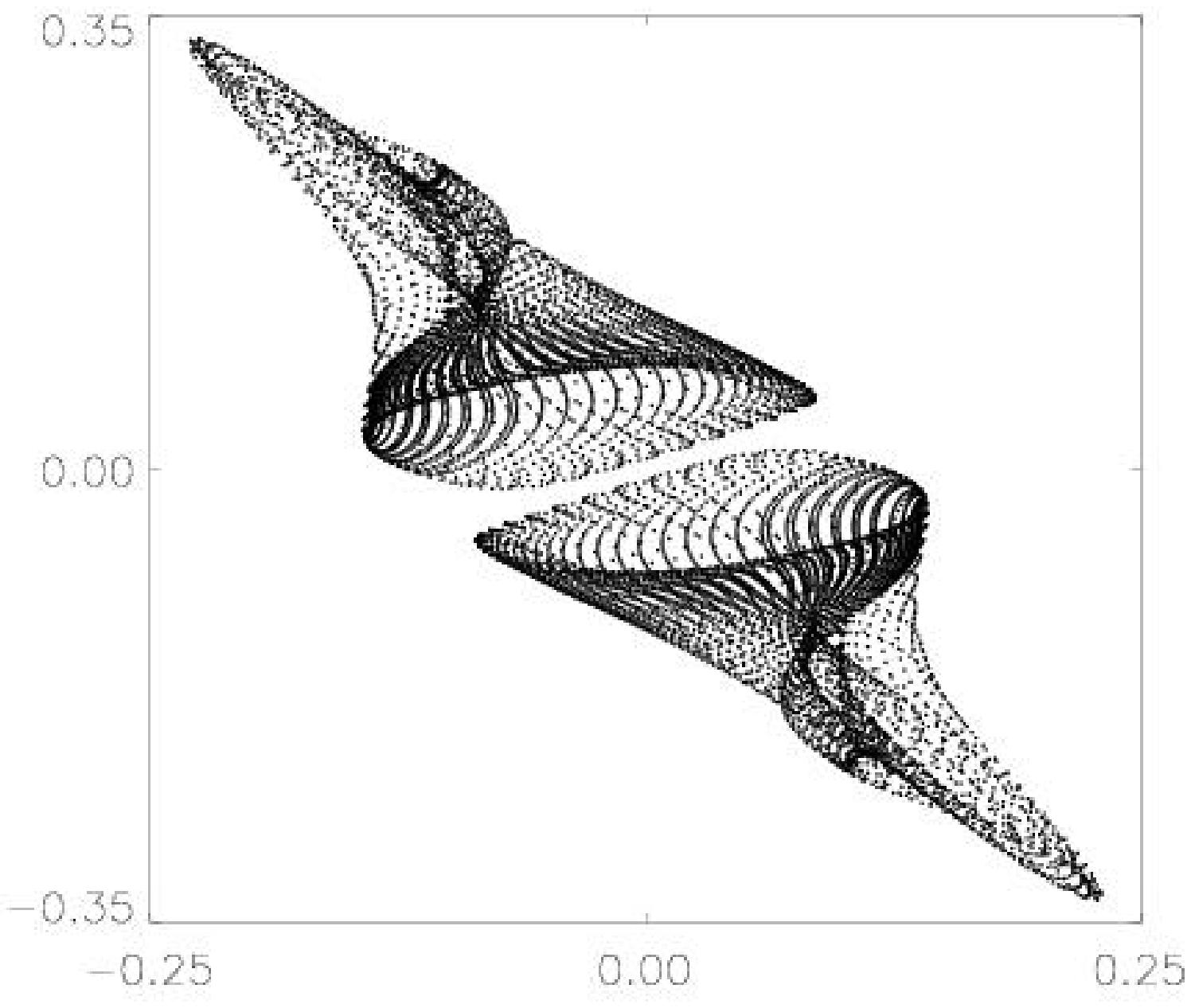,width=55mm,clip=}}
\hspace{55mm}(g)\hspace{51mm}(h)
\caption{Poincar\'e sections of regimes in the family F$^{\rm TW}$
on the (${\rm Re}\hat{v}^x_{-1,2,1},{\rm Re}\hat{v}^y_{-1,2,1})$ plane
(horizontal and vertical axes, respectively) defined by the condition
Re$\hat{v}^z_{-1,2,1}=0$:
$Ta=65$ (a), $Ta=70$ (b), $Ta=71.5$ (c), $Ta=72$ (d), $Ta=74$ (e),
$Ta=77$ (f), $Ta=78.5$ (g), $Ta=80$ (h).\label{fig:poincare}}
\end{figure}

\begin{figure}
\centerline{
\psfig{file=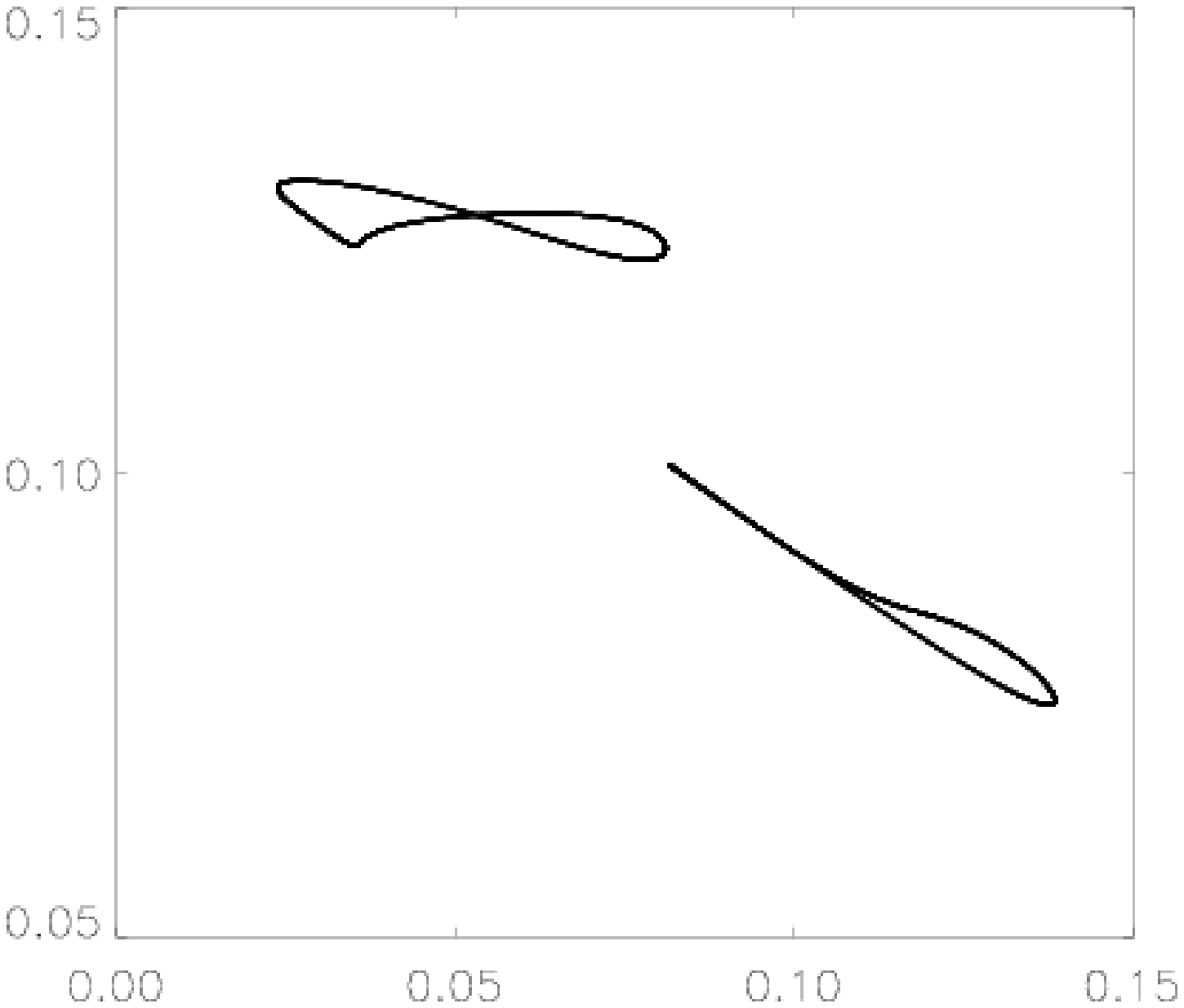,width=55mm,clip=}
\psfig{file=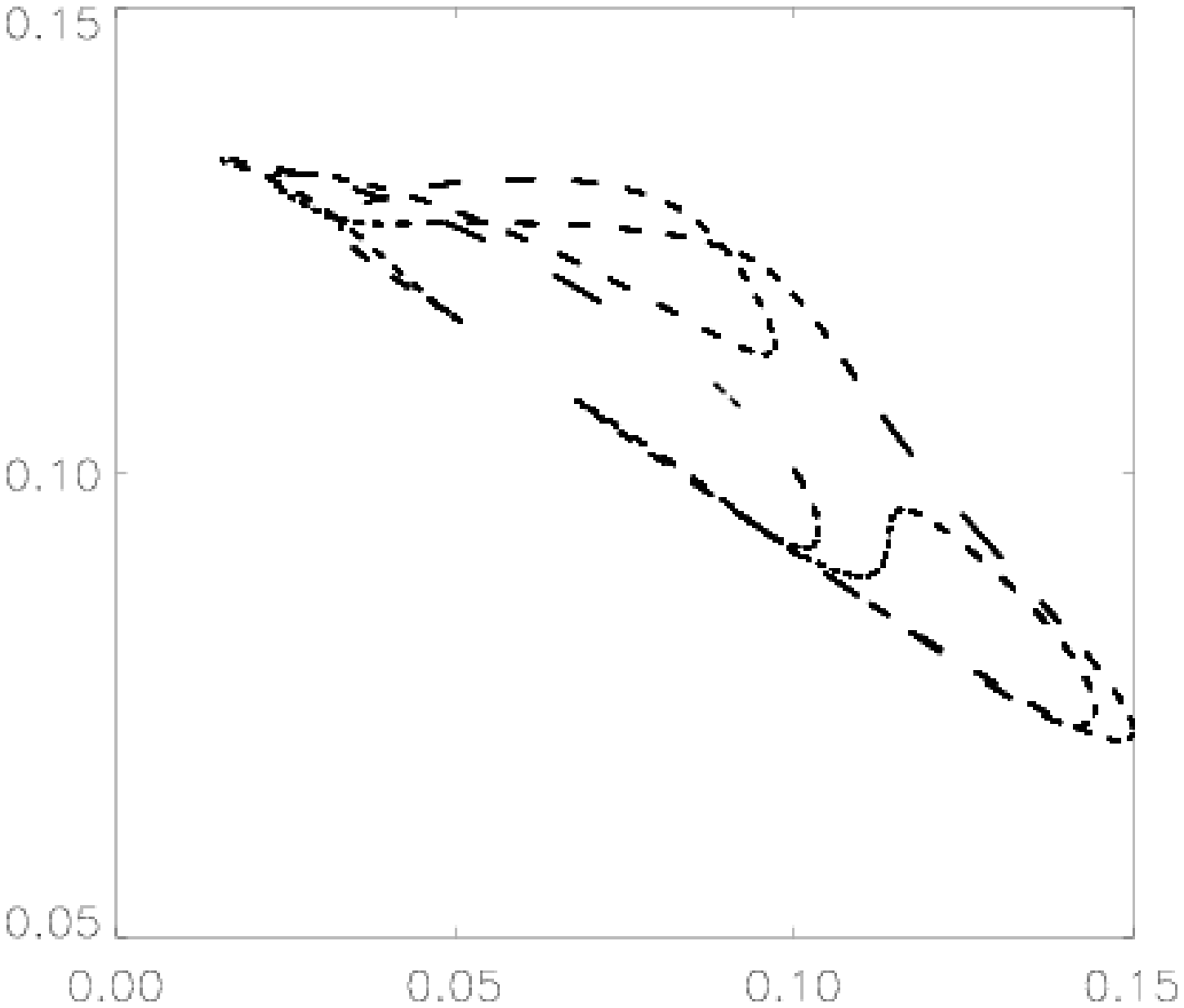,width=55mm,clip=}
\psfig{file=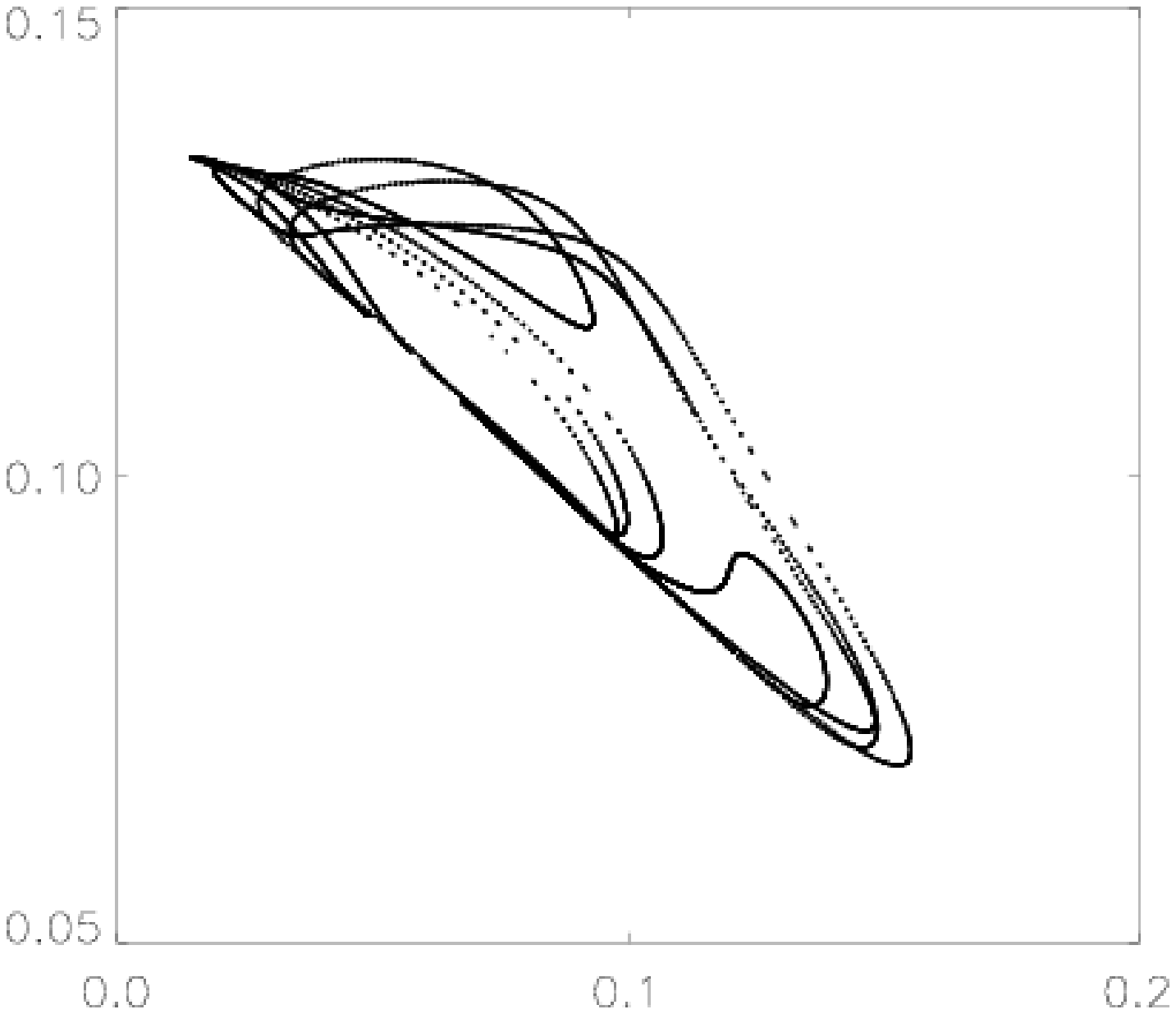,width=55mm,clip=}}
\hspace{27mm}(a)\hspace{51mm}(b)\hspace{51mm}(c)

\vspace*{2mm}
\centerline{
\psfig{file=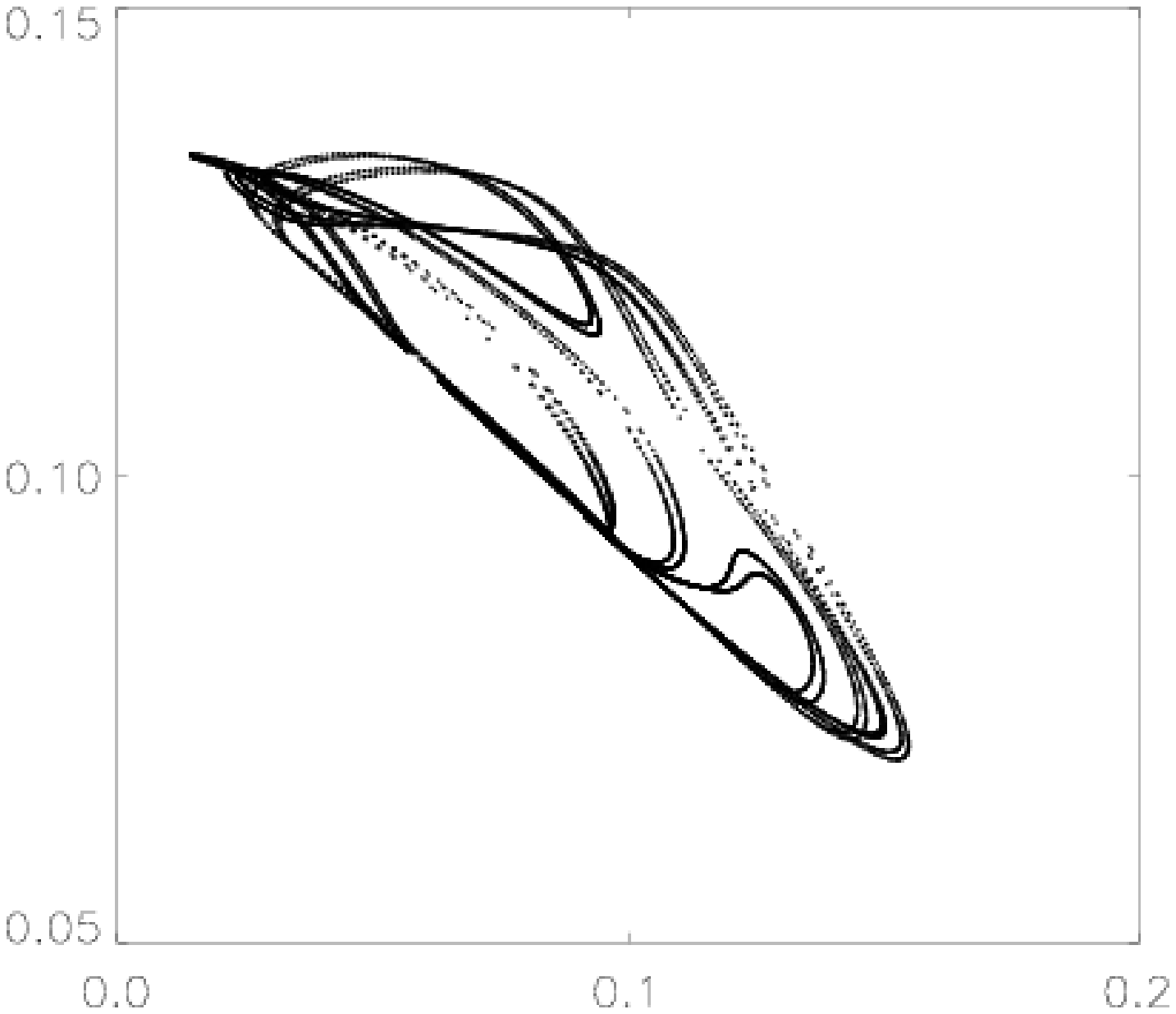,width=55mm,clip=}
\psfig{file=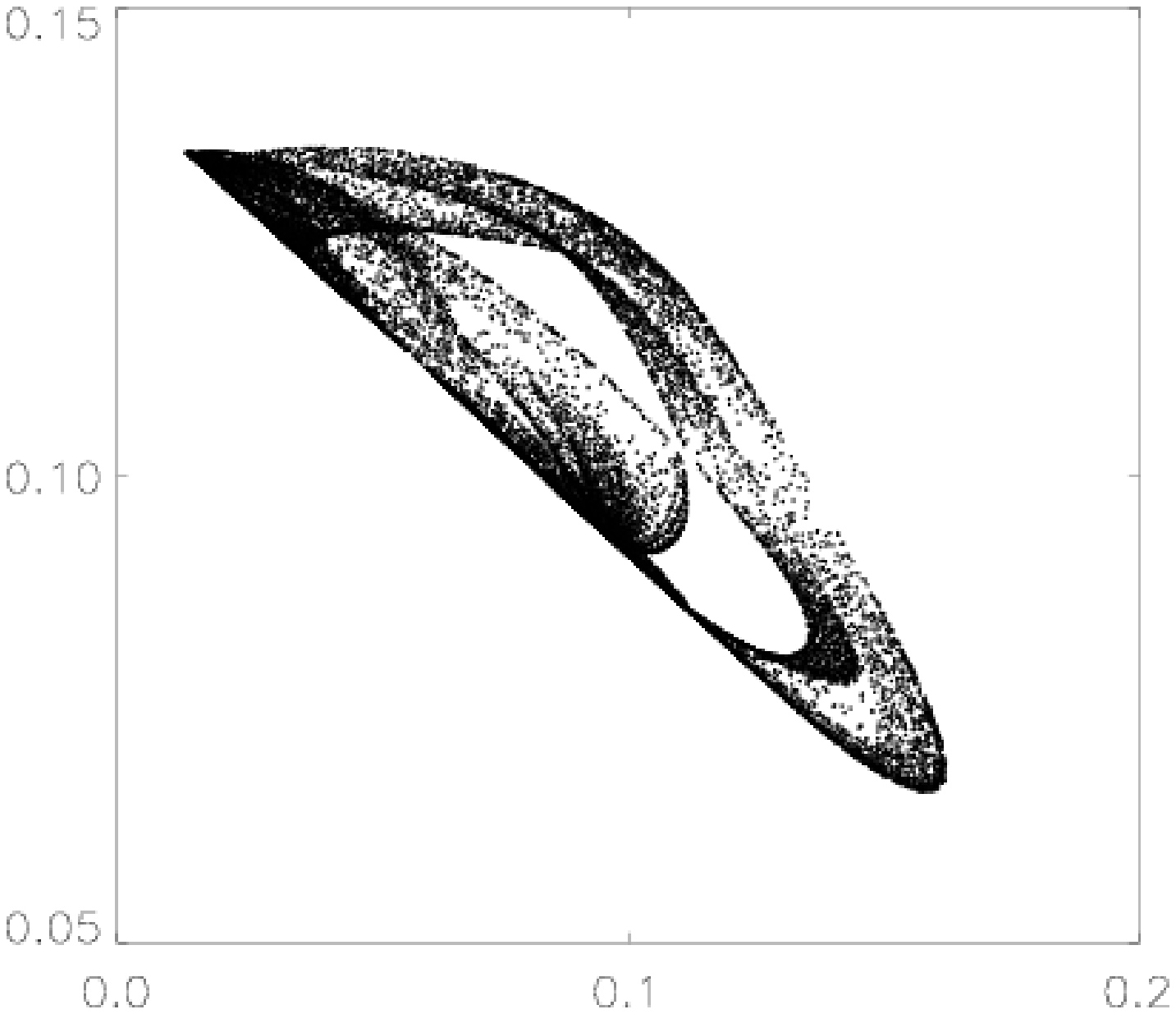,width=55mm,clip=}
\psfig{file=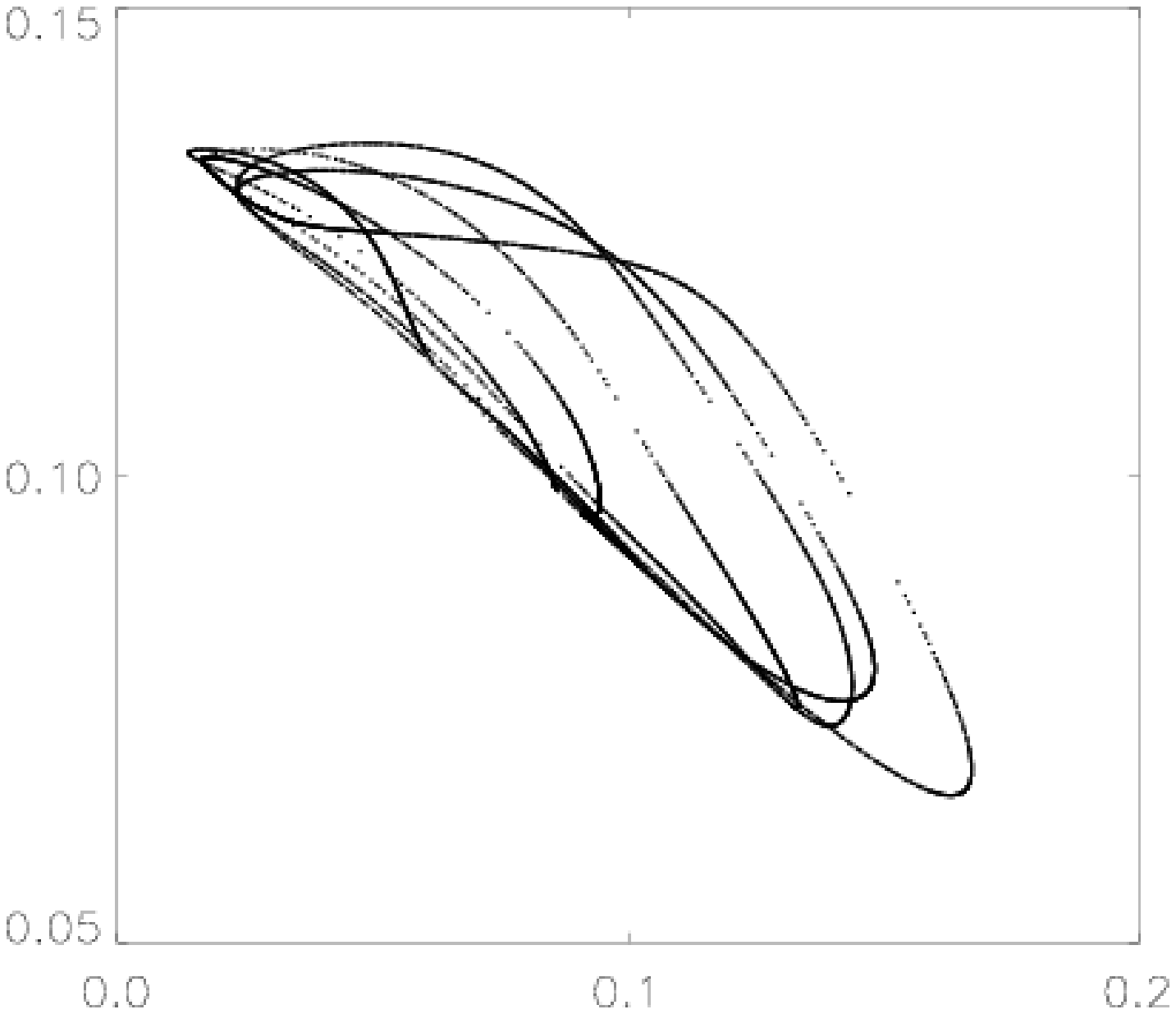,width=55mm,clip=}}
\hspace{27mm}(d)\hspace{51mm}(e)\hspace{51mm}(f)

\vspace*{2mm}
\centerline{
\psfig{file=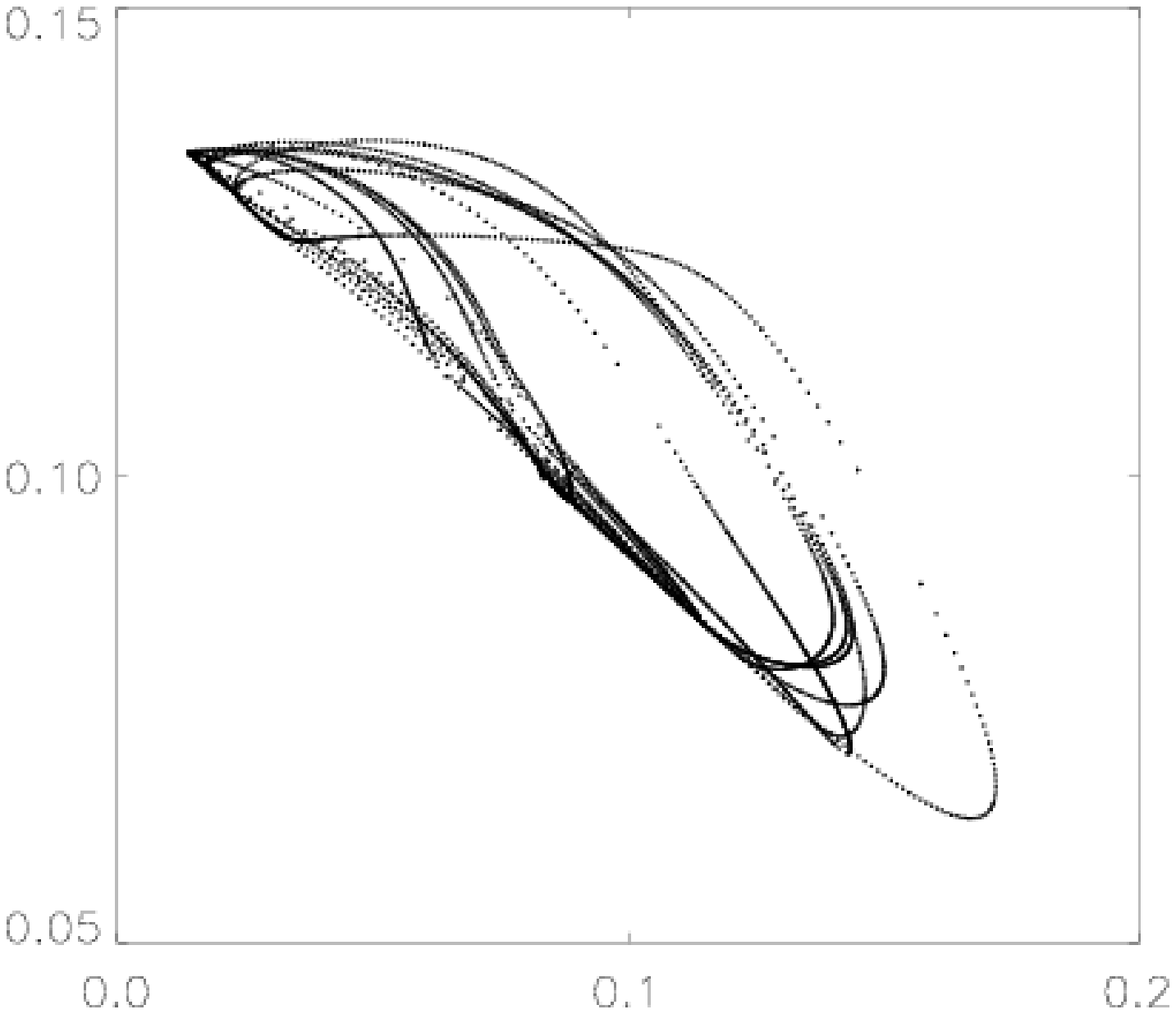,width=55mm,clip=}
\psfig{file=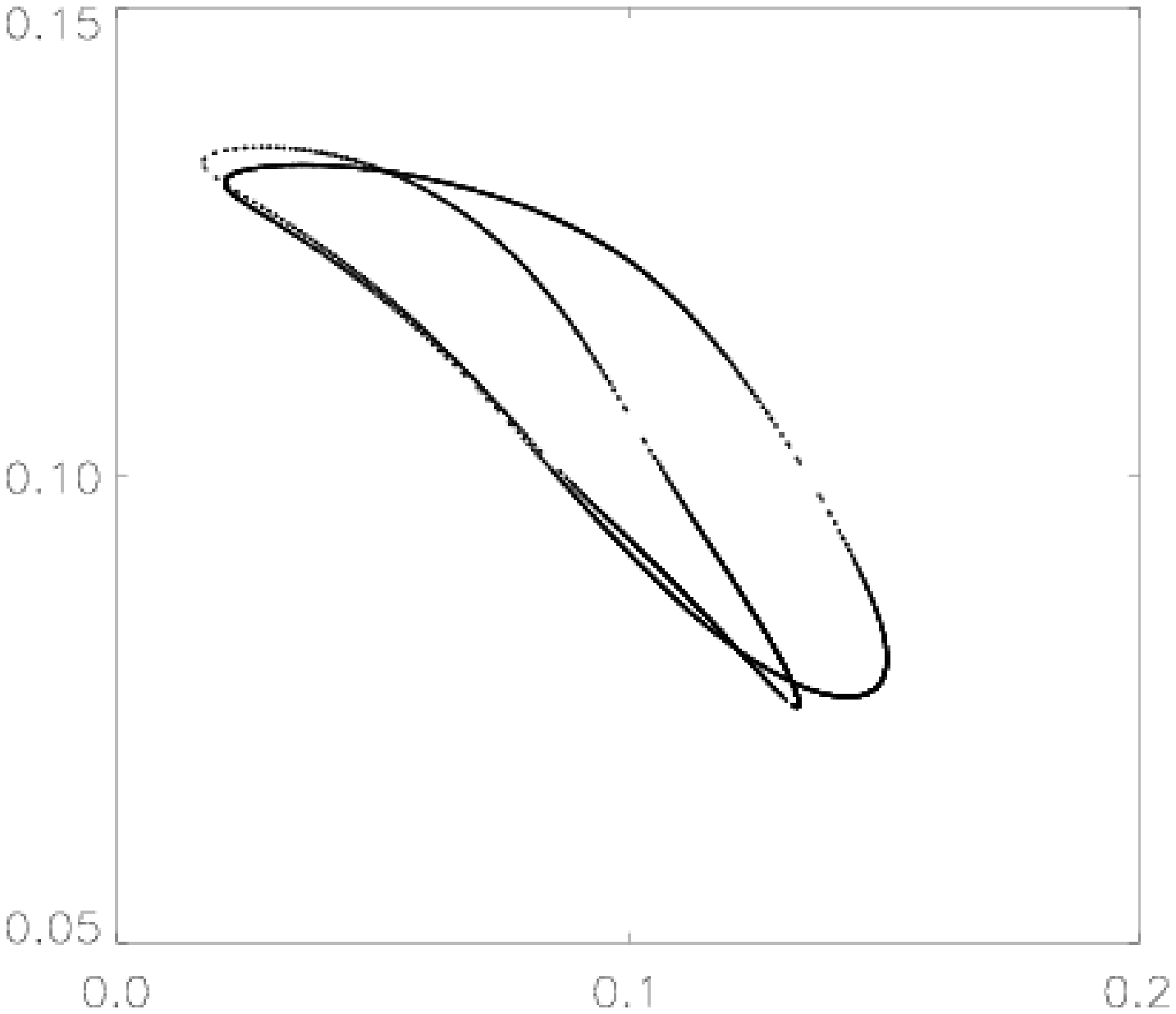,width=55mm,clip=}}
\hspace{55mm}(g)\hspace{51mm}(h)
\caption{Poincar\'e sections for regimes in the family F$^{\rm TW}$
on the ($|\hat{v}^x_{-1,2,1}|,|\hat{v}^y_{-1,2,1}|)$ quadrant (horizontal and
vertical axes, respectively) defined by the condition $|\hat{v}^z_{-1,2,1}|=0.2$:
$Ta=65$ (a); $Ta=70$ (b); $Ta=71.5$ (c); $Ta=72$ (d); $Ta=74$ (e); $Ta=77$ (f);
$Ta=78.5$ (g); $Ta=80$ (f).\label{fig:poincare_abs}}\end{figure}

Plots of the energies (\xrf{fig:qtw1_en}), not affected by drift frequencies,
clearly display the frequency halving. By contrast, plots of the time
dependencies of real or imaginary parts of individual Fourier coefficients (not
shown) are marred by the extra drift frequencies, making these plots by far more
obscure. A similar effect is observed in Poincar\'e sections. Frequency halving
for tori usually shows itself unambiguously on Poincar\'e sections, however,
the standard Poincar\'e sections are not particularly enlightening when
additional drift frequencies are present. Their influence is eliminated in
Poincar\'e sections, which are constructed with the use of absolute values of
Fourier coefficients (cf. Figs.~\ref{fig:poincare} and \ref{fig:poincare_abs}).

The family C$^{\rm SW}$ exists for $78\le Ta\le 126$. It is comprised of
chaotic attractors, alternating with windows of periodic and quasiperiodic
regimes (\xrf{fig:poincare_abs2}).
The family starts with the periodic regime P$^{\rm SW}$, which is
stable for $120\le Ta\le 126$. The orbit possesses a symmetry,
which is a combination of $s_2$ with a shift by a half of its temporal
period, hence, in agreement with Krupa (1990), the periodic orbit does not
drift. As $Ta$ is decreased, the symmetry is lost, the emerging non-symmetric
orbit and subsequent attractors of the family have two drifting frequencies
(not discussed here). Subsequently, a second frequency appears in a Hopf
bifurcation, which is halved afterwards. On a further decrease of $Ta$,
we observe an intermittency of chaotic and quasiperiodic attractors.
The sequence ``a periodic orbit, a torus, chaos'' is standard; it is usually
explained by appearance of the third basic frequency, which makes the system
structurally unstable and results in the onset of a chaotic behaviour
(Ruelle and Takens, 1971). The observed windows of quasiperiodicity can be
attributed to frequency locking (see, e.g., Ott, 2002).

Notably, coexistence of attractors of different types is observed: two types
of MHD attractors (F$^{\rm TW}$ and P$^{\rm TW}_1$) in the interval
$36\le Ta\le 81$, and a (magnetically stable) hydrodynamic attractor (TW) with
an MHD attractor (C$^{\rm SW}$) in the interval $81.80<Ta<87$\,. Coexistence of
a hydrodynamic and MHD attractors can be described in physical terms, as
{\it stiff excitation} of a magnetic field: {\it small} magnetic perturbations of
TW decay to the hydrodynamic state, while {\it large} magnetic perturbations
give rise to MHD regimes. Furthermore, in the interval $78\le Ta\le80$ three
MHD attractors coexist: P$^{\rm TW}_1$, F$^{\rm TW}$ and C$^{\rm SW}$.
From the point of view of the theory of dynamical systems
there is nothing extraordinary in coexistence of attractors in a
nonlinear system, however, to the best of our knowledge coexistence
of three convective MHD attractors was never observed before.

\begin{figure}[t]
\centerline{
\psfig{file=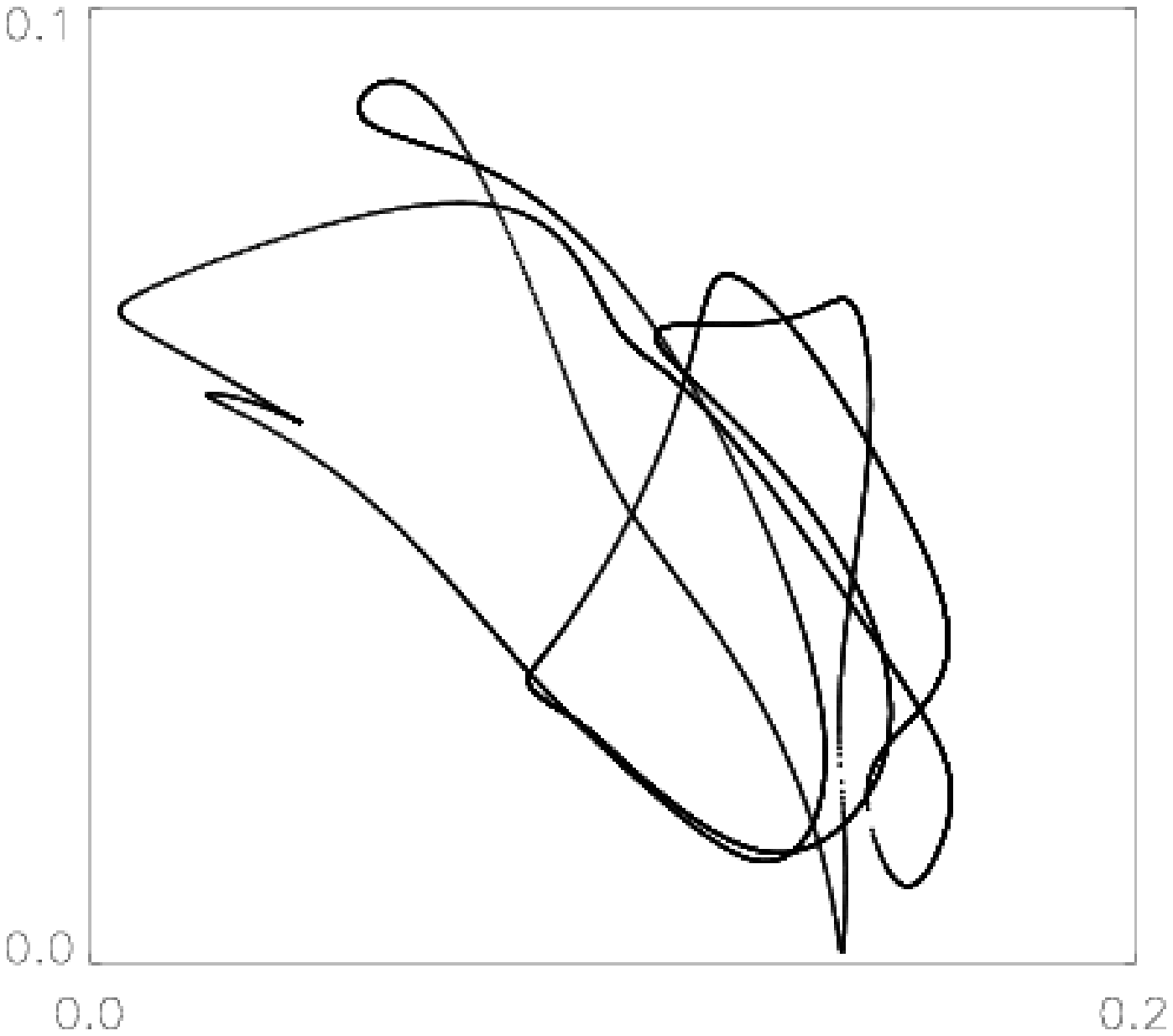,width=55mm,clip=}
\psfig{file=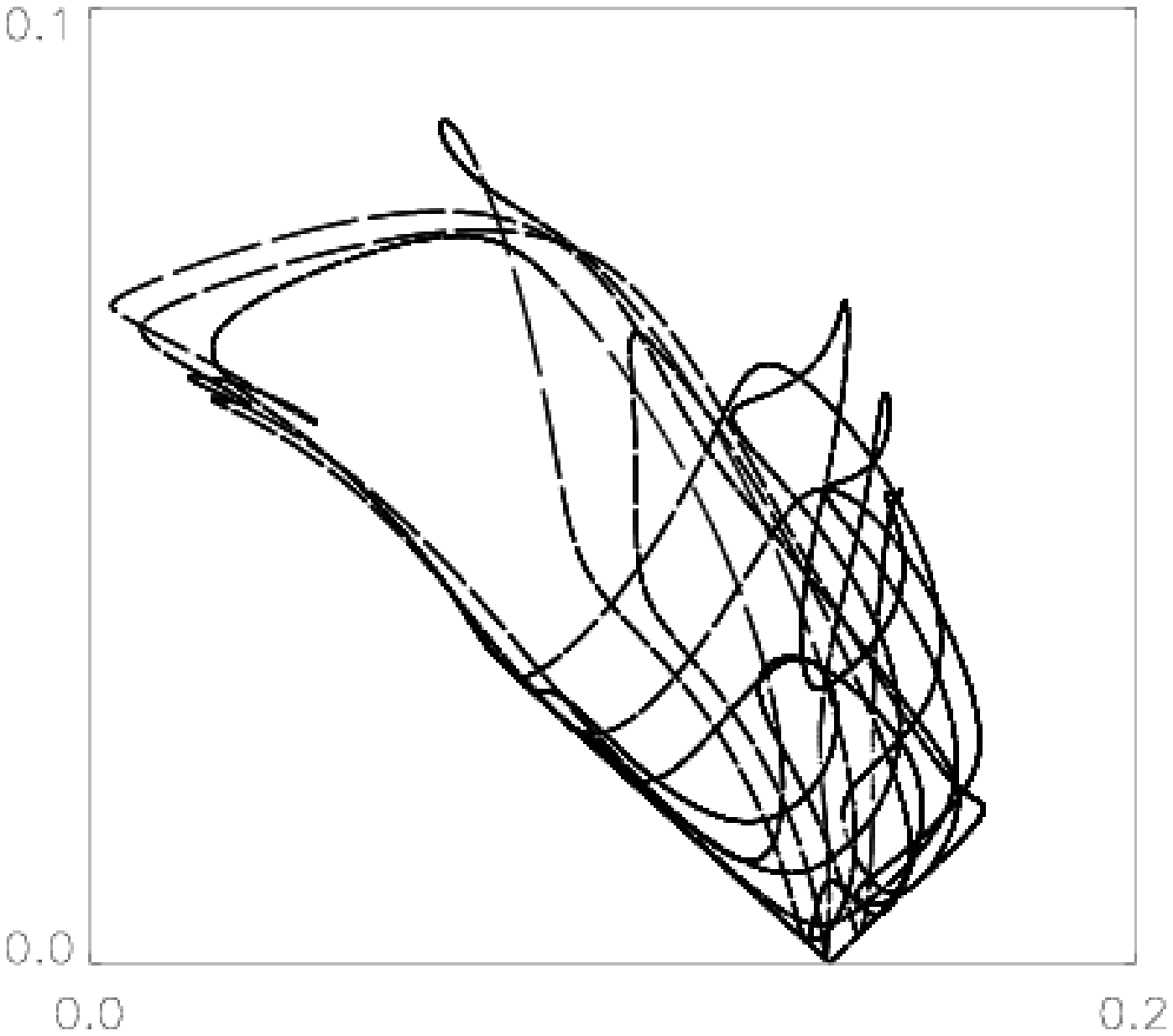,width=55mm,clip=}
\psfig{file=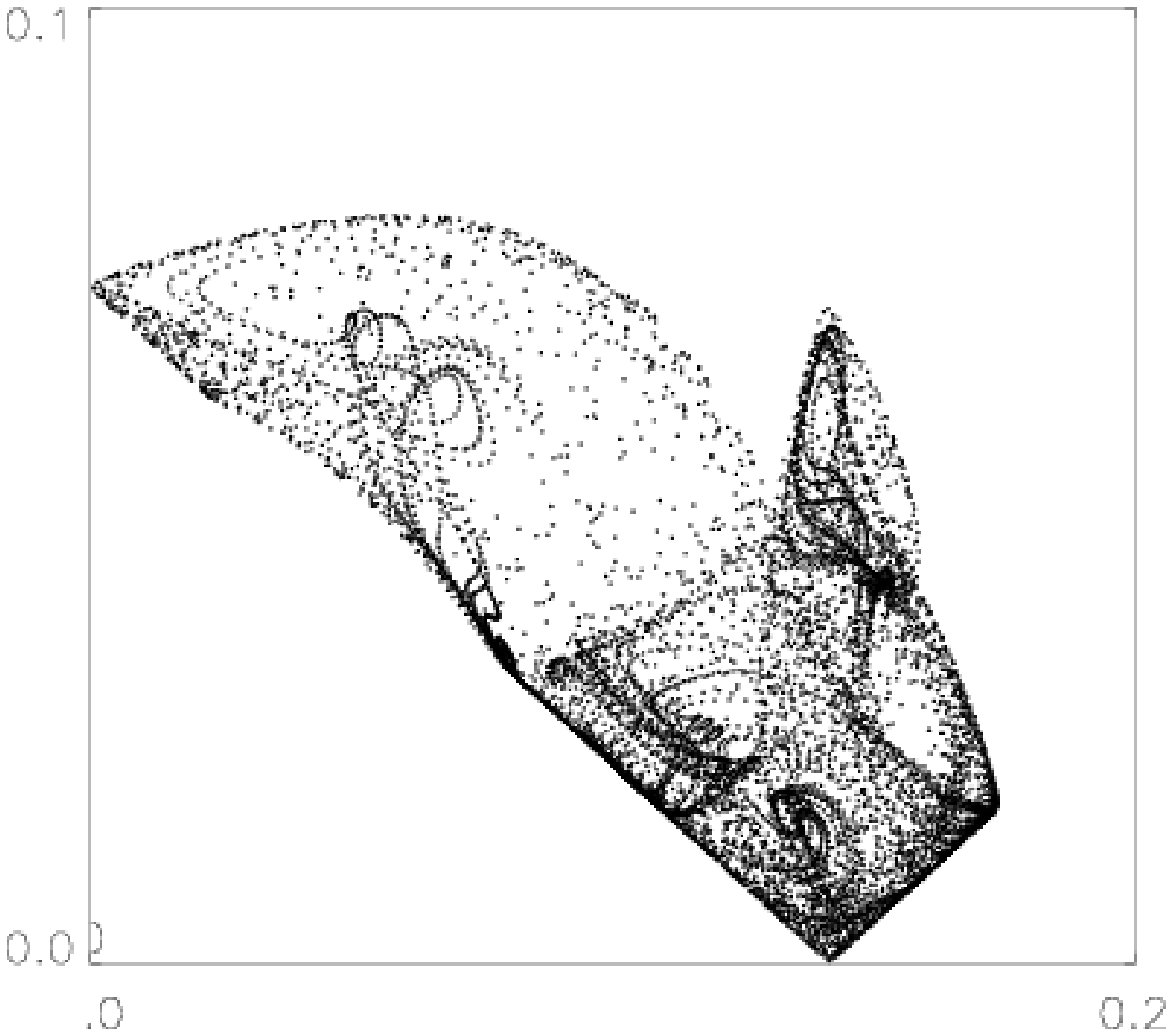,width=55mm,clip=}}
\hspace{27mm}(a)\hspace{51mm}(b)\hspace{51mm}(c)

\vspace*{2mm}
\centerline{
\psfig{file=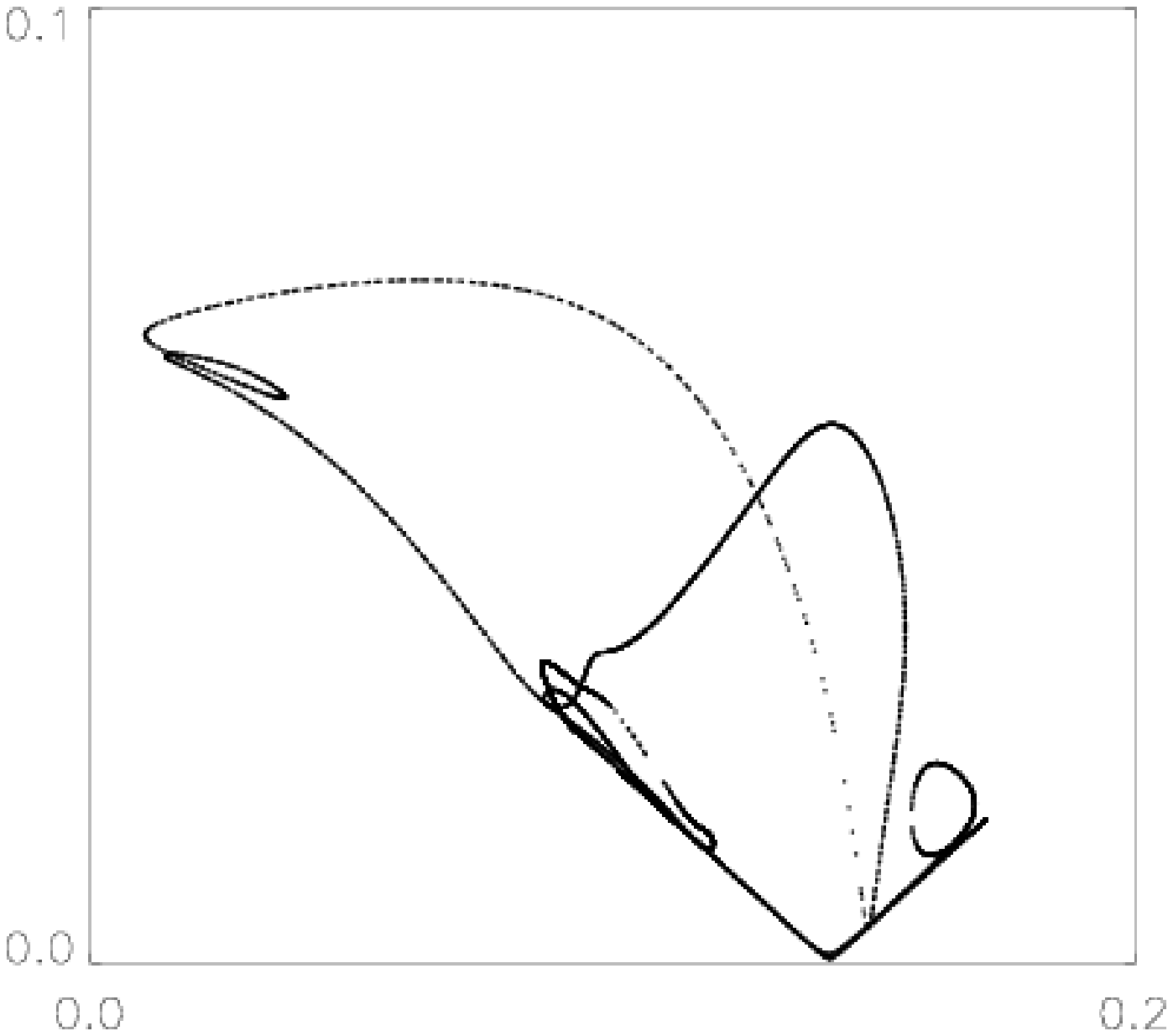,width=55mm,clip=}
\psfig{file=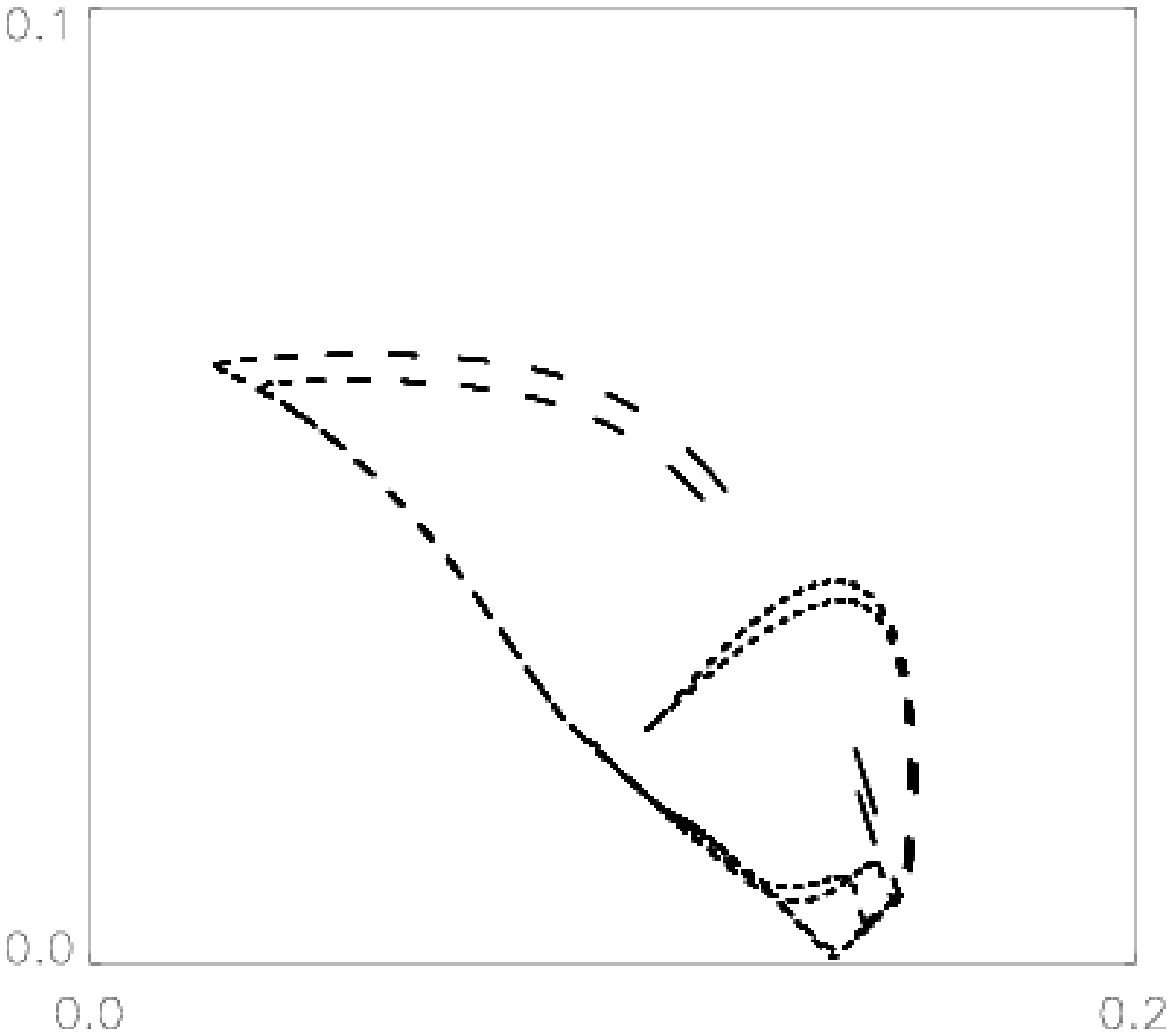,width=55mm,clip=}
\psfig{file=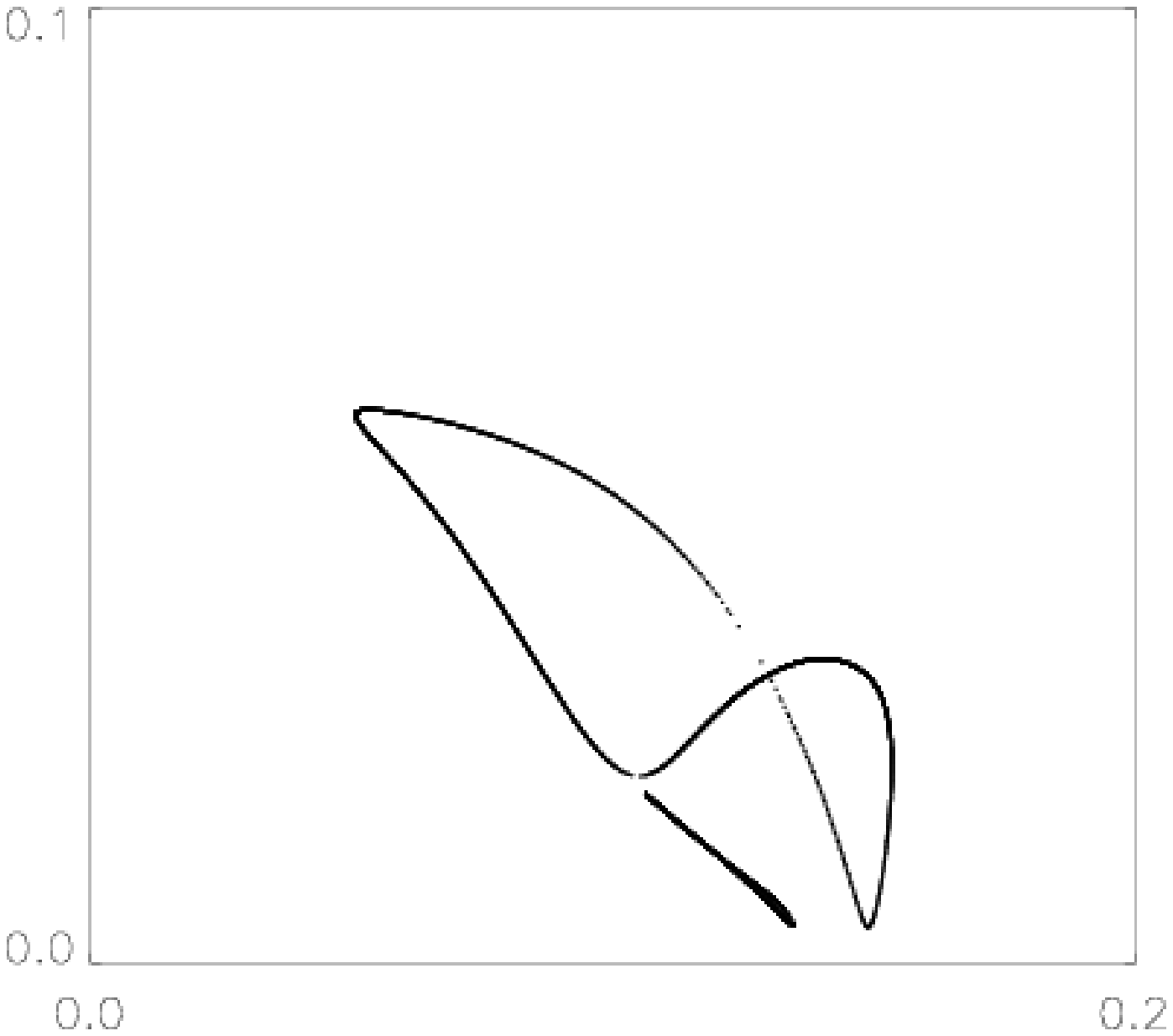,width=55mm,clip=}}
\hspace{27mm}(d)\hspace{51mm}(e)\hspace{51mm}(f)
\caption{Poincar\'e sections for regimes in the family C$^{\rm SW}$
on the ($|\hat{v}^x_{-1,2,1}|,|\hat{v}^y_{-1,2,1}|)$ quadrant (horizontal and
vertical axes, respectively) defined by the condition $|\hat{v}^z_{-1,2,1}|=0.1$:
$Ta=80$ (a); $Ta=85$ (b); $Ta=90$ (c); $Ta=98$ (d); $Ta=102$ (e); $Ta=105$
(f).\label{fig:poincare_abs2}}\end{figure}

\subsection{MHD attractors emerging from $\rm R_1$, $127\le Ta<506$}

Dominant magnetic modes of R$_1$ are of the same type for all $Ta$.
As attested by symmetries, the primary MHD attractor is the steady state
S$^{\rm R1}_2$ detaching from R$_1$ at $Ta=725.3$ in a subcritical pitchfork
bifurcation. It exists (i.e. the MHD states are stable) for $170\le Ta\le216$.
On decreasing $Ta$, a steady state S$^{\rm R1}_1$ with a smaller group
of symmetries emanates in a pitchfork bifurcation;
it becomes unstable and bifurcates to P$^{\rm TW}_2$ in a Hopf bifurcation.

\begin{figure}[t]
\centerline{\psfig{file=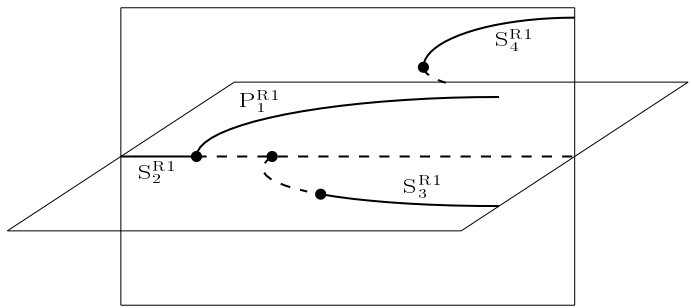,width=9cm,height=4cm}}
\caption{Phase portrait of the MHD system on the interval $216<Ta<224$.}
\label{tbdiag}\end{figure}

On increasing $Ta$, in a small interval $216<Ta<224$ four bifurcations
occurring in the vicinity of S$^{\rm R1}_2$ (in the phase space) are observed
(see \xrf{fig:nlin_ev} (b)\,). They take place in two distinct invariant
subspaces, Fix($rs_2$) and Fix($s_2$) (Fix($s$) denotes the set of fixed
points under the action of a symmetry $s$). Developments in each invariant
subspace are mutually independent (see a sketch of geometry of the phase space
shown on \xrf{tbdiag}; the vertical plane represents Fix($rs_2$) and
the horizontal one Fix($s_2$)\,).

The first (on increasing $Ta$) one is a Hopf bifurcation in Fix($rs_2$)
at $Ta=216.13$, in which a periodic orbit, P$^{\rm R1}_1$, emanates. Another one
is a saddle-node bifurcation at $Ta=223.1$ in the same invariant subspace,
in which a branch of steady states, S$^{\rm R1}_4$, terminates. Near the point
of bifurcation the basin of attraction of the steady states becomes vanishingly
small, requiring short steps when continuing the branch in $Ta$ so that
computed trajectories were not attracted by P$^{\rm R1}_1$. On decreasing
$Ta$ from, say, 224, eigenvalues of stability modes are in the beginning
imaginary, but both the real and imaginary parts decrease in magnitude,
apparently tending to zero. At $Ta=Ta_c$, $223.101<Ta_c<223.10075$,
the discriminant of the quadratic characteristic equation, defining the two
eigenvalues, changes its sign. As a result, the dominant eigenvalue exhibits
a counterintuitive behaviour: Coefficients of the characteristic equation
depend almost linearly on $Ta$ near this value $Ta_c$, but the change of sign
of the discriminant implies, that the dominant real eigenvalue is continuous,
but non-smooth (see \xrf{fig:rat}). The graph of the growth rate
of the dominant mode bends at $Ta_c$ and for $Ta<Ta_c$ behaves like
$\sqrt{Ta_c-Ta}$. The largest real eigenvalue starts to grow much faster, and
a saddle-node bifurcation occurs between $Ta=223.09999$ and 223.1\,.
An almost simultaneous vanishing of both real and imaginary parts
of eigenvalues implies, that two parameters are necessary to describe this
bifurcation, i.e. we are in a vicinity of a codimension two bifurcation.
Variation of $Ta$ near the left end of the branch S$^{\rm R1}_4$ is equivalent
to a motion along an one-dimensional curve on the plane of the parameters.

Vanishing of a pair of complex eigenvalues of linearisation of a dynamical
system is an attribute of the Takens-Bogdanov bifurcation (see Guckenheimer
and Holmes, 1988). There are further indications that the bifurcation occurring
in Fix($rs_2$) close (in the parameter space) to the values, which are fixed
in our simulations, might be a Takens-Bogdanov bifurcation: in this interval,
$216<Ta<224$, a periodic orbit emerges and afterwards
becomes unstable or disappears and a steady state emerges, and they are close
neighbours in the same symmetric subspace. However, a more attentive
inspection of the system convinces that this conjecture is wrong.

Only S$^{\rm R1}_2$ might be the trivial steady state suffering the bifurcation
(although the behaviour of the eigenvalues reminiscent of the Takens-Bogdanov
bifurcation is registered for S$^{\rm R1}_4$, this branch turns back
in a saddle-node bifurcation and thus can not serve as the trivial steady state,
existing in the case of Takens-Bogdanov bifurcation for all parameter values
in the vicinity of the point of bifurcation). It is quite possible that a pair
of complex eigenvalues of linearisation of S$^{\rm R1}_2$ simultaneously become
zero upon a variation of $Ta$ and a second parameter: S$^{\rm R1}_2$ does have
small in magnitude complex eigenvalues in the interval $216\le Ta\le220$ for the
parameter values that we have employed; also it is not far from S$^{\rm R1}_4$,
suggestive of a similar behaviour of eigenvalues of the two branches.
The structure of our system is compatible only with the Takens-Bogdanov
bifurcation (with the ${\bf Z}_2$ symmetry group generated by $s_2$, which
S$^{\rm R1}_4$ does not possess) involving a stable periodic orbit and stable
steady states, distinct from the trivial one; such a diagram is shown
on Figs. 7.3.7 and 7.3.9 {\it ibid.} On these Figures, the trivial steady
state (the analogue of S$^{\rm R1}_2$) is stable for $\mu_1<0,\ \mu_2<0$ (using
the parameter notation {\it ibid.}), the other two steady states (the analogues
of S$^{\rm R1}_4$ and its symmetric counterpart) for $\mu_1>\max(0,\mu_2)$, and
a periodic orbit for $\mu_2>\max(0,c\mu_1),\ c\approx0.752$\,. Hence variation
of $Ta$ on the interval $216\le Ta\le224$ must be equivalent to a motion along
a curve on the $(\mu_1,\mu_2)$ plane, beginning in the quadrant
$\mu_1<0,\ \mu_2<0$, following to the quadrant $\mu_1<0,\ \mu_2>0$, and passing
to the region $\mu_1>0, \ \mu_2<c\mu_1$. Consequently, the curve must cross
the line $\mu_1=0$, where in the Takens-Bogdanov bifurcation a branch of steady
states (analogues of S$^{\rm R1}_4$), stable or unstable depending on the sign
of $\mu_1$, emanates from the trivial steady state in a pitchfork bifurcation.
This is inconsistent with the behaviour of eigenvalues of linearisation of
S$^{\rm R1}_2$ (we have computed these unstable steady states after the Hopf
bifurcation, giving rise to P$^{\rm R1}_1$, by imposing the symmetries
$r\gamma^x_{L/2}$ and $s_2$) -- the dominant and subdominant eigenvalues do
become real between $Ta=221$ and 221.5, but they remain positive and do not
become small on the interval $221.5\le Ta\le224$. Therefore we conclude, that
bifurcations in our system are not induced by a Takens-Bogdanov bifurcation
(a finite perturbation of a bifurcation of this type, in which more than two
parameters are essential for the description of the bifurcation, is not ruled
out). Further analysis is necessary to understand it theoretically.

\begin{figure}[t]
\centerline{\psfig{file=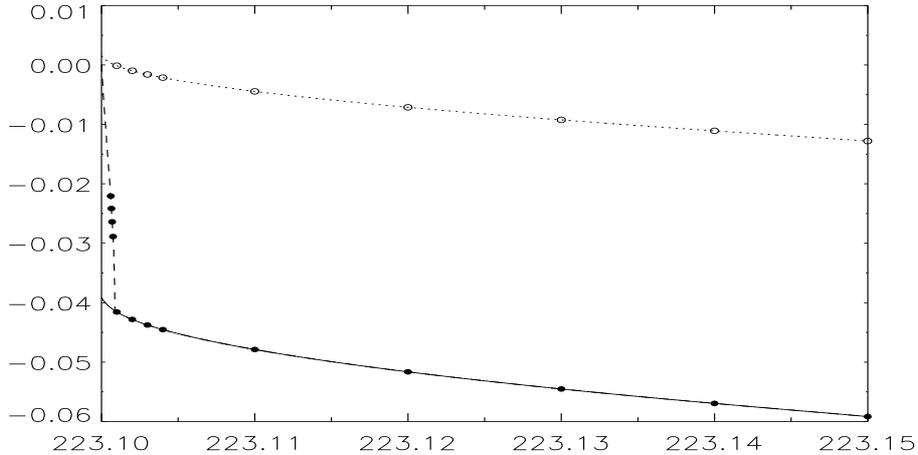,width=12cm,height=6cm}}
\caption{Computed mean of the two dominant eigenvalues (black circles; coincides
with the real part of the eigenvalues for $Ta>Ta_c\approx223.101$, where the
two eigenvalues are complex conjugate) and the discriminant of the quadratic
characteristic equation (open circles) for the dominant eigenvalues
of the operator of linearisation versus $Ta$ (horizontal axis) near the left end
of the interval of existence of S$^{\rm R1}_4$. Pad\'e extrapolation of
the mean of the two eigenvalues (solid line), the discriminant of the
characteristic equation (dotted line) and the resultant extrapolated dominant
real eigenvalue for $Ta<Ta_c$ (dashed line).\label{fig:rat}}\end{figure}

At $Ta=216.75$ an eigenvalue of linearisation of (1) near S$^{\rm R1}_2$ with
the associated eigenvector in Fix($s_2$) (the dominant eigenvalue of the
restriction of the operator of linearisation to this subspace) becomes positive.
In this supercritical pitchfork bifurcation a steady state S$^{\rm R1}_3$
emerges. Since at the bifurcation S$^{\rm R1}_2$ is unstable with
respect to perturbations in Fix($rs_2$), the branching steady
state is also unstable to such perturbations near the point of bifurcation.
S$^{\rm R1}_3$ gains stability at $Ta=218.64$ in a subcritical Hopf
bifurcation. The branch is stable up to a subcritical pitchfork
bifurcation at $Ta=271.66$. S$^{\rm R1}_3$ is the only MHD steady state that we
have found, whose group of symmetries does not involve superpositions of
the symmetry $r$ with any other symmetries. As a result, magnetic patterns
on top and bottom of the layer are different (cf. \xrf{fig:nlin_isosb} (c)
and other panels on this figure).

We again encounter coexistence of three distinct MHD attractors in
the interval $223.11\le Ta\le223.3$: S$^{\rm R1}_3$, S$^{\rm R1}_4$ and
P$^{\rm R1}_1$. The steady state S$^{\rm R1}_4$ becomes unstable at $Ta=466.5$
in a subcritical pitchfork bifurcation.

The symmetry group of S$^{\rm R1}_5$ is a subgroup of the one of S$^{\rm R1}_2$
(see Table \ref{tab:nlin_en}\,). That the steady states are related and
some symmetries are lacking, is seen on \xrf{fig:nlin_isosb} (b) and (e)
displaying magnetic patterns of the steady states. These facts suggest that
S$^{\rm R1}_5$ bifurcates from the unstable S$^{\rm R1}_2$. At both ends
of the interval of stability, S$^{\rm R1}_5$ undergoes supercritical Hopf
bifurcations with stable periodic orbits, P$^{\rm R1}_2$ and P$^{\rm R1}_3$,
emerging. The periodic orbit P$^{\rm R1}_2$ with the same group of symmetries
as that of S$^{\rm R1}_5$ is observed for $343\le Ta\le377$.

The other periodic orbit bifurcating from S$^{\rm R1}_5$, P$^{\rm R1}_3$,
exists in a small interval $505.1\le Ta\le 506$ and it has a much smaller
symmetry group than S$^{\rm R1}_5$.

\subsection{MHD attractors emerging from $\rm R_1$, $Ta$ near 506}

At the right end of the interval of existence of the periodic orbit
P$^{\rm R1}_3$, it terminates near $Ta=506$ on a steady state S$^{\rm R1}_6$
in a bifurcation, which appears to be similar to a saddle-node bifurcation on
invariant circle (SNIC; see Izhikevich, 2006). However, since the symmetry
group of the system is
non-trivial, the details of the bifurcation in our case differ from those of
the canonical SNIC. In a generic system, a periodic orbit emanates in a SNIC
bifurcation subsequent upon a saddle-node bifurcation of a steady state.
The SNIC occurs under the condition that for any parameter value before
the saddle-node bifurcation (we only consider a small neighbourhood of the
critical value of the bifurcation parameter), both parts of the one-dimensional
unstable manifold of the unstable steady state terminate on the stable one.
At the point of bifurcation, the two steady states collide and a homoclinic
trajectory emerges. After the bifurcation, a periodic orbit is formed from
this homoclinic trajectory.

\begin{figure}[t]
\centerline{\psfig{file=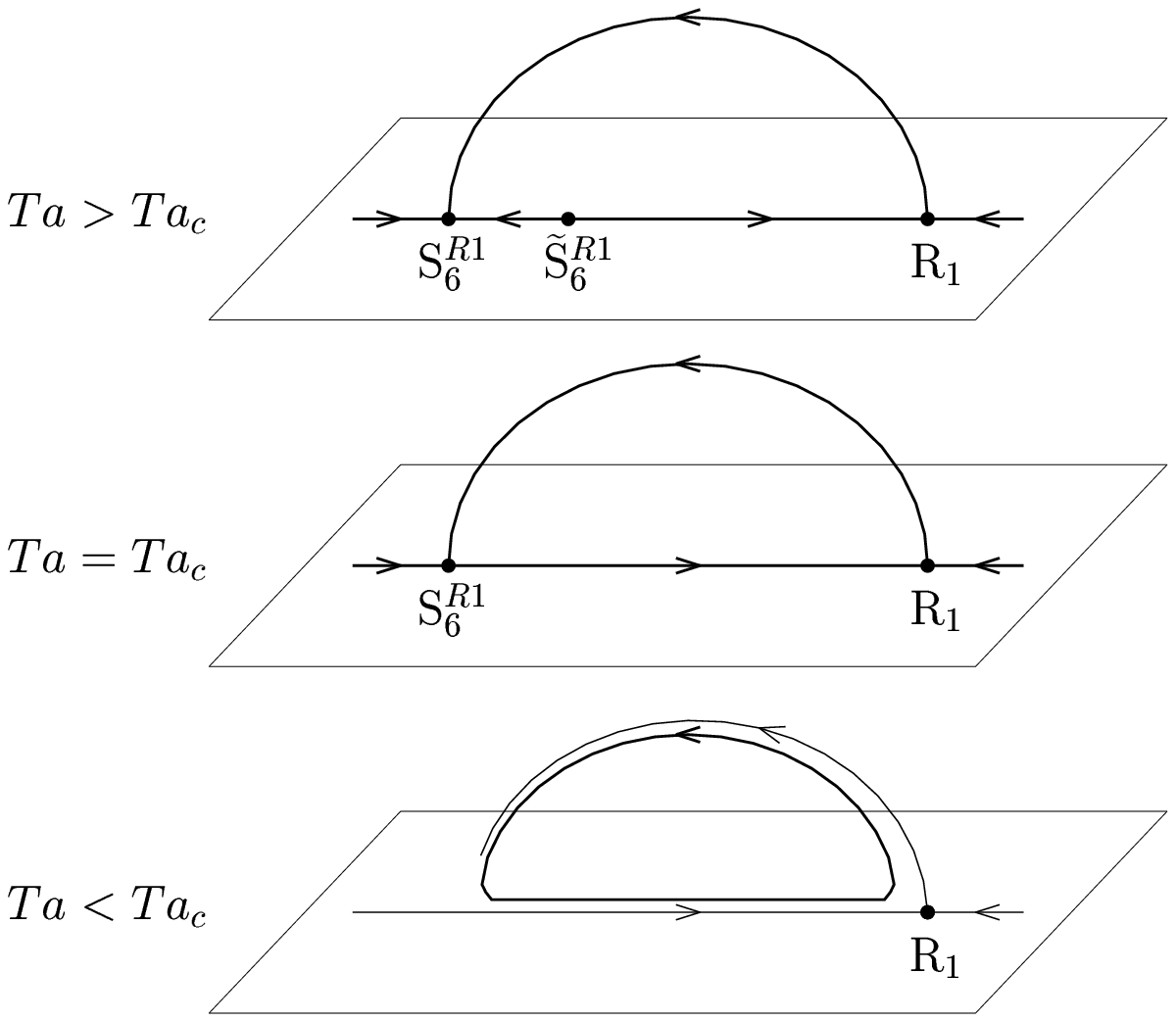,width=9cm}}
\caption{Phase portrait of the MHD system near $Ta=Ta_c\approx506.07$\,.}
\label{fig:SNIC}\end{figure}

SNIC is generic in one-parameter dynamical systems and it is often observed
in simulations, where a parameter is varied. At $Ta=Ta_c\approx506.07$
the periodic orbit P$^{\rm R1}_3$ terminates on a heteroclinic cycle; occurrence
of this upon variation of just one scalar parameter requires some
degeneracy of the system, since
a heteroclinic (homoclinic) connection from an equilibrium to an unstable
equilibrium in a generic system occurs only for a singular parameter value
(such connections are structurally unstable). For formation of a heteroclinic
cycle, several such connections must happen simultaneously.

Existence of a heteroclinic orbit at $Ta=Ta_c$ becomes possible due to the
presence of a non-trivial symmetry group. Structural stability of homoclinic and
heteroclinic connections to an unstable equilibrium in symmetric systems relies
on the presence of symmetry-invariant subspaces (Guckenheimer and Holmes,
1988). A sketch of geometry of the phase space of our system is shown on \xrf{fig:SNIC}.
The steady state S$^{\rm R1}_6$ possesses a symmetry group isomorphic to
${\bf D}_4\ltimes{\bf Z}_2$. The plane represents the fixed point subspace for
the steady state symmetry group, the vertical direction the antisymmetric
complement. We describe the bifurcation starting from larger values of
$Ta$ slightly above the saddle-node bifurcation of S$^{\rm R1}_6$. Denote by
$\tilde{\rm S}^{\rm R1}_6$ the unstable counterpart of S$^{\rm R1}_6$; it
belongs to Fix(${\bf D}_4\ltimes{\bf Z}_2$). The one-dimensional unstable
manifold of $\tilde{\rm S}^{\rm R1}_6$ also belongs to this subspace.
A part of the unstable manifold terminates on the stable S$^{\rm R1}_6$,
another one on R$_1$ (this has been checked numerically), which is stable
within this subspace. The connections from $\tilde{\rm S}^{\rm R1}_6$ to
S$^{\rm R1}_6$ and R$_1$ are structurally stable. The one-dimensional unstable
manifold of R$_1$ terminates on the stable S$^{\rm R1}_6$ (we have also checked
this numerically).

At $Ta=Ta_c$ S$^{\rm R1}_6$ and $\tilde{\rm S}^{\rm R1}_6$ collide, thus
creating the structurally unstable heteroclinic cycle
S$^{\rm R1}_6\to{\rm R}_1\to{\rm S}^{\rm R1}_6$ (it exists only for a single
value of $Ta$). For $Ta<Ta_c$, S$^{\rm R1}_6$ disappears and we observe
a periodic orbit in place of the heteroclinic cycle. The period of this orbit
tends to infinity, as $Ta$ approaches the point of bifurcation $Ta_c$.

Kinetic and magnetic energies for the periodic orbit for $Ta=506$, which is
close to the point of the saddle-node bifurcation (hence the period of the
orbit is large) is shown on \xrf{fig:homo}. The minima of magnetic energy are
attained when the trajectory is close to R$_1$; from these
steady states it jumps rapidly to the former S$^{\rm R1}_6$ (kinetic energy
becoming close to 161), from where it slowly moves back towards R$_1$.

Notably, close to the critical $Ta$ we observe not a periodic,
but rather a chaotic behaviour (see \xrf{fig:homo}). The non-periodicity
(the behaviour is, loosely speaking, periodic, but ``periods'' significantly
vary) can be caused by reasons of numerical nature, such as round-off errors
or other numerical noise accumulating during the long periods. As noticed
by Busse and Heikes (1980), round-off errors are comparable with the amplitude
of the integrated field near stagnation point, significantly affecting
trajectories approaching a heteroclinic cycle: Numerical noise can result
in emergence of quasi-periodic regimes with randomly varying ``periods''
(Stone and Holmes, 1990), like we observe here, as well as strictly periodic
cycles of a large period (Nore\al, 2003), or alter the amplitude of cycles
(Stone and Armbruster, 1999). Numerical noise can be expected to affect
similarly large-period orbits near the critical parameter value.

\begin{figure}[t]
\centerline{\psfig{file=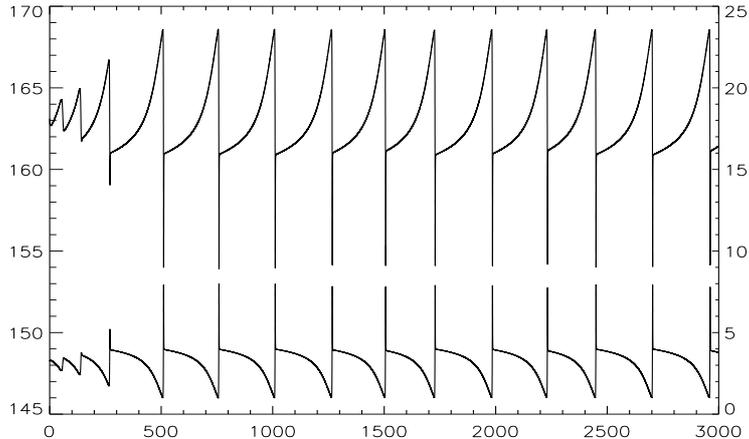,width=10cm,height=6cm}}
\caption{Kinetic (upper curve, left vertical axis) and magnetic (lower curve,
right vertical axis) energies for $Ta=506$, P$^{\rm R1}_3$ as a function of
time (horizontal axis) for a trajectory starting at the steady state
S$^{\rm R1}_5$, the MHD attractor for $Ta=505$.
\label{fig:homo}}\end{figure}

\subsection{MHD attractors emerging from $\rm R_1$, $506<Ta\le 718$}

The primary steady state S$^{\rm R1}_6$, connected with the S$^{\rm R1}_5$
branch by the periodic orbit P$^{\rm R1}_3$, is different from S$^{\rm R1}_5$
and other S$^{\rm R1}$ steady states considered above in that it is related
not to the dominant but a subdominant magnetic mode of R$_1$. The symmetry
group of the steady state S$^{\rm R1}_6$ is not a subgroup of the symmetry
group of S$^{\rm R1}_2$ (coinciding with the symmetry group of the
dominant magnetic mode) and it has a twice smaller period in the direction
along the axis of rolls, indicating that S$^{\rm R1}_6$ is unrelated to the
dominant magnetic mode (cf.~\xrf{fig:conv_isosb} (c) and \xrf{fig:nlin_isosb}
(f)\,). We have computed the subdominant magnetic mode by restricting the
kinematic dynamo problem for R$_1$ on the subspace Fix($q\gamma^y_{L/4}$).
The subdominant magnetic mode, which is dominant in Fix($q\gamma^y_{L/4}$),
has the same group of symmetries as that of S$^{\rm R1}_6$ and a spatial
structure similar to that of the magnetic field of S$^{\rm R1}_6$ (see
\xrf{fig:nlin_isosb} (f)\,). The associated eigenvalue is real and positive
in the interval $509\le Ta\le 682$ and admits the maximum 0.4 at $Ta=598$.
On increasing $Ta$, S$^{\rm R1}_6$ becomes unstable in a pitchfork bifurcation
with emergence of the steady state S$^{\rm R1}_7$, becoming unstable in its
turn in a subcritical pitchfork bifurcation at $Ta=682.26$\,.

The symmetry group of S$^{\rm R1}_8$ is a subgroup of the symmetry group of
S$^{\rm R1}_2$, indicating that the former state has possibly bifurcated from
the latter one. On decreasing $Ta$, S$^{\rm R1}_8$ becomes unstable in a
subcritical pitchfork bifurcation; in the interval $674\le Ta\le 682$ the steady
state coexists with S$^{\rm R1}_7$. When $Ta$ is increased, a stable periodic
orbit P$^{\rm R1}_4$ emanates in a Hopf bifurcation from S$^{\rm R1}_8$.
On increasing $Ta$ further,
this periodic orbit terminates on a structurally unstable heteroclinic cycle.
We discuss this bifurcation in detail in the following subsection.

\begin{figure}[t]
\centerline{\psfig{file=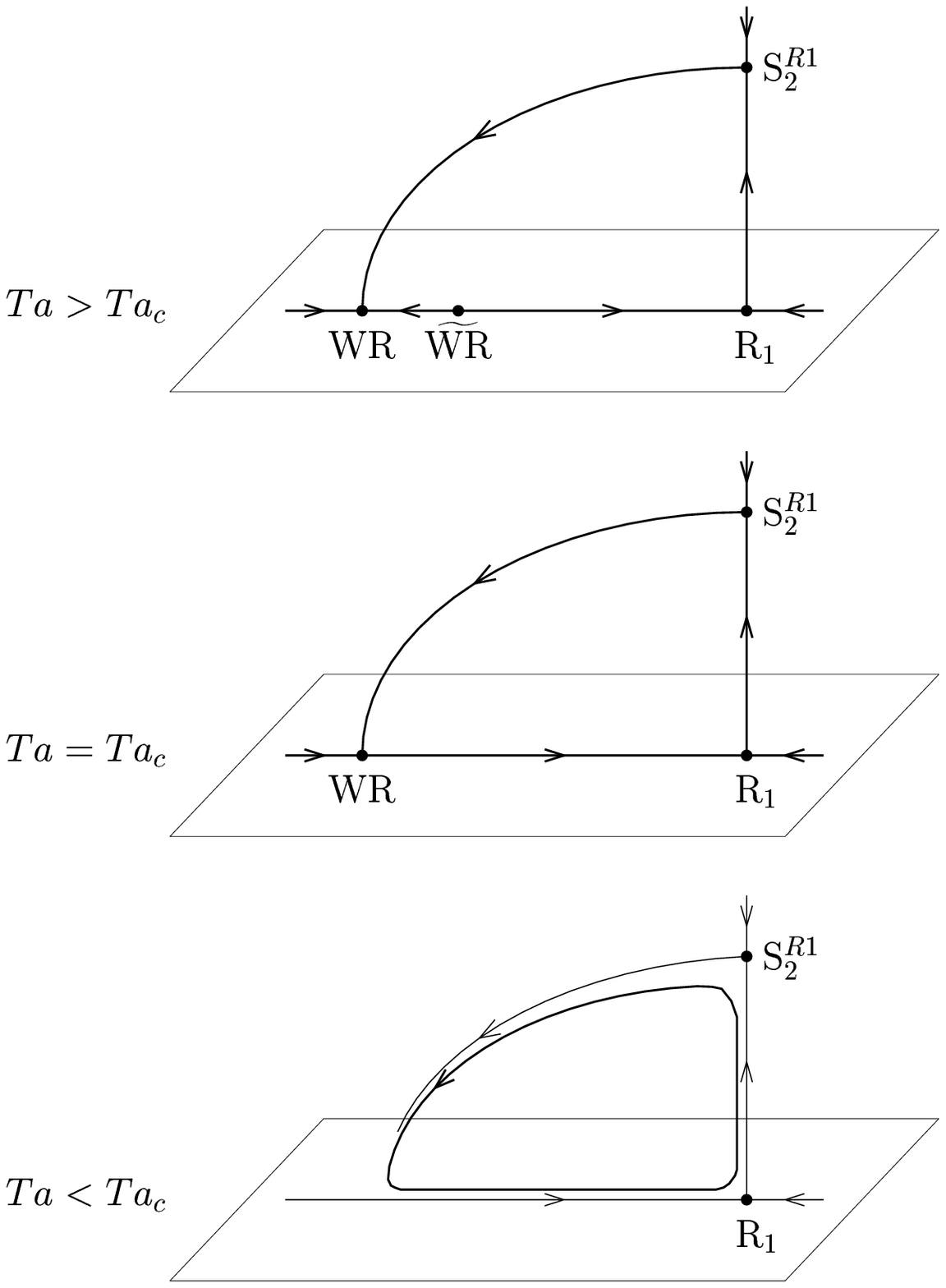,width=9cm}}
\caption{Phase portrait of the MHD system near $Ta=Ta_c\approx718.16$\,.}
\label{fig:SNIC2}\end{figure}

\subsection{MHD attractors emerging from $\rm R_1$, $Ta$ near 718}

We have discussed in subsection 4.3 a periodic orbit terminating on
a structurally unstable heteroclinic cycle involving connections between
two steady states. At $Ta=Ta_c\approx718.16$ a similar bifurcation takes place.
The periodic orbit P$^{\rm R1}_4$ terminates on a structurally unstable
heteroclinic cycle with three steady states involved, when
one of the steady states undergoes a saddle-node bifurcation.

A sketch of geometry of the phase space is shown on \xrf{fig:SNIC2}. The plane
represents the hydrodynamic subspace invariant under the symmetry $q$, and
the vertical direction the complementary magnetic subspace. We describe
the bifurcation starting from the larger values of $Ta$ slightly above
the point $Ta=Ta_c$ of the saddle-node
bifurcation of WR. For such $Ta$, in the hydrodynamic subspace we observe
R$_1$, stable in the hydrodynamic subspace, stable WR and its unstable
counterpart, $\tilde{\rm WR}$. A part of the one-dimensional unstable manifold
of $\tilde{\rm WR}$ terminates on R$_1$, the other one on WR (since
$\tilde{\rm WR}$ emerges in a subcritical bifurcation from R$_1$ and it
disappears in a saddle-node collision with WR). Connections from
$\tilde{\rm WR}$ to R$_1$ and WR are structurally stable, because R$_1$ and WR
are stable in the hydrodynamic subspace. The connection from R$_1$ to
S$^{\rm R1}_2$ belongs to Fix($q\gamma^y_{L/2}$) and it is structurally
stable, because S$^{\rm R1}_2$ is stable in this subspace (this has been
checked numerically). The unstable manifold of S$^{\rm R1}_2$ belongs to
Fix($s_2$) and it terminates on WR, stable in this subspace.

At $Ta=Ta_c$, WR and $\tilde{\rm WR}$ collide creating the heteroclinic
cycle $\rm WR\to R_1\to S^{R1}_2\to WR$ in the subspace Fix($s_2$).
The cycle is asymptotically stable within this subspace, but not
in the entire phase space, because WR possesses a growing magnetic mode.
The cycle is structurally unstable; it exists only for $Ta=Ta_c$. For $Ta<Ta_c$,
WR disappears and we observe a periodic orbit in place of the heteroclinic
cycle. The period of the orbit tends to infinity, as $Ta$ approaches the point
of bifurcation. Surprisingly, the orbit is asymptotically stable, despite it has
bifurcated from an asymptotically unstable heteroclinic cycle.

We illustrate the behaviour described above by plots of kinetic and magnetic
energies and discrepancies for the symmetries $s_2$ and $q\gamma^y_{L/2}$
for a periodic orbit at $Ta=717.9$ (\xrf{fig:hetero}), near the point of
bifurcation, where the orbit is similar to the structurally unstable
heteroclinic cycle. Plateaux of constant values of kinetic energy on
\xrf{fig:hetero} (such as $1870<t<1930$ and $1990<t<2060$) represent time
intervals when the trajectory is near the steady states R$_1$ and S$^{\rm R1}_2$,
respectively. The inflection point near $t=2090$ shows where the trajectory
is in the vicinity of the former steady state WR. Magnetic energy
is small during the transition from the former WR to R$_1$, attaining a minimum
at $t=2110$. The symmetry $q\gamma^y_{L/2}$ is present during the transition
from R$_1$ to S$^{\rm R1}_2$. An exponential growth of the symmetry discrepancy
takes place during the departure from S$^{\rm R1}_2$. The symmetry $s_2$ is
slightly broken near the former WR.

The discrepancy of the symmetry $s_2$ is shown for an axis changing
position in time (the position is optimised to minimise the discrepancy).
The $y$ coordinate of the moving axis is plotted on \xrf{fig:hetero} (c); the
displacement in the $x$ direction is by several orders of magnitude smaller,
and it is described by a function of a similar shape.
The motion of the axis can be regarded as a drift of the cycle P$^{\rm R1}_3$
along a group orbit. In accordance with the theory of Krupa (1990), the axis is
steady when the symmetry $s_2$ is present; it moves, when the discrepancy
is maximal (for instance, at the time interval $2140<t<2160$).

For $Ta=719$ a trajectory starting near S$^{\rm R1}_2$ visits the following
steady states: $\rm S^{R1}_2\to WR\to R_D\to S^{RD}_3\to S^{RD}_2$.

\begin{figure}
\hspace{2cm}\raisebox{1cm}{(a)}\hspace{12.1mm}\psfig{file=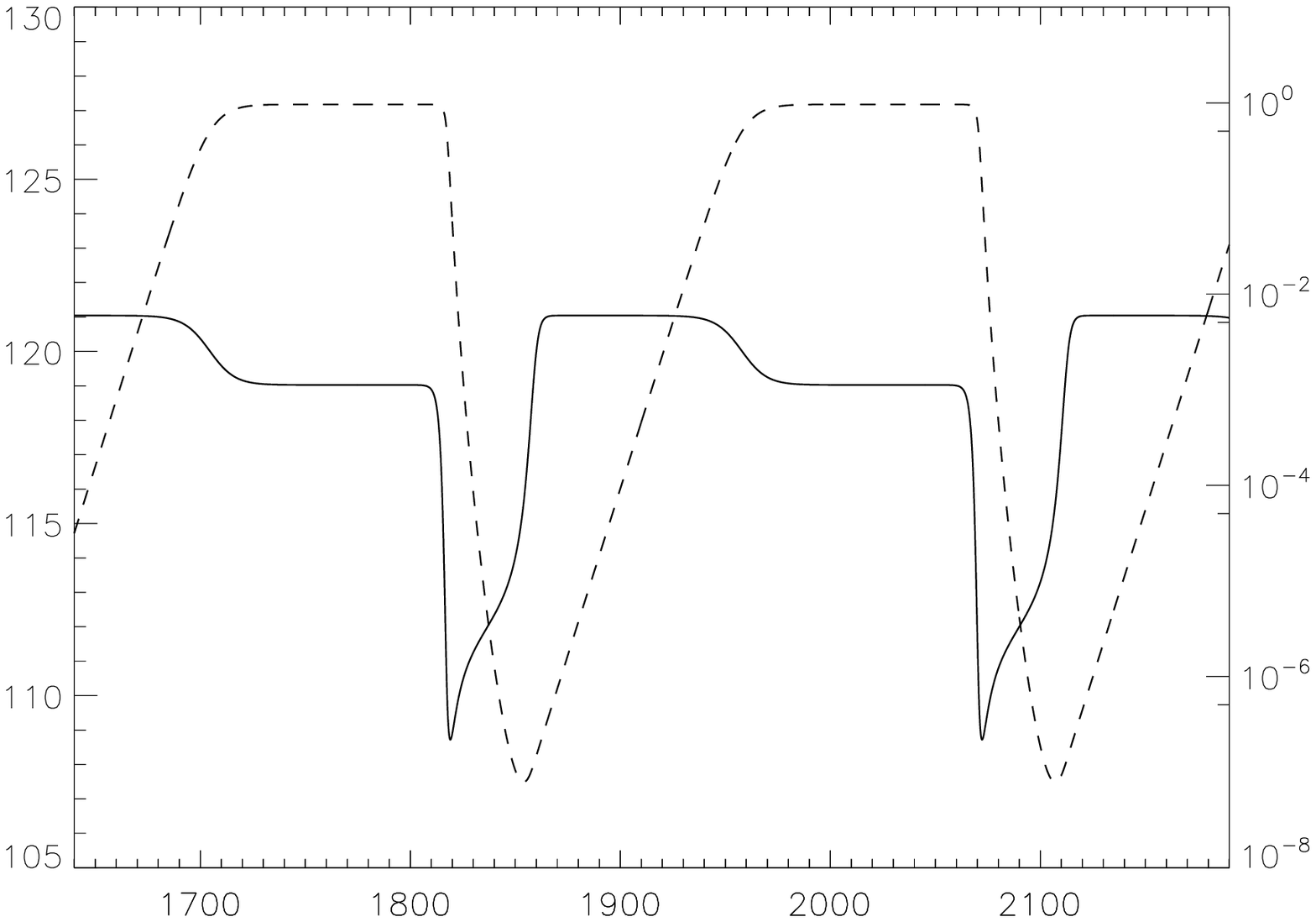,width=106.3mm,height=6cm}

\vspace*{2mm}
\hspace{2cm}\raisebox{1cm}{(b)}\hspace{11mm}\psfig{file=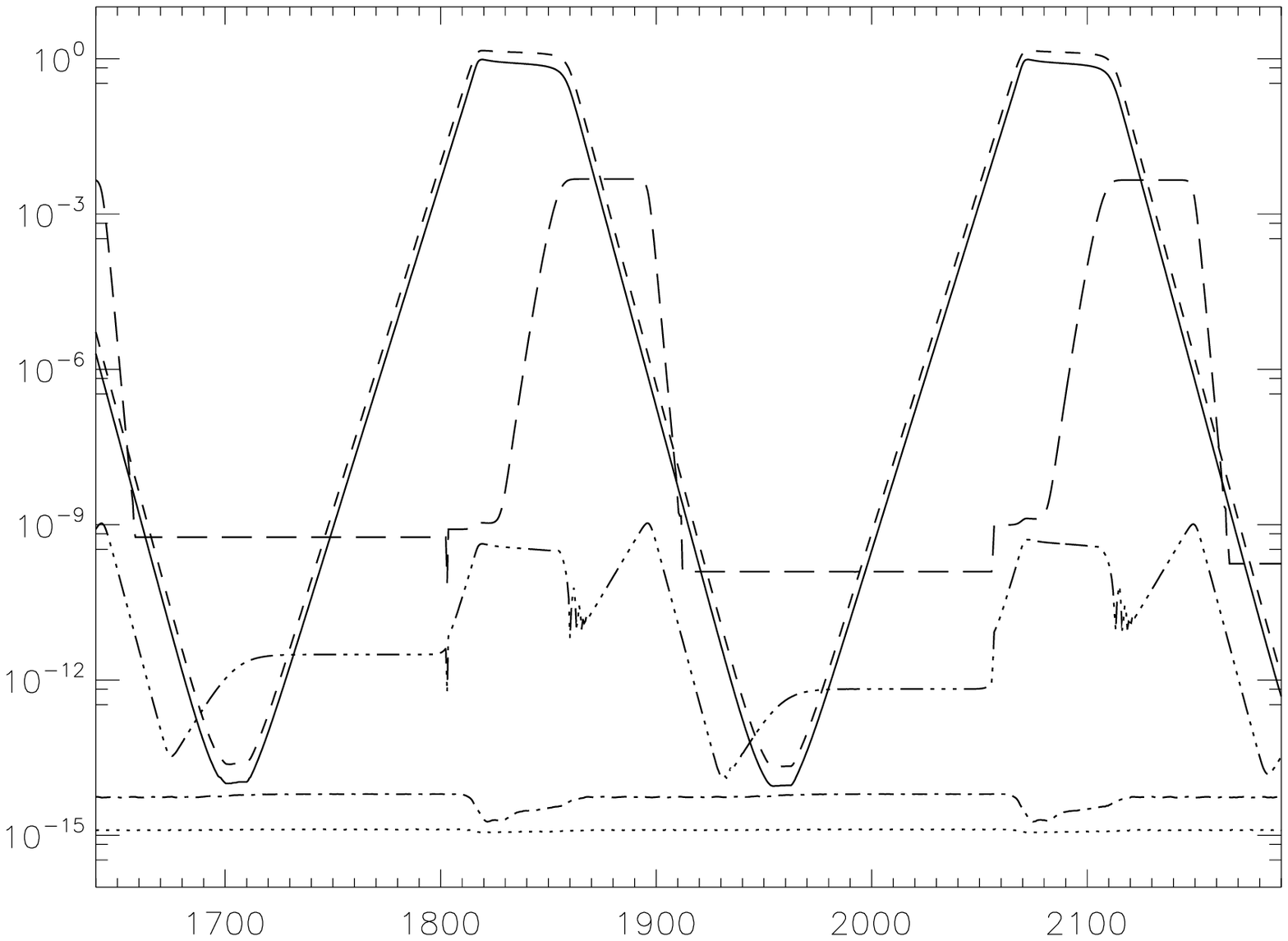,width=102.5mm,height=6cm}

\vspace*{2mm}
\hspace{2cm}\raisebox{1cm}{(c)}\hspace{9.5mm}\psfig{file=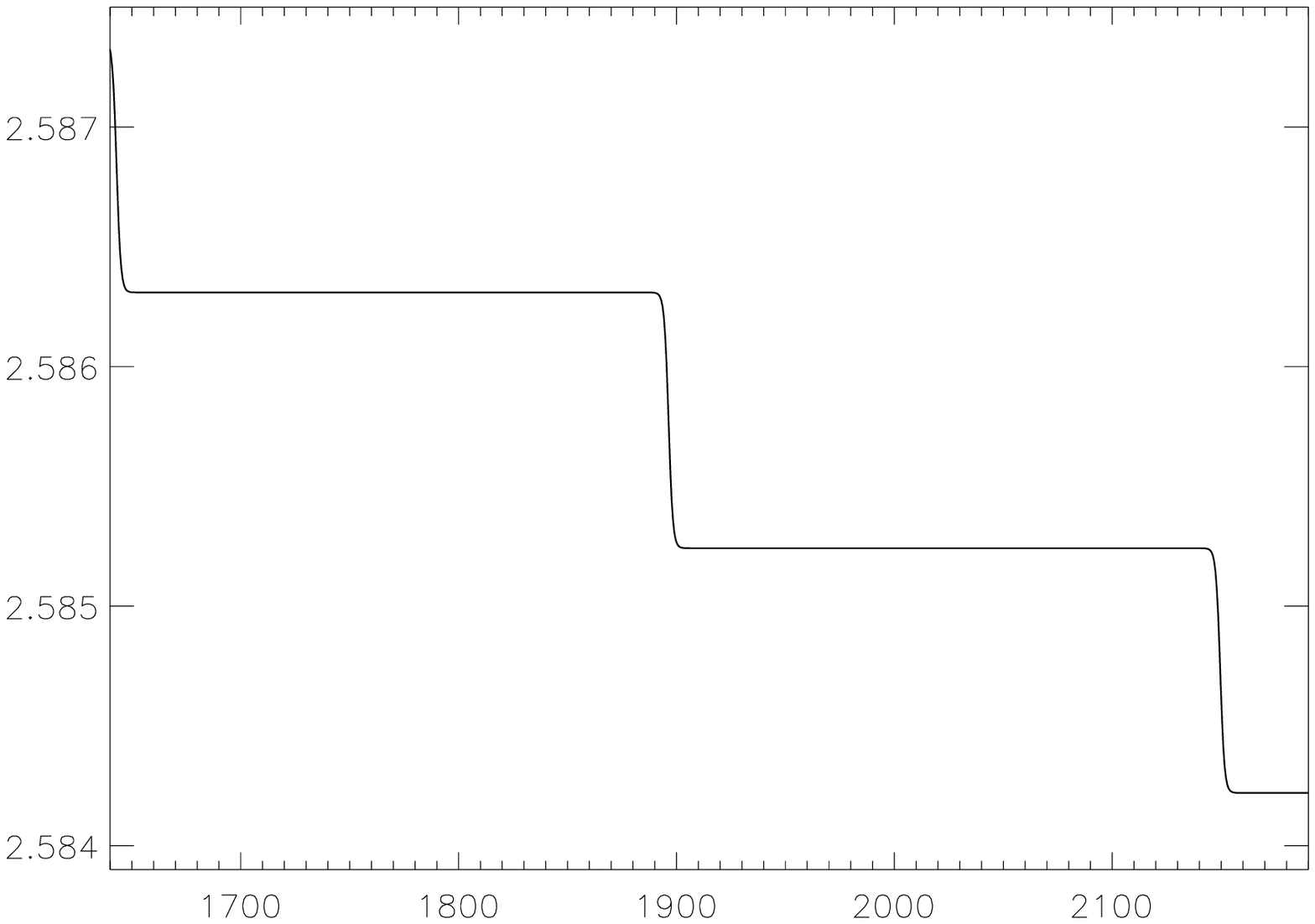,width=103mm,height=6cm}
\caption{Kinetic (solid line) and magnetic (dashed line) energies (a),
symmetry discrepancies (b) and the $y$ coordinate of the axis of the symmetry
$s_2$ (c) versus time (horizontal axis) for $Ta=717.9$, P$^{\rm R1}_4$.
On (b), the following symmetry discrepancies are traced: in the flow,
$\gamma^y_{L/2}$ (solid line), $r\gamma^x_{L/2}$ (dotted line), $s_2$
(dash-and-three-dots line); in magnetic field: $\gamma^y_{L/2}q$ (short-dash
line), $r\gamma^x_{L/2}$ (dash-and-dot line), $s_2$ (long-dash line).
Discrepancy of a symmetry $s$ in a field $\bf f$ is measured as
$\sqrt{\int|{\bf f}-s({\bf f})|^2d{\bf x}/\int|{\bf f}|^2d{\bf x}}$, where
integration over a periodicity cell is assumed.
\label{fig:hetero}}\end{figure}

\subsection{MHD attractors emerging from {\rm WR}, $721\le Ta\le 725$}

The primary steady state S$^{\rm WR}$ emerges from WR in a pitchfork
bifurcation at $Ta\approx725.71$, when the magnetic growth rate
becomes positive on decreasing $Ta$. It is stable in a short interval of $Ta$
and disappears in a saddle-node bifurcation at $Ta\approx720.25$.

\subsection{MHD attractors emerging from $\rm R_D$, $685\le Ta\le 787$}

We have found numerically that $\rm R_D$ possesses two types of dominant magnetic
modes (see Table~\ref{tab:conv_lambda} and \xrf{fig:conv_lambda}\,).
In the interval of $Ta$ where $\rm R_D$ exists there are two windows of kinematic
dynamo action, resulting in two intervals of nonlinear dynamos;
the lower one is $684\le Ta\le 787$.

The first (on increasing $Ta$) attractor S$^{\rm RD}_1$ is primary, but
it is related to a subdominant magnetic mode. The mode has the same spatial
structure as the respective nonlinear steady state displayed on
\xrf{fig:nlin_isosb} (k). The mode has the period $L/3$ along the axes
of rolls (in contrast with periods $L/2$ and $L$ for two other magnetic modes,
\xrf{fig:nlin_isosb} (m) and (n)\,). The associated eigenvalue is real and
positive for $685\le Ta\le711$, the maximum is 0.1\,.

Another primary MHD attractor is S$^{\rm RD}_3$ bifurcating from RD
in a pitchfork bifurcation when the respective eigenvalue of the
kinematic dynamo problem crosses the imaginary axis. The two branches
of primary steady states are connected by the branch S$^{\rm RD}_2$,
with a smaller symmetry group (cf.~\xrf{fig:nlin_isosb} (k), (l), (m)\,).

\subsection{MHD attractors emerging from $\rm R_D$, $1118\le Ta\le 1355$}

The third primary branch emanating from $\rm R_D$, the steady state S$^{\rm RD}_4$,
is always stable when exists. Both ends of the branch terminate on $\rm R_D$ at
$Ta=1117.47$ and $Ta=1355.20$\,.

\section{Conclusion}

Our results demonstrate that the influence of rotation on magnetic field
generation by thermal convection is non-monotonic and in no way simple.
Its nature can only be understood by a careful investigation
of attractors of the underlying dynamical system and bifurcations of branches
of convective MHD regimes. Such an investigation is made more exciting
by the presence of a large group of symmetries of the dynamical system. Even
close to the critical point for the onset of convective motion, in the terms
of the Taylor number, the bifurcation diagram is quite complex. There are
several intervals of coexistence of two or three convective MHD attractors
(also sometimes with a non-generating hydrodynamic one).

An overall picture that we have obtained fits well the usual beliefs, that up
to a point an increase of the rate of rotation from zero benefits magnetic
field generation both in linear and nonlinear regimes, and after attaining
a maximum of the mean magnetic energy (in our simulations, $E_m=32.30$ in the
regime S$^{\rm R1}_2$ for $Ta=216$) on a further increase of the rotation rate
$E_m$ gradually falls off till magnetic field generation ceases, and at still
higher angular velocities the fluid flow is arrested. However, the presence
of many individual branches and windows of nonlinear dynamo action (e.g.,
between the MHD steady states S$^{\rm RD}_3$ and S$^{\rm RD}_4$) adds a rich
``small-scale structure'' to this otherwise ``smooth'' general picture, on
some intervals of $Ta$ even reversing it. For instance, an increase
of the rate of rotation from zero inhibits magnetic field generation
in the periodic regime P$^{\rm TW}$, rather than enhances it.

Apparently the conjecture, that increasing the Taylor number first helps,
afterwards hinders and finally halts magnetic field generation, has not yet
been proved, but it appears to be correct for all parameter values (provided
the Rayleigh number is sufficiently large). In the absence of magnetic field
the critical Rayleigh number for the onset of convection grows with the Taylor
number (see Chandrasekhar, 1961); thus, in the hydrodynamic case fast rotation
is guaranteed to arrest convective flows. If a magnetic field is present,
the system is more complex and various phenomena --- such as subcritical loss
of stability of the basic state with the fluid at rest --- are not ruled out
for sufficiently strong initial magnetic fields; this can result in extension
of the interval of the Taylor number, for which convective MHD states persist
for the given other parameter values. However, it is unlikely that
this interval becomes infinite even for $P_m\gg P$, since --- in physical terms
--- magnetic field generation relies on advection by the fluid,
and not vice versa. This question awaits further investigation.

We have observed a number of interesting bifurcations. To the best of our
knowledge, two global bifurcations were not observed before. They are
similar to the SNIC (saddle-node on invariant circle) bifurcation, and
their peculiarity stems from the presence of a non-trivial symmetry group
in the convective MHD system. In these bifurcations, at $Ta\approx506.07$
and $Ta\approx718.16$, a periodic orbit terminates not on a homoclinic
(as in the SNIC), but a heteroclinic cycle, whose existence relies
on the presence of the symmetry group. Among more common, albeit rather seldom
observed in natural systems families, that we have encountered, an incomplete
Feigenbaum sequence of ``period doubling bifurcations of a torus'',
F$^{\rm TW}$ (occurring between $Ta=57$ and 80) is notable, as well as
an intermittent sequence of chaotic and quasiperiodic regimes, C$^{\rm SW}$
(occurring for $78\le Ta\le 126$), in which existence of the quasiperiodic
regimes can be apparently linked with frequency locking.

Another unusual set of bifurcations observed on the interval $216<Ta<224$ is
associated with a pair of complex eigenvalues of the linearisation of
the dynamical system, which are simultaneously vanishing. Numerical examination
of eigenvalues of the supposed trivial steady state S$^{\rm R1}_2$ suggests
that the observed bifurcations in our convective MHD system
are not linked with a Takens-Bogdanov bifurcation occurring for nearby parameter
values, but we do not rule out a finite perturbation of a bifurcation of this
type. We plan to perform a further investigation of the bifurcation in order
to gain a mathematical understanding of its nature.

We have found a number of branches of MHD steady states, which are
parity-invariant (i.e. have the symmetry $rs_2$) or possessing
the symmetry about a vertical axis, $s_2$. Certain periodic orbits ($Ta=57$,
the family F$^{\rm TW}$; $Ta=123$, the family C$^{\rm SW}$; $216.13\le Ta\le223.3$,
branch P$^{\rm R1}_1$) are not pointwise symmetric, but symmetric
on average, i.e. the symmetry with a time shift by a half of temporal period
is present. Possession of such a symmetry is an important property of attractors,
since in convective MHD systems with this symmetry the global $\alpha-$effect
is zero. Whilst in the presence of the global $\alpha-$effect
the system is inherently unstable to large-scale perturbations, in its absence
the instability, when present, develops on time scales of a larger order.
In the latter case it is described by a highly complex nonlinear mixed system
of PDE's of the second and third order (see Zheligovsky, 2009b), incorporating
such physical effects, as combined eddy diffusivity and eddy advection. We are
planning to follow this line of research.

For the chosen values of the Rayleigh and magnetic Prandtl numbers, which are
not particularly high, the convective system is not far from the onset of
convection and magnetic field generation, and the collection of hydrodynamic
and magnetic structures that we encounter is not rich (in the MHD regimes
the flows take the form of perturbed rolls, and magnetic field concentrates
in half-ropes located near the boundaries). More complex structures
are produced in a more vigorous convection for higher values of
the Rayleigh and the magnetic Prandtl numbers.

A natural extension of our investigation consists of continuation of branches
of regimes, that we have found, in other parameters, which would
significantly contribute for gaining an insight into the mathematics and
physics of convective dynamos (i.e., of the geometry of branches of attractors
in the parameter space). This project is, obviously, quite resource demanding,
but in our opinion this approach is more promising for the deeper understanding
of convective MHD systems, than analysis of individual runs for extreme
parameter values.

\section*{Acknowledgements}

RC is supported by ``Funda\c{c}\~ao para a Ci\^encia e a Tecnologia''
(Portugal), grant SFRH/BD/23161/2005. Part of this research was
carried out during the visits of OP and VZ to the School of Engineering,
Computer Science and Mathematics, University of Exeter (UK) in January -- April
2008. OP and VZ are grateful to the Royal Society for their financial support.
Research visits of OP and VZ to Observatoire de la C\^ote d'Azur (France)
in the autumns of 2007 and 2008 were supported by the French Ministry of
Education. OP and VZ are partially supported by the grants ANR-07-BLAN-0235
OTARIE from Agence nationale de la recherche (France) and
07-01-92217-CNRSL{\Large\_}a from the Russian foundation for basic research.
Computations were carried out on the computer ``M\'esocentre SIGAMM
(Simulations Interactives et Visualisation en G\'eophysique, Astronomie,
Math\'ematique et M\'ecanique)'' hosted by Observatoire de la C\^ote d'Azur,
and on the supercomputer MILIPEIA at the Laboratory for Advanced Computing at
the University of Coimbra (Portugal).

\end{document}